\documentclass[12pt]{article}
\usepackage{amsmath, bm, physics, multirow, BOONDOX-cal,mathrsfs, float,tikz, tocloft,pdflscape, nccmath}
\usetikzlibrary{positioning}
\usepackage{amssymb}
\usepackage{lscape}

\usepackage{xr}
\DeclareMathOperator*{\argmin}{\arg\!\min}

\DeclareMathOperator*{\plim}{p\!\lim}

\usepackage{graphicx,psfrag,epsf}
\usepackage{enumerate}
\usepackage{natbib}
\usepackage{url} 
\usepackage{amsfonts}

\newtheorem{corollary}{Corollary}

\newtheorem{lemma}{Lemma}
\newtheorem{proposition}{Proposition}

\newtheorem{theorem}{Theorem}
\newtheorem{assumption}{Assumption}
\newtheorem{example}{Example}

\numberwithin{corollary}{section}
\numberwithin{definition}{section}
\numberwithin{equation}{section}
\numberwithin{lemma}{section}
\numberwithin{proposition}{section}
\numberwithin{remark}{section}
\numberwithin{theorem}{section}

\begin{document}

\def\spacingset#1{\renewcommand{\baselinestretch}%
{#1}\small\normalsize} \spacingset{1}

 \begin{titlepage}
   \begin{center}
{\Large \textbf{Higher-order Expansions and Inference\\for Panel Data Models}}

    \bigskip
    
$^{\ast}${\sc Jiti Gao} and $^{\ast}${\sc Bin Peng} and $^{\ddag}${\sc Yayi Yan}\footnote{Emails: \url{jiti.gao@monash.edu}; \url{bin.peng@monash.edu}; \url{yanyayi@mail.shufe.edu.cn}.}
\bigskip

$^{\ast}$Department of Econometrics and Business Statistics,  Monash University\\
\medskip   
    
$^{\dag}$School of Statistics and Management, Shanghai University of Finance and Economics
    
 \bigskip
    
\today
\end{center}
  \medskip

\begin{abstract}
In this paper, we propose a simple inferential method for a wide class of panel data models with a focus on such cases that have both serial correlation and cross-sectional dependence. In order to establish an asymptotic theory to support the inferential method, we develop some new and useful higher-order expansions, such as Berry-Esseen bound and Edgeworth Expansion, under a set of simple and general conditions. We further demonstrate the usefulness of these theoretical results by explicitly investigating a panel data model with interactive effects which nests many traditional panel data models as special cases. Finally, we show the superiority of our approach over several natural competitors using extensive numerical studies.

\medskip


\medskip


\end{abstract}
\end{titlepage}

\spacingset{1.15} 

\section{Introduction}\label{Section1}

As we are embracing the era of data rich environment, the literature of  panel data modelling starts shifting its focus to large $N$ and $T$ cases, and then nicely falls in the category of high-dimensional data analyses. Mathematically, we may denote a panel dataset as
\begin{eqnarray}\label{eq.1.1}
	\{u_{it}\, | \,  i\in [N] , \  t\in [T]\}\quad \text{or}\quad \{U_t=(u_{1t},\ldots, u_{NT})^\top  \, | \,  t\in [T]\},
\end{eqnarray}
in which $u_{it}$'s are scalers, and $[L] =\{1, 2,\ldots, L\}$ for a positive integer $L$. With a special focus, panel data modelling aims to capture the dependence along both dimensions of $u_{it}$ (e.g., \citealp{HG2006,Petersen2009}). However, how to establish valid inference has not been satisfactorily addressed to the best of our knowledge. In this article, we aim to contribute along this line of research. To formulate our concern, suppose that
\begin{eqnarray}\label{eq.1.2}
	E[u_{it}]=0 \quad \text{and}\quad E[u_{it}u_{js}]=\alpha_{ij,ts},
\end{eqnarray}
where $\sum_{i,j=1}^{N} \sum_{t,s=1}^T |\alpha_{ij,ts}| =O(\mathbb{N})$ with $\mathbb{N}=NT$ for short.  Note that \eqref{eq.1.2} does not only  allow for correlation over both dimensions of $u_{it}$, but also permits heteroskedasticity. On this matter, Figure \ref{fig1} provides a graphical representation:
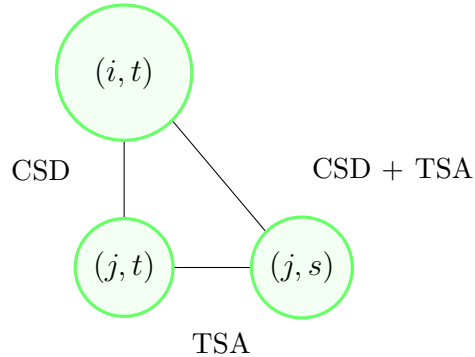
\begin{figure}[H]
	\centering
	\begin{tikzpicture}[
		roundnode/.style={circle, draw=green!60, fill=green!5, very thick},
		]
		\node[roundnode, minimum size=0.5cm]      (maintopic)          {$(j, t)$};
		\node[roundnode, minimum size=1.8cm]        (uppercircle)       [above=of maintopic] {$(i, t)$};
		\node[roundnode, minimum size=0.5cm]      (rightcircle)       [right=of maintopic] {$(j, s)$};
		\node[text width=1cm] at (-1,1.3) {\small CSD};
		\node[text width=1cm] at (1.4,-1) {\small TSA};
		\node[text width=3cm] at (4,1.3) {\small CSD + TSA};
		
		\draw[-] (uppercircle.south) -- (maintopic.north);
		\draw[-] (maintopic.east) -- (rightcircle.west);
		\draw[-] (uppercircle.south east) -- (rightcircle.north west);
	\end{tikzpicture}
	\caption{\small Heteroskedasticity and Dependence along the Two Dimensions} \label{fig1}
\end{figure}
\noindent Here, CSD and TSA stand for cross-sectional dependence and time series autocorrelation respectively. In addition, one usually assumes the following Central Limit Theorem (CLT) holds:
\begin{eqnarray}\label{eq.1.3}
	S_{\mathbb{N}}=\frac{1}{\sqrt{\mathbb{N}}}\sum_{i=1}^N \sum_{t=1}^T u_{it} =\frac{1}{\sqrt{\mathbb{N}}}\sum_{t=1}^TU_{t}^\top 1_{N}\to_D N(0,\sigma_u^2),
\end{eqnarray}
where $\sigma_u^2 =\lim_{(N, T)\rightarrow (\infty, \infty)}\frac{1}{\mathbb{N}}\sum_{t,s=1}^T 1_{N}^\top E[U_{t} U_{s}^\top] 1_{N}>0$, and  $1_{N}$ is a $N\times 1$ vector of ones. The conditions \eqref{eq.1.2} and \eqref{eq.1.3} are typical assumptions in the existing literature of panel studies (e.g., Chapter 2 of \citealp{HG2006}; Chapter 3 of \citealp{FLW2011}; Assumptions C and E of \citealp{Bai}). However, some fundamental questions have been left unanswered, e.g., (i) how to ensure \eqref{eq.1.2} and \eqref{eq.1.3} from a set of primitive conditions  in view of the presence of correlation over both dimensions of $u_{it}$ ? (ii) does the convergence of  \eqref{eq.1.3} achieve the usual Berry-Esseen bound ? and (iii) is the quantity $\sigma_u^2$ estimable ?

Among the aforementioned questions, the estimation of $\sigma_u^2$ is especially important for practical applications. Similar concerns have also been raised in the field of financial studies more than a decade ago. For example, \cite{Petersen2009} writes ``\textit{Although the literature has used an assortment of methods to estimate standard errors in panel data sets, the chosen method is often incorrect and the literature provides little guidance to researchers as to which method should be used. In addition, some of the advice in the literature is simply wrong}." Over the past couple of decades, although a variety of  panel data models have been investigated, not much work has been done to improve the estimation of a quantity like $\sigma_u^2$. More often than not, one has to assume independence along at least one dimension of the dataset (e.g., Assumption 2 of \citealp{Pesaran2006}, Proposition 2 of \citealp{Bai}, Assumption 2.1 of \citealp{Menzel2021}, among others).  

The studies sharing a similar concern with ours are \cite{goncalves_2011} and \cite{BAI2020}. Specifically,  \cite{goncalves_2011} studies a fixed effect panel data model, and proposes using the moving blocks bootstrap (MBB) technique, which allows for the error terms to have  correlation over both dimensions. However, how to select the optimal block size is left unanswered,  and to what extent CSD can be allowed is stated vaguely using some high level conditions (Assumptions 3 and 4 of the paper).  \cite{BAI2020} consider a setting similar to \cite{goncalves_2011}, and propose using a combination of the heteroskedasticity and autocorrelation consistent (HAC) covariance estimation approach and the thresholding technique, which then involves two tuning parameters --- one is the bandwidth of the HAC approach, and the other one is the threshold for penalizing entries of the covariance matrix. Such a procedure can be computationally expensive, and the optimal choices of the two tuning parameters are even more daunting. In addition, \cite{BAI2020} require the true covariance matrix of error terms to be sparse in a suitable sense, which is hard to justify in practical applications (\citealp{giannone2021economic}).

In this paper, we propose a simple inferential method for a wide class of panel data models, with a focus on the cases with correlation presenting in both dimensions.  We first establish some higher-order expansions, such as Berry--Esseen bound and Edgeworth Expansion, under a set of simple and general conditions, which extend the  results on time series (e.g., \citealp{Jirak16, jirak2021sharp}) to the panel data models.  These results with sufficiently fast rates validate the use of distributional approximation for valid inference in finite sample studies. Then we develop a simple dependent wild bootstrap (DWB) procedure for valid inferences in practice.  The DWB method is initially proposed by \cite{shao2010} to mimic the autocorrelation of time series, is easy to implement, and  requires only one tuning parameter. Accordingly, we establish the necessary asymptotic properties, and conduct extensive numerical studies to examine the theoretical findings. In addition, we propose a data driven procedure to guide the selection of the optimal tuning parameter. It is noteworthy that the DWB covariance estimator is identical to a panel HAC covariance estimator, which is able to consistently estimate the true covariance matrix without requiring truncating cross-sectional dimension (e.g., p. 1252 in \citealp{Bai}) or penalising a large covariance matrix (e.g., \citealp{BAI2020}).  Also, compared to the MBB, the DWB can better handle the missing values. The reason is that the MBB  shuffles the blocks randomly, which then destroys the data structure. By contrast, the DWB method preserves the original information of the dataset in a natural manner.  Last but not least, we demonstrate the usefulness of the DWB by explicitly investigating a panel data model with interactive effects which nests many traditional panel data models as special cases. As a by-product, we provide a solution to deal with bias correction and inference issues within one framework, which, to our knowledge, is the first result that has successfully addressed both issues.

The structure of the rest paper is as follows. Section \ref{Sec2} provides the setup and the methodology, and establishes the necessary asymptotic properties. In Section \ref{Sec3}, we conduct extensive simulations to examine the finite sample properties of the theoretical findings. Section \ref{Sec4} applies the proposed DWB approach to a real dataset by evaluating the aggregated mutual fund performance. Section \ref{Sec5} concludes. Proofs, secondary results in nature, and extra simulations are given in the online supplementary appendices.

Before proceeding further, we introduce some mathematical symbols which will be repeatedly used in the article. $|\cdot|$ denotes the absolute value of a scalar or the spectral norm of a matrix;  $\|\cdot\|$ denotes the Euclidean norm of a vector or the Frobenius norm of a matrix; for a matrix $A=\{a_{ij}\}$, let $|A|_1 = \max_{j}\sum_{i}|a_{ij}| $ and $|A|_\infty =\max_{i}\sum_{j}|a_{ij}|$; for a random variable $v$, let $\|v\|_q= (E|v|^q )^{1/q}$ for $q\ge 1$; $=_D$ denotes equality in distribution; $E^*[\cdot]$ and $\text{Pr}^*(\cdot)$ stand for the expectation and probability operations induced by the bootstrap procedure; $\to_P$ and $\to_D$ stand for convergence in probability and  convergence in distribution respectively; $\lfloor q\rfloor$ stands for the largest integer not larger than $q$; for two numbers $a$ and $b$, $a\asymp b$ stands for $a=O(b)$ and $b=O(a)$; for two random variables $c$ and $d$, we write $c \simeq d$ if $c/d \to_P 1$; let $\Psi(\cdot)$ and $\psi(\cdot)$ be the cumulative distribution function (CDF) and the probability density function (PDF) of the standard normal distribution respectively; $M_A  =I -A(A^\top A)^{-1}A^\top $ denotes the projection matrix for any matrix $A$ with full column rank; $1_N$ and $0_N$ stand for a $N\times 1$ vector of ones and a $N\times 1$ vector of zeros, respectively.

\section{The Setup and Asymptotic Theory}\label{Sec2}

In what follows, we present the basic framework and introduce some preliminary results without specifying any model in Section \ref{Sec2.1}, and propose the DWB method in Section \ref{Sec2.2}. In Section \ref{Sec2.3}, we demonstrate the usefulness of the newly proposed method by considering a panel date model with interactive effects. In Section \ref{Sec2.4}, we discuss how to handle the unbalanced panel dataset.

\subsection{The Setup}\label{Sec2.1}

We now focus on the dataset of \eqref{eq.1.1}. Instead of assuming \eqref{eq.1.2} and \eqref{eq.1.3}, we introduce a set of general conditions that can ensure the validity of both of them.

\begin{assumption}\label{Assumption1}
Let $\overline{U}_t = \frac{1}{\sqrt{N}}U_t^\top 1_{N}$, in which $U_t$ follows a process $U_t = g(\varepsilon_t, \varepsilon_{t-1}, \ldots)$ with $\varepsilon_t = (\varepsilon_{1t},\ldots, \varepsilon_{Nt})^\top$ being a sequence of independent and identically distributed (i.i.d.) random vectors, $E[U_t]=0_N$, and $g(\cdot)$ is a measurable function. In addition,  let $ \overline{U}_t^* =  \frac{1}{\sqrt{N}}U_t^{*\top} 1_{N}$, where  $U_t^* = g(\varepsilon_t,\ldots,\varepsilon_{1}, \varepsilon_{0}^\prime,\varepsilon_{-1}^\prime,\ldots)$ is  the coupled version of $U_t $, and $\{\varepsilon_t^\prime\}$ is an independent copy of $\{\varepsilon_t\}$. Suppose that $\sum_{t=0}^{\infty}t^2 \lambda_{t,\delta}^U < \infty$ for $\delta \ge 4$, where $\lambda_{t,\delta}^{U} = \|\overline{U}_t - \overline{U}_t^* \|_\delta$.
\end{assumption}

Assumption \ref{Assumption1} generalizes the nonlinear system discussed in \cite{Wu2005} to a panel data setting, and covers a wide range of data generating processes (DGPs) in the literature. Observe that $\lambda_{t,\delta}^{U}$ measures the dependence of $\overline{U}_t$ using the inputs $\varepsilon_0,\varepsilon_{-1},\ldots$ indicating that the cumulative impacts of $\varepsilon_0,\varepsilon_{-1},\ldots$ on future values are bounded in a suitable sense.

Below, we provide two examples to show Assumption \ref{Assumption1} is fulfilled by some well known DGPs, and the theoretical justification is given in the online appendices of the paper. 

\begin{example}\label{Exam1}
Consider a high-dimensional MA($\infty$) process of the form: $U_t = \sum_{j=0}^{\infty} B_j\varepsilon_{t-j}$, where $B_j$'s are $N\times N$ matrices. Suppose that (a). $|B_j| = O(\rho^j)$ for some $|\rho|<1$; (b). $\{\varepsilon_{it}\}$ is independent over $i$; and (c). $E|\varepsilon_{it}|^\delta <\infty$.  Then

\noindent 1.  $\{U_t\}$  fulfil Assumption \ref{Assumption1}.

\noindent 2.  Suppose further that  $|B_j|_1 = O(\rho^j)$, $|B_j|_\infty = O(\rho^j)$, and $E(\varepsilon_{it}^8)<\infty$. Then $\{U_t\}$ also satisfy Assumption C of \cite{Bai}, which regulates the correlation along both dimensions of $u_{it}$.

\end{example}
Obviously, Example \ref{Exam1}.1 nests Assumption 2 of \cite{Pesaran2006} (i.e., $u_{it}=  \sum_{j=0}^{\infty} b_{ij}\varepsilon_{i,t-j}$) as a special case, in which TSA presents, and heteroskedasticity over $i$ is allowed.  Assuming $B_j\equiv 0$ for $j \ge 1$, the CSD matters only. 

Example \ref{Exam1}.2 imposes more structure on the off-diagonal elements of $B_j$'s, as a result, both of CSD and TSA exist. Although it is not our focus, it is worth mentioning that some DGPs for network models (such as \citealp{zhu2017network}) are also covered by Assumption \ref{Assumption1}. Intuitively, the condition on $|B_j|_{\infty}$ implies that the cumulative impacts of each $\varepsilon_{it}$ on $\{u_{1t},\ldots,u_{Nt}\}$ is bounded, which is consistent with the notion of weak CSD, while $|B_j|_1$ suggests that the cumulative impacts of $\{\varepsilon_{1t},\ldots,\varepsilon_{Nt}\}$ on each $u_{it}$ should be bounded.

\begin{example}\label{Exam2}
Consider a high-dimensional GARCH process: $U_t = \Omega^{1/2}V_{t}$, where $|\Omega|$ is bounded, $V_{t}=\left(v_{1t},\ldots,v_{Nt}\right)^\top$, $v_{it} = h_{it}^{1/2}\varepsilon_{it}$, and $h_{it} = c_i + \sum_{j=1}^{p}C_{ij}v_{i,t-j}^2 + \sum_{j=1}^{q} D_{ij}h_{i,t-j}$. For $\forall i\in [N]$, let (a). $c_i > 0, C_{i1},\ldots,C_{ip},D_{i1},\ldots,D_{iq}\geq 0$; (b). $\sum_{j=1}^{r}\|C_{ij} + D_{ij}\varepsilon_{i,0}^2\|_{\delta/2} < 1$ with $r = \max\{p,q\}$ for some $\delta \geq 4$.
Then

\noindent 1.  $\{U_t\}$  fulfil Assumption \ref{Assumption1}.

\noindent 2. Suppose further that  $|\Omega|_1 = O(1)$, $|\Omega|_\infty = O(1)$, and $E(\varepsilon_{it}^8)<\infty$. Then $\{U_t\}$ also satisfy Assumption C of \cite{Bai}, which regulates the correlation along both dimensions of $u_{it}$.
\end{example}
Example \ref{Exam2} presents an example involving the second moment induced by $\{U_t\}$, and infers that our framework can also be used to study those models focusing on conditional heteroskedasticity.

In fact, we can further prove that $U_t\otimes V_t=(U_{ti} V_{tj})_{i,j\le N}$ satisfies Assumption 1 provided that both $U_t$ and $V_t$ satisfy Assumption 1. By Minkowski inequality and Cauchy-Schwarz inequality, we have
\begin{eqnarray*}
	&& \left\|\frac{1}{N}1_{N^2}^\top (U_t\otimes V_t) - \frac{1}{N}1_{N^2}^\top (U_t^*\otimes V_t^*)\right\|_{\delta/2}\\
	&\leq&\|\overline{U}_t - \overline{U}_t^*\|_\delta \|\overline{V}_t\|_\delta + \|\overline{U}_t^*\|_\delta \|\overline{V}_t - \overline{V}_t^*\|_\delta,
\end{eqnarray*}
which implies that $\sum_{t=1}^{\infty}t^2\left\|\frac{1}{N}1_{N^2}^\top (U_t\otimes V_t) - \frac{1}{N}1_{N^2}^\top (U_t^*\otimes V_t^*)\right\|_{\delta/2} < \infty$. In connection with Examples \ref{Exam1} and \ref{Exam2}, we can conclude that squared linear processes, squared high dimensional GARCH processes, and their cross-product processes satisfy Assumption \ref{Assumption1}, it therefore demonstrates the generality of our framework.

\medskip

We are now ready to present the first theorem of this article.

\begin{theorem}\label{Theo2.1}
Under Assumption \ref{Assumption1}, as $(N,T)\to (\infty,\infty)$,

\noindent 1. $S_{\mathbb{N}} \to_D N(0, \sigma_u^2)$,

\noindent 2. $ \sup_{w\in \mathbb{R}}|\Pr(S_{\mathbb{N}} \le  w) - \Psi_{\mathbb{N}}(w)| = O(T^{-1}(\log T)^5)$,

\noindent where $ \Psi_{\mathbb{N}}(w) = \Psi\left(\frac{w}{s_{\mathbb{N}}}\right)+\frac{1}{6}\kappa_{\mathbb{N}}^3\left(1-\frac{w^2}{s_{\mathbb{N}}^2}\right)\psi \left(\frac{w}{s_{\mathbb{N}}}\right)$,  $s_{\mathbb{N}}^2 = E[S_{\mathbb{N}}^2]$, and $\kappa_{\mathbb{N}}^3 = E[S_{\mathbb{N}}^3]$.
\end{theorem}
The first result of Theorem \ref{Theo2.1} infers that the CLT holds under Assumption \ref{Assumption1}, while the second result presents Edgeworth Expansion up to the second order. By Lemma \ref{L1}.3 of the online Appendix B (i.e., $E[S_{\mathbb{N}}^3] = O(1/\sqrt{T})$), we can simplify the second result to obtain Berry-Esseen bound:
\begin{eqnarray}\label{Eq.2.2}
	\sup_{w\in \mathbb{R}}\left|\Pr(S_{\mathbb{N}} \le  w) -\Psi\left(\frac{w}{s_{\mathbb{N}}}\right)\right| = O\left(\frac{1}{\sqrt{T}}\right).
\end{eqnarray}

By imposing more specific structures on the CSD, we can show $E[S_{\mathbb{N}}^3] = O(1/\sqrt{NT})$ and thus the rate of Berry-Esseen bound is given as follows.

\begin{corollary}\label{Cor2.1}
Suppose that (a). the conditions of Example \ref{Exam1}.2 hold;  or (b). the conditions of Example \ref{Exam2}.2 hold. Then 
\begin{eqnarray}\label{Eq.2.3}
\sup_{w\in \mathbb{R}}\left|\Pr(S_{\mathbb{N}} \le  w) -\Psi\left(\frac{w}{s_{\mathbb{N}}}\right)\right| = O\left(\max\left\{\frac{1}{\sqrt{NT}}, \frac{(\log T)^5}{T} \right\}\right).
\end{eqnarray}
\end{corollary}
Looking at \eqref{Eq.2.2} and \eqref{Eq.2.3}, in order to recover the distribution of $\Pr(S_{\mathbb{N}} \le  w)$, one will need only to focus on the population quantity $E[S_{\mathbb{N}}^2]$.  What's more, these results with sufficiently fast rates validate the use of the distributional approximation for valid inference in finite sample studies. In practice, the limiting distribution may not be useful if the approximation rate such as those involved in \eqref{Eq.2.2} and \eqref{Eq.2.3} is extremely slow (e.g., \citealp{zhou2010simultaneous}). 

Up to this point, we have fully demonstrated the generality and applicability of Assumption \ref{Assumption1}.

\subsection{The DWB Method}\label{Sec2.2}

In this subsection, we recover the asymptotic distribution $N(0,\sigma_u^2)$ of Theorem \ref{Theo2.1} using the DWB approach. Specifically, for each bootstrap replication, we draw an $\ell$-dependent time series $\{\xi_t\, |\, t\in [T]\} $ satisfying the following condition.

\begin{assumption}\label{Assumption2}
	Let $E[\xi_t ]=0$, $E|\xi_t |^2 =1$, $E|\xi_t|^4 <\infty$, $E[\xi_t \xi_s] =a (\frac{t-s}{\ell} )$,
	where $(\frac{1}{\ell}, \frac{\ell}{T})\to (0, 0)$ as $(\ell, T)\rightarrow (\infty, \infty)$, and $a(\cdot)$ is a symmetric kernel function defined on $[-1,1]$ satisfying that $a(\cdot)$ is Lipschitz continuous on $[-1,1]$, $a(0)=1$, and $K_a(x)=\int_{-\infty}^{\infty}a(u)e^{-iux}\, du \geq 0$ for $ x\in \mathbb{R}$. 
\end{assumption}

The condition of $ K_a(x)$ ensures the semi-positive definiteness of the covariance matrix of $\{\xi_t\} $, while the restrictions on $a(\cdot)$ are satisfied by a number of commonly used kernels, such as the Bartlett and Parzen kernels. In practice, one may generate $\xi =(\xi _1,\ldots, \xi _T)^\top$ using $ N(0_T, \Sigma_{\xi})$ with $\Sigma_{\xi}= \{a\left(\frac{t-s}{\ell} \right) \}_{T\times T}$, while the normal distribution is not required in theory.

Accordingly, the bootstrap version of $S_{\mathbb{N}}$ is constructed as follows:
\begin{eqnarray}\label{Eq.2.4}
	S_{\mathbb{N}}^* = \frac{1}{\sqrt{\mathbb{N}}}\sum_{t=1}^T U_t^\top 1_{N}\xi_t ,
\end{eqnarray}
which, in connection with Assumptions \ref{Assumption1} and \ref{Assumption2}, yields the following theorem of the paper.

\begin{theorem}\label{Theo2.2}
Under Assumptions \ref{Assumption1} and \ref{Assumption2}, as $(N,T)\to (\infty,\infty)$, we have

\noindent 1. $\sup_{w\in \mathbb{R}}\left| \text{\normalfont Pr}^*(S_{\mathbb{N}}^* \le w) - \Pr(S_{\mathbb{N}}\le w) \right| = o_P(1)$,

\noindent 2. $\sup_{w\in \mathbb{R}}\left| \text{\normalfont Pr}^*(S_{\mathbb{N}}^* \le w) - \Psi\left(\frac{w}{s_{\mathbb{N}}^*}\right) \right| = O_P\left(\sqrt{\frac{\ell}{T}}\right)$ with $s_{\mathbb{N}}^{*2} = E^*[S_{\mathbb{N}}^{*2}]$,

\noindent 3. $\sup_{w\in \mathbb{R}^+}\left| \text{\normalfont Pr}^*(|S_{\mathbb{N}}^*| \le w) - \Pr(|S_{\mathbb{N}}| \le  w) -  2\left(\Psi\left(\frac{w}{s_{\mathbb{N}}^*}\right) -\Psi\left(\frac{w}{s_{\mathbb{N}}}\right)\right) \right| = O_P\left(\frac{\ell}{T}\right)$.
\end{theorem}


The first result of Theorem \ref{Theo2.2} shows that  the bootstrap procedure can fully recover the asymptotic distribution  $N(0,\sigma_u^2)$ of Theorem \ref{Theo2.1}. As a consequence, one can  use the bootstrap draws to establish the corresponding confidence interval while allowing for both of CSD and TSA.  

The second result of Theorem \ref{Theo2.2} establishes the Berry--Esseen bound for the DWB procedure. To the best of our knowledge, the rate $\sqrt{\frac{\ell}{T}}$ is optimal and cannot be further improved (\citealp{shergin1980convergence}). This is because $\{\xi_t\}_{t=1}^T$ is a sequence of strongly $\ell$-dependent random variables (e.g.,  $E^*(\xi_{t}\xi_{t+\lfloor \ell/2\rfloor}) = a(1/2)$ as $\ell\to \infty$). As a result, the approximation rate of normal distribution is of the same order of $|s_{\mathbb{N}}^{*2} - s_{\mathbb{N}}^{2}|$ (see Theorem 2.3 below). 

In the third result, by utilizing the fact that the first term of Edgeworth expansion is an even function of $w$, we provide a faster rate than that given in the second result. It then sheds light on how to select the optimal $\ell$.  To see this point, note that the bootstrap draws offer a sample version of the form:
\begin{eqnarray}\label{Eq.2.5}
	E^*[S_{\mathbb{N}}^{*2}] =\frac{1}{T}\sum_{t,s=1}^T\overline{U}_t\overline{U}_s a \left(\frac{t-s}{\ell} \right)
\end{eqnarray}
to consistently estimate $E[S_{\mathbb{N}}^2]$.  Therefore, to select the optimal $\ell$ below, we minimise the mean squared error (MSE) between $E^*[S_{\mathbb{N}}^{*2}] $ and $E[S_{\mathbb{N}}^2]$, which   is equivalent to  minimise the MSE of coverage rates of the bootstrap confidence intervals\footnote{This point can be seen by setting $w = q_{\alpha}^*$, where $q_{\alpha}^*$ denotes the $\alpha$-th quantile of $|S_{\mathbb{N}}^*|$ such that $\text{\normalfont Pr}^*(|S_{\mathbb{N}}^*| \leq q_{\alpha}^*) = \alpha$. Hence, we have $\Pr(|S_{\mathbb{N}}| \le  q_{\alpha}^*) = \alpha -  \Psi\left(\frac{q_{\alpha}^*}{s_{\mathbb{N}}}\right)\frac{q_{\alpha}^*}{s_{\mathbb{N}}}(s_{\mathbb{N}}^{*2}-s_{\mathbb{N}}^2) + O_P(|s_{\mathbb{N}}^{*2}-s_{\mathbb{N}}^2|^2)+O_P(\ell/T)$.} according to the third result of Theorem 2.2, and  is also a widely adopted criterion in the literature of HAC method (e.g., \citealp{Andrews1991}). 

Before proceeding further, we impose one more condition on  $a(\cdot)$.

\begin{assumption}\label{Assumption3}
	For $q \in [2]$, suppose that $\lim_{|x|\to 0}\frac{1 - a(x)}{|x|^q} = c_q$ for some real number $0 < c_q < \infty$.
\end{assumption}

Assumption \ref{Assumption3} is standard in the literature. For example, for the Bartlett kernel, $q=1$ and $c_1= 1$; for the Parzen, Tukey-Hanning, QS kernels, and the trapezoidal functions, $q=2$ and the values of $c_2$ vary but all satisfy  $c_2<\infty$. We refer interested readers to \cite{KV2002}  and \cite{PP2001} for the properties of Bartlett kernel and trapezoidal functions respectively, and to \cite{Andrews1991} for discussions on the other kernel functions.

\begin{theorem}\label{Theo2.3}
Under Assumptions \ref{Assumption1}-\ref{Assumption3}, as $(N,T)\to (\infty,\infty)$,
	
\noindent 	{\normalfont {\bf Bias}:} $E\left(E^*[S_{\mathbb{N}}^{*2}]\right) - E[S_{\mathbb{N}}^2]= - \frac{c_q}{\ell^q} \Delta_1 + o(\ell^{-q})$,
	
\noindent	{\normalfont {\bf Variance}:} $\text{\normalfont Var}(E^*[S_{\mathbb{N}}^{*2}]) = \frac{2\ell}{T}\Delta_2 + o(\ell/T)$,
	
\noindent	where $\Delta_1= \sum_{k = -\infty}^{\infty}|k|^qE[\overline{U}_0\overline{U}_{k}]$ and $\Delta_2 =  (E[S_{\mathbb{N}}^2])^2\, \int_{-1}^{1}a^2(x)\mathrm{d}x$.
\end{theorem}
By Theorem \ref{Theo2.3}, the MSE is minimized at
\begin{eqnarray}\label{Eq.2.6}
	\ell_{\text{opt}} = \left(qc_q^2\Delta_1^2/\Delta_2\right)^{1/(2q+1)} T^{1/(2q+1)} \asymp \left\{\begin{array}{ll}
		T^{1/3}, & \text{if }q=1 \\
		T^{1/5}, & \text{if }q=2 
	\end{array} \right. .
\end{eqnarray}
Theorem \ref{Theo2.3} indicates that our DWB covariance estimator is  a panel HAC covariance estimator, which is able to consistently estimate the true covariance matrix and does not require any cross-sectional parameter truncation (e.g.,  \citealp[p. 1252]{Bai}) or regularization (e.g., \citealp{BAI2020}). 

Up to this point, no specific model has been investigated. In what follows, we specifically apply the DWB method to a panel data model with interactive effects, which has attracted considerable attention since the seminal papers of \cite{Pesaran2006} and \cite{Bai}, and nests lots of classic panel data models as special cases. Although a variety of  extensions have been published in the past decade or so, to our knowledge, no work  has successfully addressed the bias and inference issues simultaneously since Theorem 3 of \cite{Bai}. That said, we shall tackle both problems together in the next subsection.

\subsection{An Application of the DWB Method}\label{Sec2.3}

From now on, we treat $u_{it}$'s as unobservable idiosyncratic errors, and consider a specific example to demonstrate the usefulness of the DWB method:

\begin{eqnarray}\label{Eq.2.7}
Y_t = X_t\theta_0 +\Gamma_0 f_t + U_t,
\end{eqnarray}
which is initially studied in \cite{Bai}, and has been substantially extended since then (e.g., \citealp{LiQianSu, Ando}, among others). Here, only $\{Y_t, X_t \}$ are observable, while $\Gamma_0$ and $\{f_t\}$ can be correlated with $\{X_t\}$. As identifying the rank of $\Gamma_0$ is not the main focus here, we follow the aforementioned works to assume $f_t$ is a $p\times 1$ vector with $p$ being fixed and known. 

Provided dependence along both dimensions of $u_{it}$, establishing valid inference for the model \eqref{Eq.2.7} requires to tackle the following two challenging issues: (1). correct the estimation bias, and (2). estimate the asymptotic covariance matrix. Although the DWB method can address the latter one, one still needs to deal with the bias. That said, we provide a valid procedure to infer $\theta_0$ in what follows.

Consider the following objective function
\begin{eqnarray}\label{Eq.2.8}
Q_{\mathbb{N}}(\theta , \Gamma) =\sum_{t=1}^T(Y_t-X_t\theta)^\top M_\Gamma (Y_t-X_t\theta),
\end{eqnarray}
where $\Gamma$ is a generic $N\times p$ matrix and satisfies $\frac{1}{N}\Gamma^\top \Gamma =I_p $ for the purpose of identification.  Accordingly, we estimate $\theta_0$ and $\Gamma_0$ by minimizing \eqref{Eq.2.8}:
\begin{eqnarray}\label{Eq.2.9}
(\widehat{\theta},\widehat{\Gamma}) =\argmin_{\theta, \Gamma}Q_{\mathbb{N}}(\theta , \Gamma).
\end{eqnarray}
Also, we estimate $f_t$ and $U_t$ by $\widehat{f}_t=\frac{1}{N}\widehat{\Gamma}^\top (Y_t-X_t\widehat{\theta})$ and $\widehat{U}_t= Y_t-X_t\widehat{\theta}-\widehat{\Gamma}\widehat{f}_t$ respectively.

To proceed, we define a few extra notations. Let

\begin{eqnarray*}
D(\Gamma) & =& \frac{1}{\mathbb{N}}\sum_{t=1}^T \widetilde{X}_t^\top M_{\Gamma} \widetilde{X}_t \quad \text{and}\quad\Omega =\frac{1}{T}\sum_{t=1}^T U_t U_t^\top ,
\end{eqnarray*}
where $\widetilde{X}_t = X_t - \frac{1}{T}\sum_{s=1}^TX_s f_s^\top ( \frac{F^\top F}{T}  )^{-1} f_t $ and $F=(f_1,\ldots, f_T)^\top$. With these notations, we present the following lemma.

\begin{lemma}\label{Lem2.1}
Consider the model \eqref{Eq.2.7}, and suppose 

\noindent (a). $E\|x_{it} \|^4<\infty$, and $\inf_{\Gamma} D(\Gamma) >0$ with $\Gamma^\top \Gamma /N =I_p$;

\noindent (b). $\frac{1}{T}F^\top F\to_P \Sigma_F>0$ with $E\|f_t \|^4<\infty$, and $\frac{1}{N}\Gamma_0^\top \Gamma_0 \to_P\Sigma_{\Gamma}>0 $ with $E\| \gamma_{0i}\|^4<\infty$;

\noindent (c). $\{U_t \}$ is independent of $\{X_t\}$, $\Gamma_0$ and $F$,

\noindent where $x_{it}^\top$ and $\gamma_{0i}^\top $ stand for the $i^{th}$ rows of $X_t$ and $\Gamma_0$ respectively. In addition, let $\{U_t \}$ satisfy the conditions of Example \ref{Exam1}.2, and let $N/T\to \rho$ with $\rho$ being a positive constant. Then
\begin{eqnarray*}
\sqrt{\mathbb{N}}(\widehat{\theta} -\theta_0) \to_D N(\rho^{1/2}\mu_B+\rho^{-1/2}\mu_C, \Sigma_1^{-1}\Sigma_2\Sigma_1^{-1}),
\end{eqnarray*}
where $\mu_B = \plim \mu_{\mathbb{N},B}$, $\mu_C = \plim \mu_{\mathbb{N},C}$, and
\begin{eqnarray*}
\mu_{\mathbb{N},B} &=& -  D(\Gamma_0 )^{-1} \frac{1}{\mathbb{N}}\sum_{t,s=1}^T \frac{\widetilde{X}_t ^\top \Gamma_0}{N}   \left(\frac{\Gamma_0^\top \Gamma_0}{N} \right)^{-1} \left( \frac{F^\top F}{T} \right)^{-1} f_s   \sum_{i =1}^N u_{it}u_{is},\nonumber \\
\mu_{\mathbb{N},C} &=& - D(\Gamma_0)^{-1} \frac{1}{\mathbb{N}}\sum_{t=1}^T X_t^\top M_{\Gamma_0} \Omega \Gamma_0 \left(\frac{\Gamma_0^\top \Gamma_0}{N} \right)^{-1} \left(\frac{F^\top F}{T} \right)^{-1}f_t.
\end{eqnarray*}
\end{lemma}
Lemma \ref{Lem2.1} repeats Theorem 3 of \cite{Bai}, but interchanges the $i$ and $t$ dimensions. The first three conditions in the body of this lemma are equivalent to Assumptions A, B and D of \cite{Bai}, while his Assumption C has been justified by Example \ref{Exam1}.2.

\medskip

Next, we deal with the two biases before adopting the DWB method. For notational simplicity, we write

\begin{eqnarray}\label{Eq.2.10}
\widehat{\theta} -\theta_0 \simeq \frac{1}{\mathbb{N}} \sum_{t=1}^T W_t^\top U_t+ \frac{1}{T}\mu_{\mathbb{N},B}+\frac{1}{N}\mu_{\mathbb{N},C},
\end{eqnarray}
where  $W_t$ is formed by $\{X_t\}$ and $\Gamma_0$ only, and $\frac{1}{\sqrt{\mathbb{N}}} \sum_{t=1}^T W_t^\top U_t\to_D  N(0, \Sigma_1^{-1}\Sigma_2\Sigma_1^{-1})$. 

Up to this point, it is worth commenting on the two-way half panel jackknife technique of \cite{Chen2021} that  partitions sample along both dimensions. We would like to point out that, under the current context, it is probably not a good idea to split sample along the cross-sectional dimension, as it will change the structure of $\Omega$ internally in this case. Also, CSD usually depends on the ``distance" among individuals implicitly which is unknown in general, so how to split the sample along the cross-sectional dimension is unclear. On the other hand, time series is naturally ordered, so it makes more sense to work with the time dimension when splitting sample. 

We now propose the following procedure. Without loss of generality, let $T$ be an even number. For the time dimension, we create another two new sets, $S_1 = \{1,\ldots, T/2 \} $ and $S_2=  \{T/2+1,\ldots, T \}$, and define a bias corrected estimator as follows:

\begin{eqnarray*}
\widehat{\theta}_{\text{bc}} = 2\widehat{\theta} -( \widehat{\theta}_{S_1} +\widehat{\theta}_{S_2})/2,
\end{eqnarray*}
where $\widehat{\theta}_{S_1} $ and $\widehat{\theta}_{S_2}$ are obtained using sample from $[N]\otimes S_1 $ and $[N]\otimes S_2$ respectively. However, $\widehat{\theta}_{\text{bc}}$ cannot fully remove two biases, which should be expected in view of \cite{Chen2021}. By \eqref{thetabc} of the online supplementary appendix, we know that

\begin{eqnarray*}
\sqrt{\mathbb{N}} (\widehat{\theta}_{\text{bc}} -\theta_0) \simeq  \frac{1}{\sqrt{\mathbb{N}}} \sum_{t=1}^T W_t^\top U_t +\sqrt{\frac{T}{N}}\mu_{\mathbb{N},C}.
\end{eqnarray*}
Thus, we further provide the following estimator to deal with $\mu_{\mathbb{N},C}$:

\begin{eqnarray*}
\widehat{\mu}_C = - D(\widehat{\Gamma})^{-1} \frac{1}{\mathbb{N}}\sum_{t=1}^T X_t^\top M_{\widehat{\Gamma}} \widehat{\Omega}  \widehat{\Gamma} \left( \frac{\widehat{F}^\top \widehat{F}}{T}\right)^{-1} \widehat{f}_t,
\end{eqnarray*}
where $\widehat{\Omega} = \frac{1}{T}\sum_{s=1}^T (Y_s- X_s\widehat{\theta})(Y_s- X_s\widehat{\theta})^\top$. Consequently, the final form of the bias corrected estimator is given below:

\begin{eqnarray*}
\widetilde{\theta}_{\text{bc}} = \widehat{\theta}_{\text{bc}} - \frac{1}{N}\widehat{\mu}_C.
\end{eqnarray*}

\medskip

Finally, in order to infer $\theta_0$, the bootstrap procedure is as follows.

\noindent 1. For each bootstrap replication, let $Y_{t}^* =X_{t}^\top \widehat{\theta} + \widehat{\Gamma}\widehat{f}_t+\widehat{U}_t  \xi_t$.

\noindent 2. Using the bootstrap sample $\{Y_T^*, X_t \}$, calculate $\widehat{\theta}^*$:
	\begin{eqnarray*}
		\widehat{\theta}^*=\left(\sum_{t=1}^T X_t^\top M_{\widehat{\Gamma}}X_t\right)^{-1}\sum_{t=1}^T X_t^\top M_{\widehat{\Gamma}}Y_t^*.
	\end{eqnarray*}
	
\noindent 3. Repeat the first two steps $R$ times.

\medskip
 
For the above procedure, the following theorem\ holds.

\begin{theorem}\label{Theo2.4}
Suppose that the conditions of Lemma \ref{Lem2.1} and Assumption \ref{Assumption2} hold. As $(N,T)\to (\infty,\infty)$,

\noindent 1. $\sqrt{\mathbb{N}}(\widetilde{\theta}_{\normalfont\text{bc}}-\theta_0) \to_D N(0,\Sigma_1^{-1}\Sigma_2\Sigma_1^{-1})$,

\noindent 2. $\sup_w \left|\text{\normalfont Pr}^*(\sqrt{\mathbb{N}}(\widehat{\theta}^*-\widehat{\theta})  \le w) - \text{\normalfont Pr}(\sqrt{\mathbb{N}}(\widetilde{\theta}_{\normalfont\text{bc}}-\theta_0) \le w)\right| =o_P(1).$
\end{theorem}
In Theorem \ref{Theo2.4}, the first result provides an unbiased estimator for $\theta_0$, while the second result recovers the  distribution $N(0,\Sigma_1^{-1}\Sigma_2\Sigma_1^{-1})$ using the bootstrap draws. Our investigation on \eqref{Eq.2.7} is now completed.

\subsection{On Unbalanced Dataset}\label{Sec2.4}

To close our investigation on the DWB method, we consider the following unbalanced panel dataset:
\begin{eqnarray}\label{Eq.2.11} 
	\{u_{it}\, | \,  i\in [N_t] \text{ for }\forall t, \  t\in [T]\},
\end{eqnarray}
where $N_t$ varies with respect to $t$, and $\mathbb{N}=\sum_{t=1}^T N_t$. The structure of \eqref{Eq.2.11} is widely adopted in the literature (e.g., Chapter 4 of \citealp{HG2006}), and also suits the mutual fund dataset of Section \ref{Sec4}.

To accommodate the missing values, we can rewrite $\overline{U}_t$ of Assumption \ref{Assumption1} as
\begin{eqnarray}\label{Eq.2.12}
	\overline{U}_t = \frac{1}{\sqrt{N_t}}U_t^\top \mathscr{L}_t, 
\end{eqnarray}
where  $\mathscr{L}_t$ is a $N\times 1$ vector with elements being $1$ and $0$ only to represent non-missing and missing respectively. By \eqref{Eq.2.11} and \eqref{Eq.2.12}, $\|\mathscr{L}_t\|=\sqrt{N_t}$. Under some trivial modification, one can show that the established results still hold. For example, we may adopt the following condition.

\begin{assumption}\label{Assumption4}
	Suppose that  $\frac{\overline{N}T}{\mathbb{N}}\to c\in (0,\infty) $, where $\overline{N} = \max_t N_t$ and $c$ is a constant.
\end{assumption}

Assumption \ref{Assumption4} allows $\underline{N}=\min_t N_t$ to be a fixed value, however,  the number of $N_t$'s being finite has to be negligible.

\begin{corollary}\label{Coro2.2}
	Under Assumptions \ref{Assumption1}, \ref{Assumption2} and \ref{Assumption4}, as $(N,T)\to (\infty,\infty)$,
	\begin{eqnarray*}
		\sup_{w\in \mathbb{R}} \left|\text{\normalfont Pr}^*(S_{\mathbb{N}}^* \le w) - \Pr(S_{\mathbb{N}}\le w) \right| =o_P(1),
	\end{eqnarray*}
	where $S_{\mathbb{N}}^*$ and $S_{\mathbb{N}}$ are defined in an obvious manner using \eqref{Eq.2.12}.
\end{corollary}
Compared to the block bootstrap based studies, one more advantage of DWB is that it can better handle the missing values. Note that the block bootstrap shuffles the blocks randomly, as a consequence the positions of missing values will be different for each bootstrap replication. In a sense, shuffling the blocks may destroy the data structure. By contrast, the DWB method preserves the original information of the dataset much better.

Up to this point, we have finished the theoretical investigation. In the next section, we examine the theoretical results using extensive simulation studies, and compare the DWB method with some existing ones.

\section{Simulations}\label{Sec3}

In this section, we conduct simulations to validate the theoretical findings of Section \ref{Sec2}. First, we explain how to select $\ell_{\text{opt}}$ practically in Section \ref{Sec31}. Then we evaluate Theorem \ref{Theo2.2} of Section \ref{Sec2.2} and the example of Section  \ref{Sec2.3} respectively. For the sake of space, we only report some selected results below, and provide the extra simulation results in the online supplementary Appendix \ref{Sectables} of this paper.

\subsection{Numerical Implementation}\label{Sec31}

We now discuss how to calculate $\ell_{\text{opt}}$. The quantity $qc_q^2\Delta_1^2/\Delta_2$ in \eqref{Eq.2.6} in fact can be consistently estimated, so there is a data-driven $\widehat{\ell}_{\text{opt}}$. To see this, note that $c_q$ is decided by the kernel function, and is therefore known. Thus, we need only to focus on $\Delta_1$ and $\Delta_2$. 

By Theorem \ref{Theo2.3}, $\frac{1}{T}\sum_{t,s=1}^T\overline{U}_t \overline{U}_s a\left(\frac{t-s}{T^{\nu_q}} \right)\to_P  E[S_{\mathbb{N}}^2]$, where $\nu_q = 1/3$ if $q=1$, and $\nu_q =1/5$ if $q=2$ by \eqref{Eq.2.6}. Thus, $\widehat{\Delta}_2 \equiv \left(\frac{1}{T}\sum_{t,s=1}^T\overline{U}_t \overline{U}_s a\left(\frac{t-s}{T^{\nu_q}} \right) \right)^2 \int_{-1}^{1}a^2(x)\mathrm{d}x \to_P\Delta_2.$

For $\Delta_1$, let $\widehat{\Delta}_1  \equiv 2\sum_{k=1}^{Q_T}  \frac{k^q}{T}\sum_{t=1}^{T-k}\overline{U}_t\overline{U}_{t+k}$, where  $Q_T \asymp T^{2/(4q+5)} $ is a truncation parameter. Since $\text{Var}\left( \frac{1}{T}\sum_{t=1}^{T-k}\overline{U}_t\overline{U}_{t+k} - \sigma(k)\right)=O(1/T)$ by Lemma \ref{L1}.4, we require $Q_T \asymp T^{2/(4q+5)} $ to ensure $\widehat{\Delta}_1\to_P \Delta_1$.

Finally, we recommend the following data-driven bandwidth: $\widehat{\ell}_{\text{opt}} =  \widehat{\ell} \vee  \ell_{\min},$ where $\widehat{\ell}=(qc_q^2\widehat{\Delta}_1^2/\widehat{\Delta}_2 )^{1/(2q+1)}T^{1/(2q+1)}$, and $\ell_{\min} $ is a fixed value (say, $\ell_{\min} = 10$). The reason for having $\ell_{\min} $ is to boost the finite sample performance when $T$ is relatively small. Note that even for $T=200$, $T^{1/5}$ only returns 2.89, and  it is also not guaranteed that the term $(qc_q^2\widehat{\Delta}_1^2/\widehat{\Delta}_2 )^{1/(2q+1)}$ will return a value greater than 1. Therefore, to avoid an unreasonably small $ \widehat{\ell}$, we use $\ell_{\min}$ to bound $\widehat{\ell}_{\text{opt}} $ from below in the numerical implementation, which does not alter any aforementioned theoretical argument. For the model considered in Section \ref{Sec2.3}, we simply replace $\{U_t\}$ with $\{\widehat{U}_t\}$.

\subsection{Examination of Theorem \ref{Theo2.2}}\label{Sec32}

We are now ready to conduct the simulation study. The DGP is as follows: $U_t^* = \rho_u U_{t-1}^*+ \epsilon_t,$ where we consider both light tail and heavy tail behaviour for $\epsilon_t$:

\begin{center}
Case 1: $\epsilon_t \sim  \Sigma_N^{\epsilon,1/2} N(0_N , I_N)$, $\quad$ Case 2: $\epsilon_t \sim \Sigma_N^{\epsilon,1/2} (t_5,\ldots, t_5)^\top$,
\end{center}
with $ \Sigma_N^\epsilon=\{\delta_\epsilon^{|i-j|} \}_{N\times N}$, and $t_5$ stands for the $t$-distribution with a degree freedom of 5.  We let $\rho_u ,\rho_\epsilon \in\{0.25, 0.5 \}$. To introduce heteroscedasticity, we further let $\mathbb{U}_i =\sqrt{1+i/N} \widetilde{U}_i$, where $\mathbb{U}_i = (u_{i1},\ldots, u_{iT})^\top$, and $\widetilde{U}_i =(U_{i1}^*,\ldots, U_{iT}^*)^\top$ with $U_{it}^*$ being the $i^{th}$ element of $U_t^*$. Thus, $u_{it}$ has weak correlation over both dimensions, and also has heteroskedasticity over $i$. 

To implement the bootstrap procedure, $\xi_t$'s are generated in the same way as mentioned under Assumption \ref{Assumption2}. We specifically consider two kernels in the following simulations:

\noindent 1. Bartlett kernel: $\psi(w) =(1-|w|)I(|w|\le 1)$,

\noindent 2. A trapezoidal function: $a(x) = \frac{\int_{-1}^{1}w(u)w(u+|x|)\mathrm{d}u}{\int_{-1}^{1}w^2(u)\mathrm{d}u}$,

\noindent where $w(u)=\frac{u}{0.43}I\left(u\in[0,0.43)\right)+I\left(u\in[0.43,0.57]\right)+\frac{1-u}{0.43}I\left(u\in(0.57,1]\right).$

The Bartlett kernel is well adopted in the literature for its simplicity (e.g., \citealp{Andrews1991, goncalves_2011, BAI2020}; among others), while the specific form of the trapezoidal function can be seen in \cite{shao2010}. Regarding \eqref{Eq.2.6}, both kernel functions represent the cases with $q=1$ and $q=2$ respectively. The bandwidths $\ell_B$ and $\ell_T$ of the Bartlett kernel and the trapezoidal function are selected as in Section \ref{Sec31}, and, for each kernel we further consider  $0.8\ell_j$ and $1.2\ell_j$ for $j\in \{B,T\}$ to examine the sensitivity. 

For every generated dataset, we record the value $S_{\mathbb{N}}$ of \eqref{eq.1.3} and the 95\% confidence interval (CI) yielded by the 399 bootstrap draws. After $R$ replications, we report
\begin{eqnarray*}
\text{Size} = \frac{1}{R}\sum_{m=1}^R  I(S_{\mathbb{N},m}\not\in \text{CI}_m),
\end{eqnarray*}
where $S_{\mathbb{N},m}$ and $\text{CI}_m$ respectively stand for the value of $S_{\mathbb{N}}$ and the 95\% CI from the $m^{th}$ replication.

For the purpose of comparison, we first consider three traditional methods to calculate the 95\% CI. Specifically, we estimate the variances as follows:
\begin{eqnarray*}
s_1^2 = \frac{1}{\mathbb{N}}\sum_{i=1}^N\sum_{t=1}^Tu_{it}^2,\quad
s_2^2 = \frac{1}{\mathbb{N}}\sum_{i=1}^N\sum_{t,s=1}^Tu_{it}u_{is},\quad
s_3^2 = \frac{1}{\mathbb{N}}\sum_{i,j=1}^N\sum_{t=1}^Tu_{it}u_{jt},
\end{eqnarray*}
where $s_1^2$ is a consistent estimator of the variance when $u_{it}$ is independent over $(i,t)$, and $s_2^2$ and $s_3^2$ are consistent estimators of the variance provided that the observed $u_{it}$ is independent over either the cross-sectional or time dimension. 

The second method considered for comparison is the MBB method of \cite{goncalves_2011}. The block length $\ell_M$ is generated in the same way as in \cite{goncalves_2011}, so we omit the details here. We further consider $\lfloor 0.8\ell_M \rfloor$ and $\lfloor1.2\ell_M\rfloor$ to examine the sensitivity. For each dataset, we also do 399 bootstrap draws to obtain the 95\% CI.  

The third method included for comparison is the approach of \cite{BAI2020} (referred to BCL below). The implementation is identical to Section 2.1 of \cite{BAI2020}. For the two tuning parameters $L$ (for HAC) and $M$ (for penalization), we use $L=3,7,11$ and $M =0.1, 0.15, 0.2, 0.25$ as in Section 3 of their paper. We do not further provide the details of their approach as it is quite lengthy. 

We let $R=1000$. Due to space limit, we only report the results of $(\rho_u ,\rho_\epsilon) = (0.25,0.5)$ in Table \ref{Tablenew1} and Table \ref{Tablenew2} of the main text, and report the extra results in Tables \ref{Table1}-\ref{Table6} of the online supplementary appendices. For the traditional methods, the size is always greater than 5\%, which is not surprising. As all of $s_j^2$ for $j=1,2,3$ just include a proportion of the asymptotic variance, we do expect the three traditional methods will over reject. As the sample size goes up, MBB seems to converge to the nominal rate (i.e., 5\%) but slower than DWB. The BCL method tends to over reject, which might be due to the fact that many weak correlations get penalized by the thresholding method. The sizes of the DWB method are very close to the nominal one, and are quite robust in terms of the tail behaviour of $\epsilon_t$. Finally, it is noteworthy that the DWB method is not very sensitive to the choice of the kernel function, and is robust to different choices of the bandwidth.  The findings are consistent across the tables. 

\begin{center}
Insert Tables \ref{Tablenew1} and \ref{Tablenew2} about here.
\end{center}

\subsection{Examination of  Section \ref{Sec2.3}}\label{Sec33}

Having shown the superiority of the DWB, in this subsection we consider the model and the approach of Section \ref{Sec2.3}. When conducting inference, we focus on the DWB method only.  The DGP is as follows: $Y_t =X_t\theta_0 +\Gamma_0 f_t + U_t,$ where $\theta_0=1$, and $U_t$ follows the identical DGP of Case 1 of Section \ref{Sec32}. For the factor structure, we let $\Gamma_0 = (\gamma_{01},\ldots, \gamma_{0N})^\top$ with $\gamma_{0i,\ell}\sim U(0.2, 2.2)$, and  $f_t \sim N(0_{p}, I_{p})$, where $\gamma_{0i,\ell}$ stands for the $\ell^{th}$ element of $\gamma_{0i}$. We let $p=2$. To introduce a correlation between the regressors and the factor structure, we let  $X_{t} = X_{t}^* + v_{t}$, where $X_{it}^* =|\gamma_{0i}'f_{t}|$, $X_{it}^* $ stands for the $i^{th}$ element of $X_{t}^* $, and $v_{t}\sim N(0_N,I_N)$.  Based on the above DGP, $\{X_{t}\}$ are correlated with both $F=(f_1,\ldots, f_T)^\top$ and $\Gamma_0$.

The estimation procedure and the bootstrap draws are obtained in exactly the same way as documented above Theorem \ref{Theo2.4}. We calculate the size as follows:

\begin{eqnarray*}
\text{Size} = \frac{1}{R}\sum_{m=1}^R  I(\sqrt{\mathbb{N}} (\widetilde{\theta}_{\text{bc},m}-\theta_0)\not\in \text{CI}_m),
\end{eqnarray*}
where $\widetilde{\theta}_{\text{bc},m}$  and  $\text{CI}_m$ stand for  the bias corrected estimate and the 95\% confidence interval based on the 399 bootstrap draws in the $m^{th}$ replication respectively.

After 1000 replications (i.e., $R=1000$), the results are reported in Table \ref{Tablenew3}. It is easy to see that as the sample size increases, the rejection rate approaches the nominal one, which infers two facts that the bias correction method works well, and  the DWB method is able to recover the asymptotic covariance reasonably well. Due to the estimation errors, the performance is slightly worse than those in Tables  \ref{Tablenew1}-\ref{Tablenew2} and Tables \ref{Table1}-\ref{Table6}, which is acceptable.   

\begin{center}
Insert Table  \ref{Tablenew3} about here.
\end{center}

\section{An Empirical Study}\label{Sec4}

In this section, we apply the proposed DWB method to a real dataset by evaluating the aggregated mutual fund performance.

A vast literature of financial economics has been devoted to evaluating the skills of the mutual fund managers. However, the existing results present many discrepancies \cite[]{berk2015measuring}, which may be due to the fact that the analyses suffer from various modelling problems. For example, the traditional approach ignores the panel nature of the dataset, so the inter-fund information of the cross-sectional dimension has been largely ignored. In the same spirit, \citet[p. 1939]{fama2010luck} suggest that the TSA of the regression residuals may also alter the size of the usual fund alpha test. In this empirical study, we apply the DWB method of Section \ref{Sec2}, and aim to settle the discrepancies by accounting for the dependences along both dimensions of the dataset.

We obtain active U.S. equity mutual funds data from the Center for Research in Security Prices (CRSP) Survivor-Bias-Free Mutual Fund database for the period over Feb 1987 -- Sep 2017, and exclude the passive index funds (e.g., \citealp{harvey2018detecting}). As the data are monthly collected, the sample size is $T=368$.  We only include the funds which have initial total net assets above 10 million, and have more than 80\% of their holdings in equity markets. To mitigate degree of the unbalanced panel data structure, we consider three datasets by removing the funds with more than 20\%, 25\%, and 30\% missing values\footnote{The thresholds 20\%, 25\%, and 30\% are set arbitrarily. After different attempts, we note that the conclusion is not sensitive to the thresholds adopted here. In addition, we regard the three choices of the threshold as one type of robustness check.} during the entire period respectively, which leave us with 97, 114,  and 132 mutual funds for different thresholds. 

We consider the following unbalanced panel data model:
\begin{equation*}
	y_{it} = \alpha + x_t^\top \beta + u_{it},
\end{equation*}
where $y_{it}$ is the net return (excluding fees and expenses) for fund $i$,  $x_t$ includes the Fama-French-Carhart four-factor (including the market excess return factor, the Small-Minus-Big size factor, the High-Minus-Low value factor, the momentum factor), $\beta$ includes the slope coefficients, and $\alpha$ measures the abnormal performance of the mutual fund industry. We are interested in inferring $\alpha$, which is usually considered as an average indicator of the managerial skill of fund managers since the seminal work of \cite{jensen1968performance}.

After running the OLS regression, we obtain the estimates of $\alpha $ and $\beta$ as follows:
\begin{eqnarray*}
	\text{When }N=97, &&(\widehat{\alpha},\widehat{\beta}^\top)=( 0.0006,\, 07573,\, 0.0224,\, 0.0319,\,  -0.0022),\nonumber \\
	\text{When }N=114, &&(\widehat{\alpha},\widehat{\beta}^\top)=(0.0005, \, 0.7542, \, 0.0295, \, 0.0170, -0.0025),\nonumber \\
	\text{When }N=132, &&(\widehat{\alpha},\widehat{\beta}^\top)=( 0.0006,\, 0.7299,\, 0.0373,\, 0.0137,\,  0.0032).
\end{eqnarray*}
The estimated residuals can then be calculated as follows:
\begin{eqnarray*}
	\widehat{u}_{it} = y_{it} - \widehat{\alpha} - x_t^\top \widehat{\beta}.
\end{eqnarray*}

To show the necessity of accounting for CSD and TSA, we first conduct the following two tests:

\noindent 1. Examine  TSA by conducting the Ljung-Box Q-test for each time series (i.e., $\{\widehat{u}_{i1},\ldots, \widehat{u}_{iT}\}$), and report the percentage of individuals having non-negligible autocorrelation;

\noindent 2. Examine   CSD by conducting the CD test\footnote{The asymptotic distribution of the CD test follows the standard normal distribution, so at the 5\% significance level, the critical values are $\pm 1.96$. We refer interested readers to \cite{Pesaran2004} for more details.} of \cite{Pesaran2004} on $\widehat{u}_{it}$'s, and report the test statistics.

As shown in Table \ref{table_test}, a non-negligible portion of individuals show evidence of TSA, while the CD test statistic always yields a significantly large value, which indicates the presence of CSD among the residuals. It is noteworthy that the computed value of the CD test increases, as the threshold (of removing individuals) becomes less restrictive, so it is a strong sign of  CSD.

\begin{center}
Insert Table  \ref{table_test} about here.
\end{center}

Below, we start reporting the 95\%  CI by using different methods.  First, in Table \ref{Table4.1}, we present the CIs using the three traditional methods as in Section \ref{Sec32}. It is clear the CIs generated by $s_1^2$ and $s_2^2$ indicate that the annualized aggregate mutual fund alpha is positively significant, which implies that the overall mutual fund industry can actually beat the market. However, the  CIs generated by $s_3^2$ tell a different story.  The results are not very surprising given that Table \ref{table_test} shows a reasonable amount of individuals fail to reject the null of the Ljung-Box Q-Test that assumes no time series autocorrelation.

\begin{center}
Insert Tables  \ref{Table4.1} and \ref{Table4.2} about here.
\end{center}

In what follows, we consider the BCL, MBB, and DWB methods and focus on the CIs associated with the annualized alpha. The implementation of these methods is identical to that of Section \ref{Sec32}.  The results are summarized in Table \ref{Table4.2}. Note that in Table \ref{Table4.2} the BCL and MBB methods show mixed conclusions, while the DWB method consistently supports the result of $\alpha=0$ regardless the different combinations of the sample size, the bandwidth parameter, and the kernel function. Also, the consistent finding from the DWB method agrees with that of \cite{fama2010luck}, in which they conclude that the mutual fund industry cannot beat the market. 

Finally, in connection with the numerical results presented in Section \ref{Sec3}, we argue that the DWB method shows strong evidence of its superiority over some natural competitors in finite sample studies. We thus think the DWB method may yield more reliable results in practice.

\section{Conclusion}\label{Sec5}

Although a variety of panel data models have been investigated over the past couple of decades, not much work has been done to improve inferences associated with the estimation of the parameters-of-interest. In this paper, we have developed a simple dependent wild bootstrap procedure to establish inferences for a wide class of panel data models, including those with interactive effects. The proposed method allows for the error components to have CSD, TSA, and heteroskedasticity. The asymptotic properties, including Berry-Esseen bound and Edgeworth Expansion, have been established under a set of simple and general conditions. In addition, the newly proposed DWB method is easy to implement, and requires only one tuning parameter. We show the superiority of our approach over some natural competitors using extensive numerical studies.  Last but not least, we demonstrate the usefulness of the DWB by explicitly investigating a panel data model with interactive effects which nests many traditional panel data models as special cases. As a by-product, we provide a solution to deal with bias correction and inference issue within one framework, which, to our knowledge, is the first result that has successfully addressed both issues.

In this paper, we have considered stationary time series for all individuals. We are aware of the growing literature on using bootstrap assisted methods to establish inferences for co-integrated time series models (e.g., \citealp{PP2003, CNR2015, RJ2022}). Along this line of research, \cite{Shao2015} provides a recent review on the bootstrap techniques frequently adopted. It would be interesting to investigate co-integrated panel data models (associated with certain cross-sectional dependence) using bootstrap methods. Such settings should be appealing in view of the increasing availability of large financial datasets over the past two decades. We leave possible extensions for future research.

{\footnotesize

\bibliography{Refs}

@article{Shao2015,
author = {Xiaofeng Shao},
title = {Self-Normalization for Time Series: A Review of Recent Developments},
journal = {Journal of the American Statistical Association},
volume = {110},
number = {512},
pages = {1797-1817},
year  = {2015}
}

@article{shergin1980convergence,
	title={On the convergence rate in the central limit theorem for m-dependent random variables},
	author={Shergin, VV},
	journal={Theory of Probability \& Its Applications},
	volume={24},
	number={4},
	pages={782--796},
	year={1980},
	publisher={SIAM}
}

@article{Jirak16,
author = {Moritz Jirak},
title = {{Berry-Esseen theorems under weak dependence}},
volume = {44},
journal = {The Annals of Probability},
number = {3},
pages = {2024-2063},
year = {2016}
}

@article{zhou2010simultaneous,
	title={Simultaneous inference of linear models with time varying coefficients},
	author={Zhou, Zhou and Wu, Wei Biao},
	journal={Journal of the Royal Statistical Society: Series B (Statistical Methodology)},
	volume={72},
	number={4},
	pages={513--531},
	year={2010},
	publisher={Wiley Online Library}
}

@article{rhee1985edgeworth,
	title={An Edgeworth expansion for a sum of m-dependent random variables},
	author={Rhee, Wan Soo},
	journal={International Journal of Mathematics and Mathematical Sciences},
	volume={8},
	number={3},
	pages={563--569},
	year={1985},
	publisher={Hindawi}
}

@article{rhee1986characteristic,
	title={On the characteristic function of a sum of M-dependent random variables},
	author={Rhee, Wansoo T},
	journal={International Journal of Mathematics and Mathematical Sciences},
	volume={9},
	number={2},
	pages={397--404},
	year={1986},
	publisher={Hindawi}
}

@article{zhu2017network,
	title={Network vector autoregression},
	author={Zhu, Xuening and Pan, Rui and Li, Guodong and Liu, Yuewen and Wang, Hansheng},
	journal = {Annuals of Statistics},
	volume = {45},
	number = {3},
	pages = {1096-1123},
	year={2017}
}

@article{PP2003,
author = {Paparoditis, Efstathios and Politis, Dimitris N.},
title = {Residual-Based Block Bootstrap for Unit Root Testing},
journal = {Econometrica},
volume = {71},
number = {3},
pages = {813-855},
year = {2003}
}

@article{giannone2021economic,
	title={Economic predictions with big data: The illusion of sparsity},
	author={Giannone, Domenico and Lenza, Michele and Primiceri, Giorgio E},
	journal={Econometrica},
	volume={89},
	number={5},
	pages={2409--2437},
	year={2021},
	publisher={Wiley Online Library}
}

@UNPUBLISHED{RJ2022,
    author = "K. Reichold and C. Jentsch",
    title  = "A Bootstrap-Assisted Self-Normalization Approach to Inference in Cointegrating Regressions",
    note   = "\url{	arXiv:2204.01373}",
    year   = "2022"
}

@article{CNR2015,
 author = {Giuseppe Cavaliere and Heino Bohn Nielsen and Anders Rahbek},
 journal = {Econometrica},
 number = {2},
 pages = {813-831},
 title = {BOOTSTRAP TESTING OF HYPOTHESES ON CO-INTEGRATION RELATIONS IN VECTOR AUTOREGRESSIVE MODELS},
 volume = {83},
 year = {2015}
}

@article{KV2002,
 author = {Nicholas M. Kiefer and Timothy J. Vogelsang},
 journal = {Econometrica},
 number = {5},
 pages = {2093-2095},
 title = {Heteroskedasticity-Autocorrelation Robust Standard Errors Using the Bartlett Kernel without Truncation},
 volume = {70},
 year = {2002}
}

@article{nagaev1979large,
  author = {S. V. Nagaev},
title = {Large deviations of sums of independent random variables},
volume = {7},
journal = {Annals of Probability},
number = {5},
pages = {745 -- 789},
year = {1979},
}

@article {Wu2005,
	author = {Wu, Wei Biao},
	title = {Nonlinear system theory: Another look at dependence},
	volume = {102},
	number = {40},
	pages = {14150-14154},
	year = {2005},
	journal = {Proceedings of the National Academy of Sciences}
}

@article{jirak2021sharp,
	title={Sharp connections between {Berry-Esseen} characteristics and {Edgeworth} expansions for stationary processes},
	author={Jirak, Moritz and Wu, Wei Biao and Zhao, Ou},
	journal={Transactions of the American Mathematical Society},
	volume={374},
	number={6},
	pages={4129-4183},
	year={2021}
}

@book{feller2008introduction,
	title={An Introduction to Probability Theory and its Applications, Vol 2},
	author={Feller, William},
	year={1970},
	publisher={John Wiley \& Sons}
}

@ARTICLE {Bai,
    author  = "J. Bai",
    title   = "Panel Data Models with Interactive Fixed Effects",
    journal = "Econometrica",
    year    = "2009",
    volume  = "77",
    number  = "4",
    pages   = "1229-1279"
}

@article{jensen1968performance,
	title={The performance of mutual funds in the period 1945-1964},
	author={Jensen, Michael C},
	journal={Journal of Finance},
	volume={23},
	number={2},
	pages={389-416},
	year={1968},
	publisher={JSTOR}
}

@article{fama2010luck,
	title={Luck versus Skill in the Cross-Section of Mutual Fund Returns},
	author={Fama, Eugene F and French, Kenneth R},
	journal={Journal of Finance},
	volume={65},
	number={5},
	year={2010},
	pages={1915-1947}
}

@article{berk2015measuring,
	title={Measuring skill in the mutual fund industry},
	author={Berk, Jonathan B and Van Binsbergen, Jules H},
	journal={Journal of Financial Economics},
	volume={118},
	number={1},
	pages={1-20},
	year={2015},
}

@article{Pesaran2004,
    author = "M. H. Pesaran",
    title  = "General diagnostic tests for cross section dependence in panels",
    journal={Empirical Economics},
	volume={60},
	pages={13-50},
    year   = "2021"
}

@ARTICLE {Pesaran2006,
    author  = "M. H. Pesaran",
    title   = "Estimation and Inference in Large Heterogeneous Panels with a Multifactor Error Structure",
    journal = "Econometrica",
    year    = "2006",
    volume  = "74",
    number  = "4",
    pages   = "967-1012"
}

@article{goncalves_2011,
title={THE MOVING BLOCKS BOOTSTRAP FOR PANEL LINEAR REGRESSION MODELS WITH INDIVIDUAL FIXED EFFECTS},
volume={27},
number={5},
journal={Econometric Theory},
author={Gon{\c c}alves, S.},
year={2011},
pages={1048-1082}}

@ARTICLE {shao2010,
    author    = "Shao, Xiaofeng",
    title     = "The dependent wild bootstrap",
    journal   = "Journal of the American Statistical Association",
    year      = "2010",
    volume    = "105",
    number    = "489",
    pages     = "218-235",
}

@article{BN2002,
author = {Bai, Jushan and Ng, Serena},
title = {Determining the Number of Factors in Approximate Factor Models},
journal = {Econometrica},
volume = {70},
number = {1},
pages = {191-221},
year = {2002}
}

@BOOK{HG2006,
     author = {Donald Hedeker and Robert D. Gibbons},
      title = {Longitudinal Data Analysis},
      year = {2006},
      publisher = {John Wiley \& Sons, Inc.},
      edition = {first}
}

@BOOK{Anderson,
     author = {T. W. Anderson},
      title = {The Statistical Analysis of Time Series},
      year = {1971},
      publisher = {John Wiley \& Sons, Inc.},
      edition = {first}
}

@BOOK{FLW2011,
     author = {G. M. Fitzmaurice and N. M. Laird and J. H. Ware},
      title = {Applied Longitudinal Data Analysis},
      year = {2011},
      publisher = {John Wiley \& Sons, Inc.},
      edition = {second}
}

@article{BAI2020,
title = {Standard errors for panel data models with unknown clusters},
journal = {Journal of Econometrics},
year = {2020},
author = {Jushan Bai and Sung Hoon Choi and Yuan Liao},
 pages   = "forthcoming"
}

@ARTICLE {Chen2021,
    title = {Nonlinear factor models for network and panel data},
journal = {Journal of Econometrics},
volume = {220},
number = {2},
pages = {296-324},
year = {2021},
author = {Mingli Chen and Iv\'an Fern\'andez-Val and Martin Weidner},
}

@article{Andrews1991,
 author = {Donald W. K. Andrews},
 journal = {Econometrica},
 number = {3},
 pages = {817-858},
 title = {Heteroskedasticity and Autocorrelation Consistent Covariance Matrix Estimation},
 volume = {59},
 year = {1991}
}

@ARTICLE {Ando,
    author  = "T. Ando and J. Bai",
    title   = "Clustering Huge Number of Financial Time Series: A Panel Data Approach With High-Dimensional Predictors and Factor Structures",
    journal = "Journal of the American Statistical Association",
    year    = "2017",
    volume  = "112",
    number  = "519",
    pages   = "1182-1198"
}

@ARTICLE {LiQianSu,
    author  = "D. Li and J. Qian and L. Su",
    title   = "Panel Data Models with Interactive Fixed Effects and Multiple Structural Breaks",
    journal = "Journal of the American Statistical Association",
    year    = "2016",
    volume  = "111",
    number  = "516",
    pages   = "1804-1819"
}

@article{PP2001,
 author = {Efstathios Paparoditis and Dimitris N. Politis},
 journal = {Biometrika},
 number = {4},
 pages = {1105-1119},
 title = {Tapered Block Bootstrap},
 volume = {88},
 year = {2001}
}

@article{Petersen2009,
    author = {Petersen, Mitchell A.},
    title = "Estimating Standard Errors in Finance Panel Data Sets: Comparing Approaches",
    journal = {Review of Financial Studies},
    volume = {22},
    number = {1},
    pages = {435-480},
    year = {2009}
}

@Article{Menzel2021,
  author={Konrad Menzel},
  title={Bootstrap With Cluster‐Dependence in Two or More Dimensions},
  journal={Econometrica},
  year=2021,
  volume={89},
  number={5},
  pages={2143-2188}
}

@article{harvey2018detecting,
  title={Detecting repeatable performance},
  author={Harvey, Campbell R and Liu, Yan},
  journal={Review of Financial Studies},
  volume={31},
  number={7},
  pages={2499-2552},
  year={2018},
}

}

{
\renewcommand{\arraystretch}{0.8}
\setlength{\tabcolsep}{5pt}
\begin{table}[H]
\scriptsize
\caption{Results of Case 1 for Theorem \ref{Theo2.2} ($\rho_u=0.25$ and $\delta_\epsilon =0.5$)}\label{Tablenew1}
\hspace*{-0.25cm}
\begin{tabular}{llrrrrrrrrrrrr}
\hline\hline
 &  & \multicolumn{6}{c}{DWB} & \multicolumn{3}{c}{\multirow{2}{*}{MBB}} & \multicolumn{3}{c}{\multirow{2}{*}{Traditional}} \\
 &  & \multicolumn{3}{c}{Bartlett} & \multicolumn{3}{c}{Trapezoidal} & \multicolumn{3}{c}{} & \multicolumn{3}{c}{} \\
$N$ & $T$ & $0.8\ell_B$ & $\ell_B$ & $1.2\ell_B$ & $0.8\ell_T$ & $\ell_T$ & $1.2\ell_T$  &$0.8\ell_M$ & $\ell_M$ & $1.2\ell_M$ & $s_1^2$ & $s_2^2$ & $s_3^2$ \\
50 & 50 & 0.022 & 0.017 & \multicolumn{1}{r|}{0.010} & 0.035 & 0.027 & \multicolumn{1}{r|}{0.018} & 0.196 & 0.164 & \multicolumn{1}{r|}{0.164} & 0.391 & 0.265 & 0.117 \\
 & 100 & 0.036 & 0.032 & \multicolumn{1}{r|}{0.032} & 0.037 & 0.033 & \multicolumn{1}{r|}{0.032} & 0.128 & 0.116 & \multicolumn{1}{r|}{0.103} & 0.379 & 0.253 & 0.108 \\
 & 200 & 0.051 & 0.043 & \multicolumn{1}{r|}{0.046} & 0.051 & 0.048 & \multicolumn{1}{r|}{0.045} & 0.118 & 0.099 & \multicolumn{1}{r|}{0.104} & 0.380 & 0.278 & 0.135 \\
 & 400 & 0.067 & 0.061 & \multicolumn{1}{r|}{0.061} & 0.066 & 0.063 & \multicolumn{1}{r|}{0.059} & 0.092 & 0.086 & \multicolumn{1}{r|}{0.082} & 0.385 & 0.268 & 0.131 \\
100 & 50 & 0.027 & 0.017 & \multicolumn{1}{r|}{0.010} & 0.035 & 0.026 & \multicolumn{1}{r|}{0.018} & 0.172 & 0.150 & \multicolumn{1}{r|}{0.153} & 0.373 & 0.240 & 0.113 \\
\hline \hline 
\end{tabular}
\end{table}}

{
\renewcommand{\arraystretch}{0.8}
\setlength{\tabcolsep}{5pt}
\begin{table}[H]
\scriptsize
\caption{Results of Case 2 for Theorem \ref{Theo2.2} ($\rho_u=0.25$ and $\delta_\epsilon =0.5$)}\label{Tablenew2}
\hspace*{-0.25cm}\begin{tabular}{llrrrrrrrrrrrr}
\hline\hline
 &  & \multicolumn{6}{c}{DWB} & \multicolumn{3}{c}{\multirow{2}{*}{MBB}} & \multicolumn{3}{c}{\multirow{2}{*}{Traditional}} \\
 &  & \multicolumn{3}{c}{Bartlett} & \multicolumn{3}{c}{Trapezoidal} & \multicolumn{3}{c}{} & \multicolumn{3}{c}{} \\
$N$ & $T$ & $0.8\ell_B$ & $\ell_B$ & $1.2\ell_B$ & $0.8\ell_T$ & $\ell_T$ & $1.2\ell_T$  &$0.8\ell_M$ & $\ell_M$ & $1.2\ell_M$ & $s_1^2$ & $s_2^2$ & $s_3^2$ \\
 50 & 50 & 0.032 & 0.025 & \multicolumn{1}{r|}{0.006} & 0.042 & 0.039 & \multicolumn{1}{r|}{0.027} & 0.220 & 0.201 & \multicolumn{1}{r|}{0.191} & 0.265 & 0.148 & 0.156 \\
 & 100 & 0.047 & 0.035 & \multicolumn{1}{r|}{0.025} & 0.051 & 0.045 & \multicolumn{1}{r|}{0.039} & 0.140 & 0.119 & \multicolumn{1}{r|}{0.122} & 0.247 & 0.129 & 0.129 \\
 & 200 & 0.043 & 0.039 & \multicolumn{1}{r|}{0.035} & 0.042 & 0.038 & \multicolumn{1}{r|}{0.032} & 0.103 & 0.088 & \multicolumn{1}{r|}{0.089} & 0.237 & 0.109 & 0.121 \\
 & 400 & 0.050 & 0.047 & \multicolumn{1}{r|}{0.050} & 0.050 & 0.048 & \multicolumn{1}{r|}{0.044} & 0.089 & 0.071 & \multicolumn{1}{r|}{0.071} & 0.234 & 0.119 & 0.121 \\
100 & 50 & 0.021 & 0.014 & \multicolumn{1}{r|}{0.006} & 0.030 & 0.025 & \multicolumn{1}{r|}{0.017} & 0.174 & 0.165 & \multicolumn{1}{r|}{0.158} & 0.229 & 0.124 & 0.121 \\
\hline \hline 
\end{tabular}
\end{table}}

{
\renewcommand{\arraystretch}{0.8}
\setlength{\tabcolsep}{5pt}
\begin{table}[H]
\scriptsize
\caption{Results for Section \ref{Sec2.3}}\label{Tablenew3}
\begin{tabular}{llrrrrrrrrrrrr}
\hline\hline
 &  & \multicolumn{6}{c}{$(\rho_u,\delta_\epsilon) =(0.25,0.25)$} & \multicolumn{6}{c}{$(\rho_u,\delta_\epsilon) =(0.25,0.5)$} \\
 &  & \multicolumn{3}{c}{Bartlett} & \multicolumn{3}{c}{Trapezoidal} & \multicolumn{3}{c}{Bartlett} & \multicolumn{3}{c}{Trapezoidal} \\
$N$ & $T$ & $0.8\ell_B$ & $\ell_B$ & $1.2\ell_B$ & $0.8\ell_T$ & $\ell_T$ & $1.2\ell_T$  & $0.8\ell_B$ & $\ell_B$ & $1.2\ell_B$ & $0.8\ell_T$ & $\ell_T$ & $1.2\ell_T$  \\
50 & 50 & 0.173 & 0.191 & \multicolumn{1}{r|}{0.205} & 0.154 & 0.171 & \multicolumn{1}{r|}{0.183} & 0.189 & 0.204 & \multicolumn{1}{r|}{0.212} & 0.164 & 0.192 & 0.193 \\
 & 100 & 0.104 & 0.106 & \multicolumn{1}{r|}{0.111} & 0.095 & 0.100 & \multicolumn{1}{r|}{0.110} & 0.106 & 0.111 & \multicolumn{1}{r|}{0.123} & 0.104 & 0.111 & 0.115 \\
 & 200 & 0.081 & 0.088 & \multicolumn{1}{r|}{0.086} & 0.075 & 0.076 & \multicolumn{1}{r|}{0.082} & 0.068 & 0.077 & \multicolumn{1}{r|}{0.081} & 0.071 & 0.076 & 0.077 \\
 & 400 & 0.055 & 0.063 & \multicolumn{1}{r|}{0.061} & 0.058 & 0.059 & \multicolumn{1}{r|}{0.064} & 0.061 & 0.060 & \multicolumn{1}{r|}{0.057} & 0.063 & 0.059 & 0.058 \\
100 & 50 & 0.178 & 0.188 & \multicolumn{1}{r|}{0.207} & 0.175 & 0.185 & \multicolumn{1}{r|}{0.194} & 0.194 & 0.209 & \multicolumn{1}{r|}{0.225} & 0.177 & 0.195 & 0.203 \\
\hline \hline 
\end{tabular}
\end{table}}

{\renewcommand{\arraystretch}{0.8}
\begin{table}[H]
\footnotesize
	\caption{The Test Results of the Ljung-Box Q-Test and the CD Test}\label{table_test}
	\centering
	\begin{tabular}{lrr}
		\hline\hline
		& Ljung-Box Q-test & CD test statistic \\
		$N=97$ & 34.02\%  & 72.3653 \\
		$N=114$ & 39.47\% & 76.0659 \\
		$N=132$ & 40.91\% & 82.0605\\
		\hline\hline
	\end{tabular}
\end{table}}

{\renewcommand{\arraystretch}{0.8}
\begin{table}[H]
	\centering 
	{\footnotesize
		\caption{Estimates of the Annualized Aggregate Mutual Fund Alpha and the Slope Coefficients}\label{Table4.1}
		\begin{tabular}{lrrrrr}
			\hline\hline
			&  & Estimates & 95\% CI ($s_1^2$) & 95\% CI ($s_2^2$) & 95\% CI ($s_3^2$) \\
			$N=97$ & Annualized Alpha & 0.72\% & (0.24\%, 1.08\%) & (0.12\%, 1.20\%) & (-0.12\%, 1.44\%) \\
			& Market Factor & 0.7573 & (0.7443, 0.7703) & (0.6811, 0.8335) & (0.7327, 0.7819) \\
			& Size Factor & 0.0224 & (0.0071, 0.0376) & (-0.0154, 0.0601) & (-0.0061, 0.0508) \\
			& Value Factor & 0.0319 & (0.0164, 0.0475) & (-0.0118, 0.0756) & (-0.0039, 0.0678) \\
			& Momentum Factor & -0.0022 & (-0.0130, 0.0086) & (-0.0163, 0.0119) & (-0.0224, 0.0180) \\
			&  &  &  &  &  \\
			$N=114$ & Annualized Alpha & 0.60\% & (0.24\%, 1.08\%) & (0.12\%,  1.20\%) & ( -0.24\%, 1.44\%) \\
			& Market Factor & 0.7542 & (0.7420, 0.7664) & (0.6833, 0.8251) & (0.7295, 0.7789) \\
			& Size Factor & 0.0295 & (0.0151, 0.0438) & (-0.0049, 0.0639) & (0.0023, 0.0567) \\
			& Value Factor & 0.0170 & (0.0024, 0.0317) & (-0.0228, 0.0569) & (-0.0188, 0.0529) \\
			& Momentum Factor & -0.0025 & (-0.0127, 0.0077) & (-0.0158, 0.0108) & (-0.0215, 0.0165) \\
			&  &  &  &  &  \\
			$N=132$ & Annualized Alpha & 0.72\% & (0.36\%, 1.08\%) & (0.24\%, 1.2\%) & (-0.12\%, 1.56\%) \\
			& Market Factor & 0.7299 & (0.7181, 0.7416) & (0.6622, 0.7976) & (0.7058, 0.7540) \\
			& Size Factor & 0.0373 & (0.0230, 0.0516) & (0.0033, 0.0714) & (0.0119, 0.0627) \\
			& Value Factor & 0.0137 & (-0.0002, 0.0276) & (-0.0236, 0.0510) & (-0.0213, 0.0488) \\
			& Momentum Factor & 0.0032 & (-0.0065, 0.0128) & (-0.0094, 0.0158) & (-0.0157, 0.0220) \\
			\hline\hline
	\end{tabular}}
\end{table}

\begin{table}[H]
	\centering
	{\footnotesize
		\caption{ The 95\% CIs for Annualized Aggregate Mutual Fund Alpha} \label{Table4.2}
		\begin{tabular}{llrrr}
			\hline\hline
			\multicolumn{5}{c}{BCL} \\
			&  & $L=3$ & $L=7$ & $L=11$ \\
			$N=97$ & $M =0.1$ & (-0.00\%, 1.32\%) & (0.00\%, 1.32\%) & (0.00\%, 1.32\%) \\
			& $M =0.15$ & (-0.00\%, 1.32\%) & (0.00\%, 1.32\%) & (0.00\%, 1.32\%) \\
			& $M =0.2$ & (-0.00\%, 1.32\%) & (0.00\%, 1.32\%) & (0.00\%, 1.32\%) \\
			& $M =0.25$ & (-0.00\%, 1.32\%) & (0.00\%, 1.32\%) & (0.00\%, 1.32\%) \\
			$N=114$ & $M =0.1$ & (-0.12\%, 1.44\%) & (-0.00\%, 1.32\%) & (-0.00\%, 1.32\%) \\
			& $M =0.15$ & (-0.12\%, 1.44\%) & (-0.00\%, 1.32\%) & (-0.00\%, 1.32\%) \\
\hline\hline
	\end{tabular} }
\end{table}}

\newpage

{\small

\setcounter{page}{1}
\begin{center}

   \large \bf Online Supplementary Appendices to  \\``Higher-order Expansions and Inference \\for Panel Data Models"

\end{center}

This documents includes Appendix A and Appendix B. Overall, the structure  is as follows. 

\medskip

\noindent In Appendix A,
\begin{itemize}
\item Appendix \ref{SecExExample} discusses the case with $E[U_t]\ne 0_N$;

\item Appendix  \ref{SecFE} provides an extra example to show the usefulness of the DWB method;

\item Appendix \ref{Sectables} provides some extra simulation results;

\item Appendix \ref{SecA.1} outlines the theoretical development, presents some notations which will be used throughout the theoretical development, and also provides some useful bounds;

\item Appendix \ref{AppA.2} presents the proofs of the main results. 
\end{itemize}

\noindent In Appendix B,

\begin{itemize}
\item Appendix \ref{SecB.1} introduces a few definitions to facilitate development of the preliminary lemmas;

\item Appendix \ref{SecB.2} summaries the preliminary lemmas;

\item Appendix \ref{SecB.3} provides the proofs of the preliminary lemmas.
\end{itemize}

\section*{Appendix A}
	
	\renewcommand{\theequation}{A.\arabic{equation}}
	\renewcommand{\thesubsection}{A.\arabic{subsection}}
	\renewcommand{\thefigure}{A.\arabic{figure}}
	\renewcommand{\thetable}{A.\arabic{table}}
	\renewcommand{\thelemma}{A.\arabic{lemma}}
	\renewcommand{\theexample}{A.\arabic{example}}
	\renewcommand{\theremark}{A.\arabic{remark}}
	\renewcommand{\thecorollary}{A.\arabic{corollary}}
	\renewcommand{\theassumption}{A.\arabic{assumption}}
	\renewcommand{\theproposition}{A.\arabic{proposition}}
	
	\setcounter{equation}{0}
	\setcounter{lemma}{0}
   \setcounter{example}{0}
	\setcounter{table}{0}
	\setcounter{figure}{0}
	\setcounter{remark}{0}
	\setcounter{corollary}{0}
	\setcounter{assumption}{0}
	\setcounter{proposition}{0}

\subsection{The Case with $E[U_t]\ne 0_N$}\label{SecExExample}

First, we consider the case with $E[U_t]\ne 0_N$. The following corollary should be obvious in view of the results provided in the main text.

\begin{corollary}\label{Cor2.2}
	Let Assumption \ref{Assumption1} hold, and let $E[U_t] = \mu $ and $\widehat{\mu} = \frac{1}{T}\sum_{t=1}^{T}U_t$. As $(N,T)\to (\infty,\infty)$,
	
	\noindent 1. $\sqrt{\mathbb{N}}1_{N}^{*,\top}(\widehat{\mu}-\mu) \to_D N(0, \sigma_u^2)$,
	
	\noindent 2. $ \sup_{w\in \mathbb{R}}|\Pr(\sqrt{\mathbb{N}}1_{N}^{*,\top}(\widehat{\mu}-\mu) \le  w) - \Psi_{\mathbb{N}}(w)| = O(T^{-1}(\log T)^5)$,
	
\noindent where $1_{N}^{*} = 1_{N}/N$, $ S_{\mathbb{N}} = \sqrt{\mathbb{N}}1_{N}^\top(\widehat{\mu}-\mu)$, $\sigma_u^2 =\lim \frac{1}{\mathbb{N}}\sum_{t,s=1}^T 1_{N}^\top E[(U_{t}-\mu) (U_{s}-\mu)^\top] 1_{N} $, and $\Psi_{\mathbb{N}}(w)$ is defined in exactly the same way as that in Theorem \ref{Theo2.1} but using $S_{\mathbb{N}}$ provided here.
\end{corollary}
Note that we can generalize Corollary \ref{Cor2.2} by replacing $\sqrt{N}1_{N}^{*} = 1_{N}/\sqrt{N}$ with any weighting vector satisfying $|\sqrt{N}1_{N}^{*}| = O(1)$.
	
\subsection{Panel Data with Fixed Effects}\label{SecFE}

The classic panel data model with fixed effects has been extensively investigated by, for example, \cite{goncalves_2011} and \cite{BAI2020}, and is usually specified as follows:

\begin{eqnarray}\label{Eq.a1}
Y_t = X_t \theta_0 + A +U_t,
\end{eqnarray}
where $Y_t$ is a $N\times 1$ vector including response variables from different individuals at time $t$, $X$ is a $N\times d$ matrix including regressors, and $A = (a_1,\ldots, a_N)^\top$ is a vector of individual specific effects and is usually unobservable. Suppose that, $d$ is finite, and there is no time-invariant column in $X_t$. As $A$ is potentially can be correlated with $X_t$, one normally gets rid of $A$ by removing time mean before carrying on regression. The demeaned model becomes

\begin{eqnarray*}
Y_t^\dag  =X_t^\dag \theta_0 +U_t^\dag,
\end{eqnarray*}
in which $Y_t^\dag = Y_t -\overline{Y}$ with $\overline{Y} = \frac{1}{T}\sum_{t=1}^T Y_t$, and  $X_t^\dag $ and $U_t^\dag $ are defined accordingly. 

The OLS estimate of $\theta_0$ is given below:

\begin{eqnarray}\label{Eq.a2}
\widehat{\theta} = \left(\sum_{t=1}^T X_t^{\dag, \top} X_t^\dag \right)^{-1} \sum_{t=1}^T X_t^{\dag, \top} Y_t^\dag,
\end{eqnarray}
which immediately yields that $\widehat{U}_{t}^\dag = Y_t^\dag - X_{t}^\dag \widehat{\theta}$. 

In this case, the bootstrap procedure can be specified as follows:

\noindent 1. For each bootstrap replication, let $Y_{t}^{\dag, *} =X_{t}^\dag \widehat{\theta} + \widehat{U}_{t}^\dag \xi_t$ for each $t$.

\noindent 2. Calculate $\widetilde{\theta}$ as in \eqref{Eq.a2} using the bootstrap samples $\{Y_{t}^{\dag, *} , X_t^\dag \}$. 

\noindent 3. Repeat the first two steps  $R$ times.

\medskip

Then the next corollary  holds.

\begin{corollary}\label{Coro1}
	Consider the model \eqref{Eq.a1}, and suppose that  (a). $\frac{1}{\mathbb{N}}\sum_{t=1}^T X_t^{\dag, \top} X_t^\dag \to_P\Sigma_1$ and $E\|x_{it} \|^4<0$, where   $\Sigma_1$ is a $d\times d$ positive definite matrix, and $x_{it}^\top$ is the $i^{th}$ row of $X_t$; (b). $\{X_t\}$ and $\{U_t\}$ are independent of each other; (c). Assumptions \ref{Assumption1} and \ref{Assumption2} hold. As $(N, T)\to (\infty,\infty)$,
	\begin{eqnarray*}
		\sup_{w\in \mathbb{R}} \left|\text{\normalfont Pr}^*(\sqrt{\mathbb{N}}(\widetilde{\theta}-\widehat{\theta}) \le w) - \text{\normalfont Pr}(\sqrt{\mathbb{N}}(\widehat{\theta}-\theta_0) \le w)\right| =o_P(1).
	\end{eqnarray*}
\end{corollary}
The extra conditions in the body of Corollary \ref{Coro1} can be easily justified, so we no longer discuss them in detail. Finally, it is noteworthy that we do not require $N/T\to 0$ here, which is adopted in Assumption 4 of \cite{goncalves_2011}. The reason is that Lemma A.3 of \cite{goncalves_2011} aims to capture the strong and weak CSD in one framework, so does not utilize the  rate yielded by the cross-sectional dimension of the error component.

\subsection{Extra Simulation Results}\label{Sectables}

In this section, we provide some extra simulation results.

\medskip

\noindent \textbf{Extra simulation results for Theorem \ref{Theo2.2}}:

We first present the extra simulation results to examine Theorem \ref{Theo2.2} of the main text. Specifically, we vary the values of $\rho_u$ and $\delta_\epsilon$ for Case 1 and Case 2 of Section \ref{Sec32}. 

\begin{center}
Insert Tables \ref{Table1}-\ref{Table6} about here
\end{center}
As shown in these tables, the DWB method is quite robust in terms of the tail behaviour of the error component and different values of  $\rho_u$ and $\delta_\epsilon$. 

\medskip
 
\noindent \textbf{Extra simulation results for Corollary \ref{Coro1}}:
 
 We now examine the results associated with the  model \eqref{Eq.a1}. The DGP is as follows: $Y_t = X_t\theta_0 +A+U_t,$ where $\theta_0=1$ for simplicity, $A_i = \frac{1}{T}\sum_{=1}^T |X_{it}|$, $A_i$ and $X_{it}$ are the $i^{th}$ elements of $A$ and $X_t$ respectively,  and $U_t$ is generated in the same way as in Case 1 of Section \ref{Sec32} of the main text. The regressor $X_t$ is generated by $X_t = t/T+ N(0_N ,\Sigma_N^x)$, where  $\Sigma_N^x=\{0.2^{|i-j|} \}_{N\times N}$. For each generated dataset, we record $\sqrt{\mathbb{N}}(\widehat{\theta}-\theta_0)$ and the 95\% CI yielded by the 399 bootstrap draws of $\sqrt{\mathbb{N}}(\widetilde{\theta}-\widehat{\theta})$. After $R$ replications, we report
\begin{eqnarray*}
\text{Size} = \frac{1}{R}\sum_{m=1}^R  I(\sqrt{\mathbb{N}}(\widehat{\theta}_m-\theta_0)\not\in \text{CI}_m),
\end{eqnarray*}
where $\widehat{\theta}_m$ and $\text{CI}_m$ respectively stand for the value of $\widehat{\theta}$ and the 95\% CI recorded in the $m^{th}$ replication. The methods used for comparison are adjusted accordingly in an obvious manner, so we omit the details. Again, we let $R=1000$. 

\begin{center}
Insert Tables \ref{Table7}-\ref{Table10}  about here
\end{center}
The overall pattern presented by Tables \ref{Table7}-\ref{Table10} is very similar to those shown in Tables \ref{Table1}-\ref{Table6}  and Tables \ref{Tablenew1}-\ref{Tablenew2} of the main text.

\subsection{Outline of the Theoretical Development, Symbols and Useful Bounds}\label{SecA.1}
	
In this section, we first  outline of the theoretical development, and then introduce some mathematical symbols and useful bounds to facilitate the development. 
	
\medskip
	
\noindent \textbf{Outline of the Theoretical Development} 

\begin{itemize}
\item Below, we first introduce a few symbols and provide some useful bounds, which facilitate the development of Proposition \ref{Prop2.1} and the first result of Theorem \ref{Theo2.1}. 

\item To derive the second result of Theorem \ref{Theo2.1}, we prepare Lemmas \ref{L0}-\ref{L7}, some basic results (e.g., the moments conditions of Lemma \ref{L1}) of which will also be used to select the optimal bandwidth (i.e.,  Theorem \ref{Theo2.3}). 

\item With the aforementioned results in hand, we develop Theorem \ref{Theo2.3}, Theorem \ref{Theo2.2}, and Corollary \ref{Coro2.2}. (Although Theorem \ref{Theo2.3} is presented after Theorem \ref{Theo2.2} in the main text, we provide its proof first in this document.)

\item After establishing the above results, we prove Theorem \ref{Theo2.4}. The proof of Corollary \ref{Coro1} is a much more simplified version, so it is given in the end of Appendix \ref{AppA.2}.
\end{itemize}

\medskip

\noindent \textbf{Notation} --- 	For $0\leq a\leq b$, we define the Berry-Esseen tail associated with $S_{\mathbb{N}}$ as follows:
	
	\begin{eqnarray*}
		\mathscr{T}_{a}^{b}(w) = \int_{a\leq |x| \leq b}e^{-ixw}E(e^{ixS_{\mathbb{N}}}) \left(1 - \frac{|x|}{b} \right)\frac{1}{x}\mathrm{d}x,
	\end{eqnarray*}
	which arises in Berry's smoothing inequality. For $a > 0$, we define the following Berry-Esseen characteristic:
	
	\begin{eqnarray*}
		\mathscr{C}_a = \inf_{b\geq a}\left(\sup_{w\in\mathbb{R}}|\mathscr{T}_{a}^{b}(w)| + 1/b \right).
	\end{eqnarray*}
	
	\medskip
	
	Let $U_t^{\prime} \equiv  g(\varepsilon_t,\ldots,\varepsilon_{1}, \varepsilon_{0}^\prime,\varepsilon_{-1},\ldots)$ and $\overline{U}_t^{\prime}  \equiv  \frac{1}{\sqrt{N}}U_t^{\prime\top} 1_{N}$,  where $\varepsilon_0^\prime$ is an independent copy of $\varepsilon_0$. Define $\theta_{t,\delta}^U = \|\overline{U}_t - \overline{U}_t^{\prime}\|_{\delta}$, which is fully bounded by $\lambda_{t,\delta}^U+ \lambda_{t+1,\delta}^U$. To see this, write  
	
	\begin{eqnarray}\label{eq.a.1}
		\theta_{t,\delta}^U &=&  \|\overline{U}_t -\overline{U}_t^{\prime}\|_{\delta} \le  \|\overline{U}_t -\overline{U}_t^*\|_{\delta}  +\|\overline{U}_t^* -\overline{U}_t^{\prime}\|_{\delta} \nonumber\\
		&=& \lambda_{t,\delta}^U +\|\overline{U}_{t+1}^* -\overline{U}_{t+1}\|_{\delta} = \lambda_{t,\delta}^U +  \lambda_{t+1,\delta}^U,
	\end{eqnarray}
	in which the second equality follows from $\overline{U}_t^* -\overline{U}_t^{\prime} =_D \overline{U}_{t+1}^* -\overline{U}_{t+1}$. 
	
	Equation \eqref{eq.a.1} infers that the conditions imposed on $\lambda_{t,\delta}^U $ in Assumption \ref{Assumption1} also apply to $\theta_{t,\delta}^U$. For the same purpose, for $0\le m\le t$, let   $U_t^{(m,\prime)} \equiv  g(\varepsilon_t, \ldots,\varepsilon_{t-m+1},\varepsilon_{t-m}^\prime,\varepsilon_{t-m-1},\ldots )$, and $ U_t^{(m,*)} \equiv g (\varepsilon_t,\ldots,\varepsilon_{t-m+1},\varepsilon_{t-m}^\prime,\varepsilon_{t-m-1}^\prime,\ldots)$. Accordingly, we have $\overline{U}_t^{(m,\prime)}$ and $\overline{U}_t^{(m,*)}$.  In addition, define $U_t^{(m,**)}$ and $\overline{U}_t^{(m,**)}$ using $\{\varepsilon_t^{\prime\prime}\}$, which is another independent copy of $\{\varepsilon_t\}$. 
	
	\medskip
	
	Define the $\sigma$-field $\mathscr{F}_{t} = \sigma(\varepsilon_t,\varepsilon_{t-1},\ldots)$ and the projection operator:
	\begin{eqnarray*}
		\mathcal{P}_t (\cdot) = E[\cdot\mid\mathscr{F}_{t}] - E[\cdot \mid\mathscr{F}_{t-1}].
	\end{eqnarray*}
	For $1\leq \delta^*\leq \delta$ and some integer $0\le m \le t$,
	
	\begin{eqnarray}\label{eq.a.2}
		\|\mathcal{P}_{t-m} (\overline{U}_t)\|_{\delta^*}  &=&\| E [\overline{U}_t\mid\mathscr{F}_{t-m}]-E[\overline{U}_t \mid \mathscr{F}_{t-m-1}]\|_{\delta^*}=\|E[\overline{U}_t\mid\mathscr{F}_{t-m}]-E[\overline{U}_t^{(m,\prime)}\mid\mathscr{F}_{t-m-1}]\|_{\delta^*}\nonumber\\
		&=&\|E [\overline{U}_t-\overline{U}_t^{(m,\prime)}\mid\mathscr{F}_{t-m}]\|_{\delta^*}\leq \|\overline{U}_t-\overline{U}_t^{(m,\prime)}\|_{\delta^*} \leq \|\overline{U}_t-\overline{U}_t^{(m,\prime)}\|_{\delta}\nonumber \\
		&=&\|\overline{U}_m-\overline{U}_m^{\prime}\|_{\delta}=  \theta_{m,\delta}^U,
	\end{eqnarray}
	where the first inequality follows from Jensen's inequality, the second inequality follows from the moments monotonicity, and the fourth equality follows from $\overline{U}_t-\overline{U}_t^{(m,\prime)} =_D \overline{U}_m-\overline{U}_m^{\prime}$.
	
	\medskip
	
	Let $\gamma \equiv \gamma_T \to \infty$ and $\gamma/T\to 0$. For $t\geq s$, define $\mathscr{F}_{t,s} = \sigma(\varepsilon_t,\ldots,\varepsilon_{s})$, $\mathscr{F}_{t,s}^* = \sigma(\varepsilon_t',\ldots,\varepsilon_{s}')$, and $\overline{U}_{t\gamma} = E[\overline{U}_t\mid \mathscr{F}_{t,t-\gamma}]$. Further, we let
	
	\begin{eqnarray*}
		\overline{U}_{t\gamma}^* =\left\{\begin{array}{ll}
			E[\overline{U}_t^*\mid  \mathscr{F}_{t,1},\mathscr{F}_{0,t-\gamma}^*], & \text{for }1\leq t \leq \gamma \\
			\overline{U}_{t\gamma} , & \text{for } t > \gamma
		\end{array} \right. .
	\end{eqnarray*}
	
	\medskip
	
	With the above definitions in hand, we are ready to present two propositions which will be used repeatedly in the following proofs.

\begin{proposition}\label{Prop2.1}
	Under Assumption \ref{Assumption1}, we have
	\begin{enumerate}
		\item $\|\overline{U}_t\|_{\delta^*} \le 2\sum_{t=0}^{\infty}\lambda_{t,\delta}^U  <\infty$ with $1\le \delta^* \le \delta$,
		
		\item $\sum_{t=1}^{\infty}t^2|E[\overline{U}_t\overline{U}_0]|\le \|\overline{U}_0 \|_2\sum_{t=0}^{\infty}t^2 \lambda_{t,2}^U<\infty$,
		
		\item $\|\sum_{t=1}^T\overline{U}_t\|_\delta = O(\sqrt{T})$,
			
			\item $\sum_{t=1}^{\infty}t^2\|\overline{U}_{t\gamma} - \overline{U}_{t\gamma}^*\|_\delta <\infty$.
	\end{enumerate}
\end{proposition}

Note that Assumption \ref{Assumption1} is not only flexible, but also regulates CSD and TSA respectively, and provides an underlying data generating process to s	atisfy the moment restriction: 
\begin{eqnarray}\label{Eq.2.1}
	E|\overline{U}_t|^4 =E\left|\frac{1}{\sqrt{N}}U_t^\top 1_{N} \right|^4<\infty,
\end{eqnarray}
which has been widely adopted in the literature of panel data analysis (see, for example, \citealp[Assumption C]{BN2002}).   To see this, in the first result of Proposition \ref{Prop2.1}, by taking $\delta^*=2$ and $\delta^*=4$ respectively, we bound CSD, and the fourth moment presented in \eqref{Eq.2.1}. In the same spirit, the second result of Proposition \ref{Prop2.1} imposes a restriction on TSA.

\subsection{Proofs of Main Results}\label{AppA.2}
	
\noindent \textbf{Verification of Example \ref{Exam1}}:
\medskip	

\noindent (1).	Without loss of generality, let $\delta=4$ in what follows. Then write
	
	\begin{eqnarray*}
		\|\overline{U}_t - \overline{U}_t^* \|_4&=&\left(E\left|\sum_{j=t}^{\infty}\frac{1}{\sqrt{N}}1_N^\top B_j(\varepsilon_{t-j} - \varepsilon_{t-j}^\prime)\right|^4\right)^{1/4}\\
		&\le &\left(\sum_{j=t}^{\infty}E\left|\frac{1}{\sqrt{N}}1_N^\top B_j(\varepsilon_{t-j} - \varepsilon_{t-j}^\prime)\right|^4\right)^{1/4} + 6 \left(\sum_{j=t}^{\infty}\frac{1}{N}1_N^\top B_j B_j^\top 1_N \right)^{1/2}\\
		&:=&A_1+A_2,
	\end{eqnarray*}
	where the definitions of $A_1$ and $A_2$ are obvious. 
	
	Consider $A_1$. For notational simplicity, let $B_j = \{B_{j,kl}\}_{k,l\in [N]}$ and $\frac{1}{\sqrt{N}}1_N^\top B_j = (B_{j,\, \centerdot 1},\ldots,B_{j,\, \centerdot N})$. As $\{\varepsilon_{it}\}$ are independent over $i$, we can write
	
	\begin{eqnarray*}
		&&E\left|\frac{1}{\sqrt{N}}1_N^\top B_j(\varepsilon_{k-j} - \varepsilon_{k-j}^\prime)\right|^4\\
		&=&E\left|\sum_{l=1}^{N}B_{j,\, \centerdot l}^2(\varepsilon_{l,k-j} - \varepsilon_{l,k-j}^\prime)^2\right|^2 + 4E\left|\sum_{l=1}^{N-1}\sum_{k=l+1}^{N}B_{j,\, \centerdot l}B_{j,\, \centerdot k}(\varepsilon_{l,t-j} - \varepsilon_{l,t-j}^\prime)(\varepsilon_{k,t-j} - \varepsilon_{k,t-j}^\prime) \right|^2\nonumber \\
		&\le &O(1)\left(\sum_{l=1}^{N}B_{j,\, \centerdot l}^2\right)^2 +O(1) \sum_{l=1}^{N-1}\sum_{k=l+1}^{N}B_{j,\, \centerdot l}^2B_{j,\, \centerdot k}^2\\
		&\le &O(1) \left(\frac{1}{N}1_N^\top B_j B_j^\top 1_N\right)^2,
	\end{eqnarray*} 
	where the first inequality follows from some direct calculation, and the second inequality follows from $\sum_{l=1}^{N-1}\sum_{k=l+1}^{N}B_{j,\, \centerdot l}^2B_{j,\, \centerdot k}^2\leq  (\sum_{l=1}^{N}B_{j,\, \centerdot l}^2)^2$. Finally, we can write
	\begin{eqnarray*}
		A_1 \le O(1)\left\{\sum_{j=t}^{\infty}\left(\frac{1}{N}1_N^\top B_j B_j^\top 1_N\right)^2\right\}^{1/4}\leq  O(1)\left\{\sum_{j=t}^{\infty}|B_j|^4\right\}^{1/4} = O(\rho^t).
	\end{eqnarray*}
	
	For $A_2$,  it is obvious that $6\left\{\sum_{j=t}^{\infty}|B_j|^2\right\}^{1/2}= O(\rho^t)$.
	
	Based on the above development, $\sum_{t=0}^{\infty}t^2\|\overline{U}_t - \overline{U}_t^*\|_4 \le O(1)\sum_{t=0}^{\infty}t^2\rho^t<\infty,$ so Assumption \ref{Assumption1} is met. 
		
		\medskip
		
\noindent (2).	First, consider Assumption C (i) of \cite{Bai}. Recall that $B_j = \{B_{j,kl}\}_{k,l\in [N]}$, and thus $u_{it} = \sum_{j=0}^{\infty}\sum_{l=1}^{N}B_{j,il}\varepsilon_{l,t-j}$. Since $|B_j| = \sup_{x} \frac{<x,B_j^\top B_j x>}{<x,x>}$, $B_j = [B_{j,\, \centerdot 1},\ldots,B_{j,\, \centerdot N}] $ and $\max_iB_{j,\, \centerdot i}^\top B_{j,\, \centerdot i} \leq \left(\sup_{x} \frac{<x,B_j^\top B_j x>}{<x,x>}\right)^2$, we have
\begin{eqnarray*}
	\|u_{it}\|_8&\leq& \sum_{j=0}^{\infty}\|\sum_{l=1}^{N}B_{j,il}\varepsilon_{l,t-j}\|_8 = O(1)\sum_{j=0}^{\infty}\left(\sum_{l=1}^{N}B_{j,il}^8 +(\sum_{l=1}^{N}B_{j,il}^2)^4 \right)^{1/8} \\
	&\leq& O(1)\sum_{j=0}^{\infty}(\sum_{l=1}^{N}B_{j,il}^2)^{1/2} = O(1)\sum_{j=0}^{\infty}\rho^j =O(1).
\end{eqnarray*}

Consider Assumption C (ii). Note that $|B_j|_1 = \max_{1\leq p\leq N}\sum_{l=1}^{N}|B_{j,lp}| =O(\rho^j)$ and 
$$
|E(u_{it}u_{js})| \leq E(\varepsilon_{it}^2) \sum_{v=0}^{\infty}\sum_{l=1}^{N}|B_{v+t-s,il}B_{v,jl}|,
$$ 
then write
\begin{eqnarray*}
	\frac{1}{NT}\sum_{i,j,t,s}|E(u_{it}u_{js})|& \leq & E(\varepsilon_{it}^2)\frac{1}{NT}\sum_{i,j,t,s} \sum_{v=0}^{\infty}\sum_{l=1}^{N}|B_{v+t-s,il}B_{v,jl}| \\
	&\leq& O(1)\sum_{v=0}^{\infty} \frac{1}{N}\sum_{l=1}^{N} \sum_{u=0}^{T} \left(\sum_{i=1}^{N}|B_{v+u,il}|\right)\left(\sum_{j=1}^{N}|B_{v,jl}|\right)\\
	&=&O(1)\sum_{v=0}^{\infty} \sum_{u=0}^{T} \rho^{v+u} \rho^{v} =O(1).
\end{eqnarray*}

Consider Assumption C (iii). Here we consider $t=s$ for notational simplicity, the derivation for $t \neq s$ can be achieved similarly, so we omit its discussion. Since $u_{it} = \sum_{j=0}^{\infty}\sum_{l=1}^{N}B_{j,il}\varepsilon_{l,t-j}$, we have
\begin{eqnarray*}
	\|N^{-1/2}\sum_{i=1}^{N}[u_{it}^2-E(u_{it}^2)]\|_4 &\leq& \|N^{-1/2}\sum_{i=1}^{N} \sum_{j=0}^{\infty}\sum_{l=1}^{N}B_{j,il}^2(\varepsilon_{l,t-j}^2-1)\|_4\\
	&&+4\|N^{-1/2}\sum_{i=1}^{N} \sum_{j=0}^{\infty}\sum_{k=1}^{\infty}\sum_{l=1}^{N-1}\sum_{p=1}^{N-l}B_{j,il}B_{j+k,i(l+p)}\varepsilon_{l,t-j}\varepsilon_{l+p,t-j-k}\|_4\\
	&=& A_1 + 4A_2.
\end{eqnarray*}
For $A_1$, write
\begin{eqnarray*}
	A_1^4 &\leq& O(1)\frac{1}{N^2}\sum_{j=0}^{\infty}\sum_{l=1}^{N} \left(\sum_{i=1}^{N}B_{j,il}^2 \right)^4\\
	&&+O(1) \frac{1}{N^2}\sum_{i_1,i_2,i_3,i_4}\sum_{j=0}^{\infty}\sum_{k=1}^{\infty}\sum_{l=1}^{N-1}\sum_{p=1}^{N-l} B_{j,i_1l}^2 B_{j,i_2l}^2 B_{j+k,i_3(l+p)}^2B_{j+k,i_4(l+p)}^2\\
	&=&A_{11} + A_{12}.
\end{eqnarray*}
Since $\max_l|\sum_{i=1}^{N}B_{j,il}^2|\leq |B_j|^2=O(\rho^{2j})$, we have $A_{11}=O(1/N)$. Similarly, we have
$$
A_{12}\leq \sum_{j=0}^{\infty}\sum_{k=1}^{\infty}O(\rho^{2j})O(\rho^{2j+2k}) =O(1).
$$
For $A_2$, by Minkowski inequality, we have
\begin{eqnarray*}
	A_2 &\leq& \sum_{j=0}^{\infty}\sum_{k=1}^{\infty}\|N^{-1/2}\sum_{l=1}^{N-1}\sum_{p=1}^{N-l}\sum_{i=1}^{N}B_{j,il}B_{j+k,i(l+p)}\varepsilon_{l,t-j}\varepsilon_{l+p,t-j-k}\|_4.
\end{eqnarray*}
Then it suffices to show $\|N^{-1/2}\sum_{l=1}^{N-1}\sum_{p=1}^{N-l}\sum_{i=1}^{N}B_{j,il}B_{j+k,i(l+p)}\varepsilon_{l,t-j}\varepsilon_{l+p,t-j-k}\|_4 = O(\rho^{j+k})$. Note that  $|B_j|_1 = \max_{1\leq p\leq N}\sum_{l=1}^{N}|B_{j,lp}| =O(\rho^j)$ and  $|B_j|_\infty = \max_{1\leq l\leq N}\sum_{p=1}^{N}|B_{j,lp}| =O(\rho^j)$, we have
\begin{eqnarray*}
	&&\|N^{-1/2}\sum_{l=1}^{N-1}\sum_{p=1}^{N-l}(\sum_{i=1}^{N}B_{j,il}B_{j+k,i(l+p)})\varepsilon_{l,t-j}\varepsilon_{l+p,t-j-k}\|_4^4\\
	&=& O(1) N^{-2}\sum_{l_1,l_2=1}^{N-1}\sum_{p_1=1}^{N-l_1}\sum_{p_2=1}^{N-l_2}(\sum_{i=1}^{N}B_{j,il_1}B_{j+k,i(l_1+p_1)})^2(\sum_{i=1}^{N}B_{j,il_2}B_{j+k,i(l_2+p_2)})^2\\
	&\leq&O(1) \left(N^{-1}\sum_{l=1}^{N}\sum_{p=1}^{N}(\sum_{i=1}^{N}|B_{j,il}|\cdot|B_{j+k,ip}|)^2\right)^2\\
	&\leq&  O(1) \left(N^{-1}\sum_{l=1}^{N}(\sum_{i=1}^{N}|B_{j,il}|\cdot\sum_{p=1}^{N}|B_{j+k,ip}|)^2\right)^2=O(\rho^{4(j+k)}).
\end{eqnarray*}
Hence, $A_2 \leq \sum_{j=0}^{\infty}\sum_{k=1}^{\infty}O(\rho^{j+k})=O(1)$.

Consider Assumption C (iv). Here, we consider $T^{-2}N^{-1}\sum_{t,s,u,v}\sum_{i,j}|\mathrm{cov}(u_{it}u_{is},u_{ju}u_{jv})| = O(1)$ since the term $T^{-1}N^{-2}\sum_{t,s}\sum_{i,j,k,l}|\mathrm{cov}(u_{it}u_{is},u_{ju}u_{jv})| = O(1)$ can be verified in a similar way. Without loss of generality, we only consider the case of $t>s>u>v$. Write
\begin{eqnarray*}
	u_{it}u_{is} - E(u_{it}u_{is}) &=& \sum_{k=0}^{\infty}\sum_{l=1}^{N}B_{t-s+k,il}B_{k,il}(\varepsilon_{l,s-k}^2-1)\\
	&& + \sum_{k=0}^{\infty}\sum_{j=t-s+k+1}^{\infty}\sum_{l,p=1,l\neq p}^{N}B_{j,il}B_{k,ip}\varepsilon_{l,t-j}\varepsilon_{p,s-k}\\
	&=&I_1 + I_2
\end{eqnarray*}
and
\begin{eqnarray*}
	u_{ju}u_{jv} - E(u_{ju}u_{jv}) &=& \sum_{k=0}^{\infty}\sum_{l=1}^{N}B_{u-v+k,jl}B_{k,jl}(\varepsilon_{l,v-k}^2-1)\\
	&& + \sum_{k=0}^{\infty}\sum_{n=u-v+k+1}^{\infty}\sum_{l,p=1,l\neq p}^{N}B_{n,jl}B_{k,jp}\varepsilon_{l,u-n}\varepsilon_{p,v-k}\\
	&=&I_3 + I_4.
\end{eqnarray*}

Next, we prove $T^{-2}N^{-1}\sum_{t,s,u,v}\sum_{i,j}|E(I_1 \times I_3)|$ and $T^{-2}N^{-1}\sum_{t,s,u,v}\sum_{i,j}|E(I_2 \times I_4)|$ since the rest terms can be done in a similar way.

Note that $\max_{1\leq j\leq l}\sum_{l=1}^{N}|B_{k,jl}|=O(\rho^{k})$ and $\max_{1\leq l\leq l}\sum_{j=1}^{N}|B_{k,jl}|=O(\rho^{k})$. Then we have
\begin{eqnarray*}
	&&T^{-2}N^{-1}\sum_{t,s,u,v}\sum_{i,j}|E(I_1 \times I_3)|\\
	&=& T^{-2}N^{-1}\sum_{t,s,u,v}\sum_{i,j}|\sum_{k=0}^{\infty}\sum_{l=1}^{N}B_{t-v+k,il}B_{s-v+k,il}B_{u-v+k,jl}B_{k,jl}E[(\varepsilon_{it}^2-1)^2]|\\
	&\leq&O(1)T^{-2}N^{-1}\sum_{t,s,u,v}\sum_{k=0}^{\infty}\sum_{l=1}^{N}\sum_{i,j}|B_{t-v+k,il}B_{s-v+k,il}B_{u-v+k,jl}B_{k,jl}|\\
	&\leq& O(1)T^{-2}N^{-1}\sum_{t,s,u,v}\sum_{k=0}^{\infty}\left(\sum_{i=1}^{N}|B_{t-v+k,il}|\sum_{l=1}^{N}|B_{s-v+k,il}|\right)\left(\sum_{j=1}^{N}|B_{u-v+k,jl}|\sum_{l=1}^{N}|B_{k,jl}|\right)\\
	&=&O(1)T^{-2}N^{-1}\sum_{t,s,u,v}\sum_{k=0}^{\infty}O(\rho^{t-v+k})O(\rho^{s-v+k})O(\rho^{u-v+k})O(\rho^{k})=O(1/N).
\end{eqnarray*}

In addition, similar to the proof of term $A_2$, we can show $T^{-2}N^{-1}\sum_{t,s,u,v}\sum_{i,j}|E(I_2 \times I_4)|=O(1)$. The proof is now completed.  \hspace*{\fill}{$\blacksquare$}
	
	\medskip

	\noindent \textbf{Verification of Example \ref{Exam2}}:
	\medskip
	
\noindent (1).	Without loss of generality, let $\delta=4$ in what follows. Write
		\begin{eqnarray*}
		\|\overline{U}_t - \overline{U}_t^* \|_4^4&=&E\left|\sum_{i=1}^{N}\Omega_{N,i} (v_{i,t}- v_{i,t}^*)\right|^4\\
		&= &E\left|\sum_{i=1}^{N}\Omega_{N,i}^2 (v_{i,t}- v_{i,t}^*)^2\right|^2 + 4E\left|\sum_{i=1}^{N-1}\sum_{k=i+1}^{N}\Omega_{N,i}\Omega_{N,k}(v_{i,t} - v_{i,t}^*)(v_{k,t} - v_{k,t}^*) \right|^2\nonumber \\
		&= &I_1 + I_2,
	\end{eqnarray*}
	where $\frac{1}{\sqrt{N}}1_{N}^\top \Omega^{1/2}  =(\Omega_{N,1},\ldots,\Omega_{N,N})$. Since $|\Omega| < \infty$, we have $\sum_{i=1}^{N}\Omega_{N,i}^2 \leq |\Omega| < \infty$. As $\{v_{i,t}\}$ are independent over $i$, we have
	$$
	I_1 \leq \max_i\|v_{i,t} - v_{i,t}^*\|_4^4 \left(\sum_{i=1}^{N}\Omega_{N,i}^2\right)^2 =O \left(\max_i\|v_{i,t} - v_{i,t}^*\|_4^4 \right),
	$$
	and
	$$
	I_2 \leq \max_i\|v_{i,t} - v_{i,t}^*\|_4^4 \left(\sum_{i=1}^{N-1}\Omega_{N,i}^2\sum_{k=1}^{N-i}\Omega_{N,k}^2\right)=O \left(\max_i\|v_{i,t} - v_{i,t}^*\|_4^4 \right).
	$$
	
	We next prove $\|v_{i,t} - v_{i,t}^*\|_\delta = O(\rho^t)$ for some $0 < \rho < 1$. In particular, $v_{i,t}$ have the representation
	$$
	v_{i,t} = \sqrt{c_i}\varepsilon_{i,t} \left(1 + \sum_{n=1}^\infty \sum_{1\leq l_1,...,l_n\leq r}\prod_{j=1}^n  (C_{i,l_j} + D_{i,l_j}\varepsilon_{i,t-l_1-\cdots l_n}^2)\right)^{1/2}.
	$$
	Define 
	$$
	v_{i,t}^* = \sqrt{c_i}\varepsilon_{i,t} \left(1 + \sum_{n=1}^\infty \sum_{1\leq l_1,...,l_n\leq r}\prod_{j=1}^n  (C_{i,l_j} + D_{i,l_j}\varepsilon_{i,t-l_1-\cdots l_n}^{*2})\right)^{1/2},
	$$
	where $\varepsilon_{i,t}^* = \varepsilon_{i,t}$ for $t\geq 1$. Then, by $|x-y|^\delta \leq |x^2-y^2|^{\delta/2}$ for $x,y\geq 0$ and $\sum_{j=1}^{r}\|C_{ij}+D_{ij}\varepsilon_{i,0}^2\|_2<1$, we have for some $0< \rho <1$
	\begin{eqnarray*}
	\|v_{i,t} - v_{i,t}^* \|_4^4 &\leq &4 \|\sum_{n=[t/r]}^{\infty} \sum_{1\leq l_1,...,l_n\leq r}\prod_{j=1}^n  (C_{i,l_j} + D_{i,l_j}\varepsilon_{i,t-l_1-\cdots l_n}^2)\|_2^2\\
	&\leq & 4 \left(\sum_{n=[t/r]}^{\infty}(\sum_{j=1}^{r}\|C_{ij}+D_{ij}\varepsilon_{i,0}^2\|_2)^n \right)^2 = O(\rho^t).
	\end{eqnarray*}
	
Based on the above development, 
\begin{eqnarray*}
	\sum_{t=0}^{\infty}t^2\|\overline{U}_t - \overline{U}_t^*\|_4 \le O(1)\sum_{t=0}^{\infty}t^2\rho^t<\infty,
\end{eqnarray*}
so Assumption \ref{Assumption1} is met.

	\medskip

\noindent (2). The proof of this part is similar with the proof of Example 1, so omitted here. \hspace*{\fill}{$\blacksquare$}
\bigskip
	
\noindent \textbf{Proof of Proposition \ref{Prop2.1}:}
	\medskip
	
 (1).  Note that $\overline{U}_t$ can be decomposed as $\overline{U}_t= \sum_{m=0}^\infty \mathcal{P}_{t-m}(\overline{U}_t)$, so we write
	
	\begin{eqnarray*}
		\|\overline{U}_t\|_{\delta^*} \le \sum_{m=0}^\infty \| \mathcal{P}_{t-m}(\overline{U}_t)\|_{\delta^*} \le \sum_{m=0}^\infty \theta_{m,\delta}^U\le \sum_{m=0}^\infty( \lambda_{m,\delta}^U+\lambda_{m+1,\delta}^U)<\infty,
	\end{eqnarray*}
	where the second inequality follows from \eqref{eq.a.2}, the third inequality follows from \eqref{eq.a.1}, and the last inequality follows from Assumption \ref{Assumption1}. Then the first result follows.
	
	\medskip
	
(2). Since $E[\overline{U}_t] = E[\overline{U}_t^*\mid \mathscr{F}_{0}] = 0$, we have
	
	\begin{eqnarray*}
		\sum_{t=1}^{\infty}t^2|E[\overline{U}_t\overline{U}_0]| &=&\sum_{t=1}^{\infty}t^2|E[E(\overline{U}_t-\overline{U}_t^*\mid\mathscr{F}_{0})\overline{U}_0]|  \le \sum_{t=1}^{\infty}t^2\|\overline{U}_t-\overline{U}_t^*\|_2 \|\overline{U}_0\|_2 =O(1),
	\end{eqnarray*}
	where the first equality follows from the independence between $\overline{U}_t^*$ and $\overline{U}_0$, and the inequality follows from Cauchy-Schwarz inequality. 
	
	\medskip
	
		(3).   Write
	
	\begin{eqnarray*}
		&&\left\|\sum_{t=1}^{T}\overline{U}_t\right\|_\delta = \left\|\sum_{t=1}^{T}\sum_{m=0}^{\infty}\mathcal{P}_{t-m} (\overline{U}_t) \right\|_\delta\leq \sum_{m=0}^{\infty}\left\|\sum_{t=1}^{T}\mathcal{P}_{t-m} (\overline{U}_t)\right\|_\delta\\
		&\leq& O(1)\sum_{m=0}^{\infty} \left( E \left[\sum_{t=1}^T|\mathcal{P}_{t-m} (\overline{U}_t)|^{2}\right]^{\delta/2} \right)^{\frac{1}{\delta}}= O(1)\sum_{m=0}^{\infty} \left( E \left[\sum_{t=1}^T|\mathcal{P}_{t-m} (\overline{U}_t)|^{2}\right]^{\delta/2} \right)^{\frac{2}{\delta}\cdot \frac{1}{2}}\\
		&\leq& O(1)\sum_{m=0}^{\infty} \left\{ \sum_{t=1}^T\left( E \left[|\mathcal{P}_{t-m} (\overline{U}_t)|^{2}\right]^{\delta/2} \right)^{\frac{2}{\delta}} \right\}^{\frac{1}{2}}=O(1) \sum_{m=0}^{\infty} \left\{ \sum_{t=1}^T\left( E  |\mathcal{P}_{t-m} (\overline{U}_t)|^{\delta}\right)^{\frac{2}{\delta}} \right\}^{\frac{1}{2}}\\
		&\le & O(1) \sum_{m=0}^{\infty} \left(T| \theta_{m,\delta}^{U}|^2 \right)^{\frac{1}{2}} =O(1) T^{\frac{1}{2}} \sum_{m=0}^{\infty}\theta_{m,\delta}^U=O(T^{1/2}) ,
	\end{eqnarray*}
	where the first inequality follows from the triangle inequality, the second inequality follows from the Burkholder's inequality, the third inequality follows from the Minkowski inequality,  the fourth inequality follows from \eqref{eq.a.2}, and the fifth equality follows from \eqref{eq.a.1} and Assumption \ref{Assumption1}.
	
	(4). Similar to $\overline{U}_{t\gamma}^{*}$, we define  $\overline{U}_{t\gamma}^{**}$ using $\{\varepsilon_t^{\prime\prime}\}$, which is another independent copy of $\{\varepsilon_t\}$. Then for $1\leq t \leq \gamma$
	
	\begin{eqnarray*}
		\|\overline{U}_{t\gamma} - \overline{U}_{t\gamma}^*\|_\delta &=& \|E[\overline{U}_{t\gamma} - \overline{U}_{t\gamma}^* + \overline{U}_{t\gamma}^* \mid \mathscr{F}_{t,t-\gamma} ]  - E[\overline{U}_{t\gamma}^* - \overline{U}_{t\gamma}^{**} + \overline{U}_{t\gamma}^{**} \mid  \mathscr{F}_{t,1},\mathscr{F}_{0,t-\gamma}^*  ] \|_\delta\\
		&=& \|E[\overline{U}_{t\gamma} - \overline{U}_{t\gamma}^* \mid \mathscr{F}_{t,t-\gamma} ]  - E[\overline{U}_{t\gamma}^* - \overline{U}_{t\gamma}^{**} \mid \mathscr{F}_{t,1},\mathscr{F}_{0,t-\gamma}^* ] \|_\delta\\
		&\leq& 2\|\overline{U}_{t} - \overline{U}_{t}^*\|_{\delta}=2\lambda_{t,\delta}^U,
	\end{eqnarray*}
	where the second equality follows from the fact that $E[\overline{U}_{t\gamma}^* \mid \mathscr{F}_{t,t-\gamma} ] =E[\overline{U}_{t\gamma}^{**} \mid \mathscr{F}_{t,1},\mathscr{F}_{0,t-\gamma}^*]$, and the inequality follows from the Jensen's inequality. For $t > \gamma$,
	\begin{eqnarray*}
		\|\overline{U}_{t\gamma} - \overline{U}_{t\gamma}^*\|_\delta =0.
	\end{eqnarray*}
	
	In connection with Assumption \ref{Assumption1}, the result follows immediately. 
	
	\hspace*{\fill}{$\blacksquare$}
\bigskip
	
	\noindent \textbf{Proof of Theorem \ref{Theo2.1}:}
	\medskip
	
	(1). Let $S_{\mathbb{N},L} = \frac{1}{\sqrt{T}}\sum_{t=1}^{T}\sum_{l=0}^{L-1}\mathcal{P}_{t-l} (\overline{U}_t)$ and $\widehat{S}_{\mathbb{N},L} = \frac{1}{\sqrt{T}}\sum_{t=1}^{T}\sum_{l=0}^{L-1}\mathcal{P}_{t} (\overline{U}_{t+l})$, in which $L\to \infty$ and $L/T\to0$. 
	
	Note that $\overline{U}_t = \sum_{l=0}^{\infty}\mathcal{P}_{t-l} (\overline{U}_t)$ and $\{\mathcal{P}_{t-l} (\overline{U}_t)\}_{t=1}^T$ is a sequence of martingale differences, and thus
	
	\begin{eqnarray*}
		\|S_{\mathbb{N},L} - S_{\mathbb{N}}\|_2 &=&\left\|\frac{1}{\sqrt{T}}\sum_{t=1}^{T}\sum_{l=L}^{\infty}\mathcal{P}_{t-l} (\overline{U}_t) \right\|_2 \leq \frac{1}{\sqrt{T}}\sum_{l=L}^{\infty}\left\|\sum_{t=1}^{T} \mathcal{P}_{t-l} (\overline{U}_t)\right\|_2 \\
		&\leq&  O(1)\frac{1}{\sqrt{T}}\sum_{l=L}^{\infty}\left\{\sum_{t=1}^{T} E\left[ \left(\mathcal{P}_{t-l} (\overline{U}_t)\right)^2\right]\right\}^{1/2} \le O(1)\sum_{l=L}^{\infty}\theta_{l,2}^U \to 0,
	\end{eqnarray*}
	where the second inequality follows from Burkholder's inequality, the third inequality follows from  \eqref{eq.a.2}, and the last step follows from $L\to \infty$. Similarly, by using Burkholder's inequality, we have as $L\to \infty$ and $L/T\to0$
	
	\begin{eqnarray*}
		\|\widehat{S}_{\mathbb{N},L} - S_{\mathbb{N},L}\|_2 &\leq& \frac{1}{\sqrt{T}}\sum_{l=0}^{L-1} \left\|\sum_{t=1}^{l}\mathcal{P}_{t-l} (\overline{U}_{t})\right\|_2  + \frac{1}{\sqrt{T}}\sum_{l=0}^{L-1}\left\|\sum_{t=T-l+1}^{T}\mathcal{P}_{t} (\overline{U}_{t+l})\right\|_2  =O(1) \frac{\sqrt{L}}{\sqrt{T}}\to 0.
	\end{eqnarray*}
	Hence, we have $\|\widehat{S}_{\mathbb{N},L} - S_{\mathbb{N}}\|_2 \to 0$.
	
	Note that $\{\sum_{l=0}^{L-1}\mathcal{P}_{t} (\overline{U}_{t+l})\}_{t=1}^{T}$ is a sequence of martingale differences subject to $\mathscr{F}_t$, so the asymptotic normality can be easily obtained by using a standard martingale central limit theory. The proof of the first result is now completed. 
	
	\medskip
	
	(2).  We first define $S_{\mathbb{N}\gamma} = \frac{1}{\sqrt{T}} \sum_{t=1}^T \overline{U}_{t\gamma}$, and establish some rates of convergence associated with $S_{\mathbb{N}} $ and $S_{\mathbb{N}\gamma}$. Define $D_{t,j} = E(\overline{U}_{t} \mid \mathscr{F}_{t,t-j}) - E(\overline{U}_{t} \mid \mathscr{F}_{t,t-j+1})$. Then $D_{t,j}$, $t=T,...,1$ form martingale differences with respect to $\mathscr{F}_{\infty,t-j}$ and $\|D_{t,j}\|_\delta \leq \theta_{j,\delta}^{U}$. By Burkholder's inequality and Minkowski's inequality, we have
	$$
	\left\|\frac{1}{\sqrt{T}} \sum_{t=1}^T D_{t,j} \right\|_{\delta}^2 \leq O(1) \sum_{t=1}^{T}\left\|\frac{1}{\sqrt{T}}D_{t,j} \right\|_\delta^2 =O(1) \theta_{j,\delta}^{U,2}.
	$$
	
	Since $\overline{U}_{t}-\overline{U}_{t\gamma}=\sum_{j=\gamma+1}^{\infty}D_{t,j}$, we have for large $\gamma>0$
	$$
	\|S_{\mathbb{N}} -S_{\mathbb{N}\gamma}\|_\delta\leq \sum_{j=\gamma+1}^{\infty} \left\|\frac{1}{\sqrt{T}} \sum_{t=1}^T D_{t,j}\right\|_\delta\leq O(1)\gamma^{-2}\sum_{j=\gamma}^{\infty}j^2\theta_{j,\delta}^{U} =O(\gamma^{-2}).
	$$

	Hence, by Cauchy-Schwarz inequality and Proposition \ref{Prop2.1}, we have
	\begin{eqnarray*}
		|E(S_{\mathbb{N}}^{2}) - E(S_{\mathbb{N}\gamma}^{2})|\leq \|S_{\mathbb{N}}-S_{\mathbb{N}\gamma} \|_2\cdot \|S_{\mathbb{N}}+S_{\mathbb{N}\gamma} \|_2=O(\gamma^{-2}).
	\end{eqnarray*}
	Similarly, by H\"older's inequality, we have
	\begin{eqnarray*}
		|E(S_{\mathbb{N}}^{3}) - E(S_{\mathbb{N}\gamma}^{3})|\leq \|S_{\mathbb{N}}-S_{\mathbb{N}\gamma} \|_3\cdot\left( \|S_{\mathbb{N}}\|_3^{2}+\|S_{\mathbb{N}}\|_3\|S_{\mathbb{N}\gamma}\|_3+\|S_{\mathbb{N}\gamma} \|_3^{2} \right)=O(\gamma^{-2}).
	\end{eqnarray*}
	
	We are now ready to start the investigation:
	
	\begin{eqnarray*}
		\sup_{w\in \mathbb{R}}|\mathrm{Pr}(S_{\mathbb{N}} \le  w) - \Psi_{\mathbb{N}}(w)| &\leq& \sup_{w\in \mathbb{R}}|\mathrm{Pr}(S_{\mathbb{N}} \le  w) - \mathrm{Pr}(\overline{Z} + \widetilde{Z} \le  w)|\\
		&&+ \sup_{w\in \mathbb{R}}|\mathrm{Pr}(\overline{Z} + \widetilde{Z} \le  w) - \Psi_{\mathbb{N}}(w)|\\
		&:=&I_{T,1} + I_{T,2},
	\end{eqnarray*}
	where $\overline{Z}$ and $\widetilde{Z}$ are defined in Section \ref{SecB.1} of the supplementary Appendix B.
	
	Consider $I_{T,1}$. By $e^{ix} = \cos(x) + i \sin(x)$ and Lipschitz continuity, we have
	
	\begin{eqnarray*}
		|E(e^{ixS_{\mathbb{N}}}) - E(e^{ixS_{\mathbb{N}\gamma}})| = O(|x|\gamma^{-2}).
	\end{eqnarray*}
	Hence, by the Berry's smoothing inequality (Lemma 2, XVI.3 in \citealp{feller2008introduction}), we have
	
	\begin{eqnarray*}
		I_{T,1} &\leq& \int_{-c\sqrt{T}}^{c\sqrt{T}} |E(e^{ixS_{\mathbb{N}}}) - E(e^{ix(\overline{Z}+\widetilde{Z})}) | \frac{1}{|x|}\mathrm{d}x + \mathscr{C}_{c\sqrt{T}}\\
		&=& \mathscr{U}_{T} + O(\sqrt{T}/\gamma^{2}) + \mathscr{C}_{c\sqrt{T}},
	\end{eqnarray*}
	where $\mathscr{U}_{T}$ is defined in Section \ref{SecB.1}, and $\mathscr{C}_{c\sqrt{T}}$ is defined with respect to $\overline{U}_{t}$ and is given in the beginning of Section \ref{SecA.1}.  Selecting $\gamma$ large enough and using Lemma \ref{L7}.2, we have
	\begin{eqnarray*}
		\sup_{w\in \mathbb{R}}|\mathrm{Pr}(S_{\mathbb{N}} \le  w) - \mathrm{Pr}(\overline{Z} + \widetilde{Z} \le  w)| = O(T^{-1}(\log T)^5) + \mathscr{C}_{c\sqrt{T}}.
	\end{eqnarray*}
	
	We now consider $\mathscr{C}_{c\sqrt{T}}$. Recall that we have defined $(\mathscr{T}_{a}^b)^\diamond$, $H_t$'s, and $S_{\mathbb{N}}^{\diamond}$ in Section \ref{SecB.1} of the online Appendix B below. First, note that selecting $a>0$ such that $c > ab$ and using $E(e^{ix S_{\mathbb{N}}^{\diamond}}) = 0$ for $|x| > \sqrt{T}|ab|$, we have $\sup_{w\in\mathbb{R}}(\mathscr{T}_{c\sqrt{T}}^{\infty})^\diamond(w)=0$. Also, note that by Taylor expansion,
	
	\begin{eqnarray*}
		e^{ixS_{\mathbb{N}}^{\diamond}} - e^{ixS_{\mathbb{N}}} = \frac{\partial e^{ixz}}{\partial z}\big|_{z=S_{\mathbb{N}}}\cdot\frac{1}{\sqrt{T}}(H_T - H_0) +  \frac{\partial^2 e^{ixz}}{\partial z^2}\big|_{z=S_{\mathbb{N}}}\cdot\frac{1}{T}(H_T - H_0)^2 + o\left(\frac{1}{T}(H_T - H_0)^2\right).
	\end{eqnarray*}
	Hence, it is easy to know that $|E(e^{ixS_{\mathbb{N}}}) - E(e^{ixS_{\mathbb{N}}^{\diamond}})| = O(T^{-1})$, which thus yields $\mathscr{C}_{c\sqrt{T}} = O(T^{-1})$. We now can conclude that $I_{T,1} = O(T^{-1}(\log T)^5)$. 
	
	\medskip
	
	Finally, we consider $I_{T,2}$. Define $\Psi_{\mathbb{N}\gamma}(w)$ in analogy to $\Psi_{\mathbb{N}}(w)$ with respect to $S_{\mathbb{N}\gamma}$. Write
	
	\begin{eqnarray*}
		I_{T,2} &\leq& \sup_{w\in \mathbb{R}}|\mathrm{Pr}(\overline{Z} + \widetilde{Z} \le  w) - \Psi_{\mathbb{N}\gamma}(w)| + \sup_{w\in \mathbb{R}}|\Psi_{\mathbb{N}\gamma}(w) - \Psi_{\mathbb{N}}(w)|=O(\gamma^{-1} +\gamma^{-2}),
	\end{eqnarray*}
	where the last step follows from Lemma \ref{L5}.2, and the facts that $|E(S_{\mathbb{N}}^{2}) - E(S_{\mathbb{N}\gamma}^{2})| =O(\gamma^{-2})$ and $ |E(S_{\mathbb{N}}^{3}) - E(S_{\mathbb{N}\gamma}^{3})|=O(\gamma^{-2})$ as shown in the beginning of the proof of this result. Selecting $\gamma = T/(\log T)^5$, we have $I_{T,2} = O(T^{-1}(\log T)^5)$. 
	
	Collecting the above results, the proof is now completed.\hspace*{\fill}{$\blacksquare$}
\bigskip
	
	\noindent \textbf{Proof of Corollary \ref{Cor2.1}:}
	\medskip
	
Given Theorem 2.1 (2), to prove this Corollary, we need to show $E(S_{\mathbb{N}}^3) = O(1/\sqrt{NT})$ under (a). the conditions of Example 1.2 or (b). the conditions of Example 2.2.

Note that 
\begin{eqnarray}
	E(S_{\mathbb{N}}^{3})&\leq& O(1) \frac{1}{T^{3/2}}\sum_{1\leq k\leq s\leq t\leq T} |E (\overline{U}_k\overline{U}_s\overline{U}_t)| \leq O(1)\frac{1}{T^{3/2}}\sum_{k=1}^{T}\sum_{s=k}^{k+\gamma} \sum_{t=s}^{s+\gamma}   |E (\overline{U}_k\overline{U}_s\overline{U}_t)|\nonumber \\
	&\leq&O(1)\frac{1}{T^{3/2}}\sum_{k=1}^{T}\sum_{s=k}^{k+\gamma} \left(\sum_{t=s}^{s+s-k-1}   |E (\overline{U}_k\overline{U}_s\overline{U}_t)| + \sum_{t=2s-k}^{s+\gamma}   |E (\overline{U}_k\overline{U}_s\overline{U}_t)|\right) \nonumber
\end{eqnarray}
By the proof of Lemma \ref{L1} (3), for $k\leq s\leq t$, we have
$$
|E (\overline{U}_k\overline{U}_s\overline{U}_t)| 
		\leq \|\overline{U}_k\|_3 \cdot \left(\| \overline{U}_s\|_3 \|   \overline{U}_t - \overline{U}_t^{(t-k,*)}\|_3 +\|\overline{U}_s- \overline{U}_s^{(s-k,*)} \|_3\|\overline{U}_t  \|_3\right).
$$
Then by the proof of Lemma \ref{L1} (3), in order to prove $E(S_{\mathbb{N}}^3) = O(1/\sqrt{NT})$ it suffices to show $\|\overline{U}_t\|_3 = O(N^{-1/6})$ and $\|   \overline{U}_t - \overline{U}_t^{(t-k,*)}\|_3 = O(N^{-1/6} \rho^{t-k})$.

We first consider the example of linear process. For $\|\overline{U}_t\|_3$, by $u_{it} = \sum_{j=0}^{\infty}\sum_{l=1}^{N}B_{j,il}\varepsilon_{l,t-j}$ and $\max_{1\leq l\leq N}|\sum_{i=1}^{N}B_{j,il}| =O(\rho^j)$, we have
\begin{eqnarray*}
E(\overline{U}_t^3)&=&E\left[\left(\frac{1}{\sqrt{N}}\sum_{j=0}^{\infty}\sum_{l=1}^{N}\sum_{i=1}^{N}B_{j,il}\varepsilon_{l,t-j}\right)^3\right]\\
&=&\frac{1}{N^{3/2}}\sum_{j=0}^{\infty}\sum_{l=1}^{N}\left(\sum_{i=1}^{N}B_{j,il}\right)^3E\left[\varepsilon_{l,t-j}^3\right]=O(1/\sqrt{N}),
\end{eqnarray*}
which implies that $\|\overline{U}_t\|_3 = O(N^{-1/6})$.

Consider $\|\overline{U}_t - \overline{U}_t^{(t-k,*)}\|_3$. Similarly, by Minkowski inequality, we have
\begin{eqnarray*}
\|\overline{U}_t - \overline{U}_t^{(t-k,*)}\|_3&=&\|\frac{1}{\sqrt{N}}\sum_{j=t-k}^{\infty}\sum_{l=1}^{N}\sum_{i=1}^{N}B_{j,il}(\varepsilon_{l,t-j}-\varepsilon_{l,t-j}^*)\|_3\\
&\leq&2\sum_{j=t-k}^{\infty}\|\frac{1}{\sqrt{N}}\sum_{l=1}^{N}\sum_{i=1}^{N}B_{j,il}\varepsilon_{l,t-j}\|_3 \\
&=& O(N^{-1/6})\sum_{j=t-k}^{\infty}\rho^{j} =  O(N^{-1/6} \rho^{t-k}).
\end{eqnarray*}

We next consider the example of high-dimensional GARCH process.  For $\|\overline{U}_t\|_3$, by $\Omega^{1/2} = (\Omega_{il})_{i,l\in[\mathbb{N}]}$ $u_{it} = \sum_{l=1}^{N}\Omega_{il}v_{l,t-j}$ and $\max_{1\leq l\leq N}|\sum_{i=1}^{N}\Omega_{il}| =O(1)$, we have
$$
E(\overline{U}_t^3) = E\left[\left(\frac{1}{\sqrt{N}}\sum_{l=1}^{N}\sum_{i=1}^{N}\Omega_{il}v_{l,t-j}\right)^3\right] = \frac{1}{N^{3/2}}\sum_{l=1}^{N}\left(\sum_{i=1}^{N}\Omega_{il}\right)^3E\left[v_{l,t-j}^3\right]= O(1/\sqrt{N}).
$$
Consider $\|\overline{U}_t - \overline{U}_t^{(t-k,*)}\|_3$. Similarly, by Minkowski inequality, we have
\begin{eqnarray*}
E\left[(\overline{U}_t-\overline{U}_t^{(t-k,*)})^3\right] &=&\frac{1}{N^{3/2}}\sum_{l=1}^{N}(\sum_{i=1}^{N}\Omega_{il})^3E\left[(v_{l,t}-v_{l,t-j}^{(t-k,*)})^3\right]\\
&=& O\left(N^{-1/2} \rho^{3(t-k)}\right).
\end{eqnarray*}

The proof is now completed.

	\hspace*{\fill}{$\blacksquare$}
	\medskip
	
	Since an important part of the proof of Theorem 2.2 follows from one part of the proof of Theorem 2.3, we give the proof of Theorem 2.3 first.
	\medskip
	
	\noindent \textbf{Proof of Theorem \ref{Theo2.3}:}
	\medskip
	
	(1). For ease of notation, let $s_{\mathbb{N}}^{*2} := E^*[S_{\mathbb{N}}^{*2}] = \frac{1}{T}\sum_{t,s=1}^{T}\overline{U}_t\overline{U}_s a\left( \frac{t-s}{\ell}\right)$, and write
	
	\begin{eqnarray*}
		\ell^q\left(E[s_{\mathbb{N}}^{*2}]-s_{\mathbb{N}}^{2}\right)&=& \ell^q \sum_{k=-\ell}^{\ell}\left[a\left(k/\ell\right)-1\right]E(\overline{U}_k\overline{U}_0) - 2\ell^q \sum_{k=1}^{\ell}\frac{k}{T}\left[a\left(k/\ell\right)-1\right]E(\overline{U}_k\overline{U}_0)\\
		&&- 2\ell^q\sum_{k=\ell+1}^{T-1}\frac{T-k}{T}E(\overline{U}_k\overline{U}_0) := I_1+I_2+I_3,
	\end{eqnarray*}
	where the definitions of $I_1$ to $I_3$ are obvious. We consider these three terms one by one.
	
	Consider $I_{1}$. By Assumption \ref{Assumption3}, for $\forall\epsilon > 0$, we choose $\nu_\epsilon > 0$ such that  
	
	\begin{eqnarray*}
		|k/\ell| < \nu_\epsilon\quad \text{and}\quad \left|\frac{1-a(k/\ell)}{|k/\ell|^q}-c_q \right| < \epsilon.
	\end{eqnarray*}
	Letting $\ell_T^* = \lfloor\nu_\epsilon\ell \rfloor$, write
	
	\begin{eqnarray*}
		I_1 &=& \sum_{k=-\ell_T^*}^{\ell_T^*}\frac{ a\left(k/\ell\right)-1}{\left|k/\ell\right|^q}|k|^qE(\overline{U}_k\overline{U}_0) + 2\sum_{k=\ell_T^*+1}^{\ell}\frac{ a\left(k/\ell\right)-1}{\left|k/\ell\right|^q}|k|^qE(\overline{U}_k\overline{U}_0).
	\end{eqnarray*}
	Then, by Proposition \ref{Prop2.1}, it is easy to see that the first term of the right-hand side converges to $- c_q \sum_{k = -\infty}^{\infty}|k|^qE(\overline{U}_0\overline{U}_{k})$. For the second term, since $|a(\cdot)|\leq M$, $\left|\frac{1-a(k/\ell)}{|k/\ell|^q} \right| \leq (M+1)/\nu_\epsilon^q$ due to the fact that $k/\ell\geq \nu_\epsilon$. Then this term is bounded by
	
	\begin{eqnarray*}
		(M+1)/\nu_\epsilon^q \sum_{k=\ell_T^*+1}^{\infty}k^2|E(\overline{U}_k\overline{U}_0)|,
	\end{eqnarray*}
	which, by Proposition \ref{Prop2.1} again, converges to 0 as $\ell\to \infty$.
	
	\medskip
	
	Consider $I_2$. As $\ell^q/T\to 0$ of Assumption \ref{Assumption2} and $\sum_{k=1}^\infty k^2|E(\overline{U}_k\overline{U}_0)|<\infty$ of Proposition \ref{Prop2.1}, we have
	
	\begin{eqnarray*}
		|I_2| \leq \frac{\ell^q}{T} 2(M+1) \sum_{k=1}^\infty k^2|E(\overline{U}_k\overline{U}_0)| \to 0.
	\end{eqnarray*}
	
	\medskip
	
	Consider $I_3$ and write
	
	\begin{eqnarray*}
		|I_3|\leq 2 \frac{\ell^q}{\ell^2}\sum_{k=\ell+1}^{\infty}k^2|E(\overline{U}_k\overline{U}_0)| \to 0,
	\end{eqnarray*}
	where the last steps follows from the facts that $\frac{\ell^q}{\ell^2}$ is bounded and $\ell \to \infty$. The proof of the first result is now completed.
	
	\medskip
	
	\noindent (2). Define $v(r,s,t)=E(\overline{U}_{0}\overline{U}_{r}\overline{U}_{s}\overline{U}_{t})$, $\sigma(r) = E(\overline{U}_{0}\overline{U}_{r})$ and
	
	\begin{eqnarray*}
		\kappa(r,s,t) = v(r,s,t) - \sigma(r)\sigma(s-t) - \sigma(s)\sigma(r-t)-\sigma(t)\sigma(r-s).
	\end{eqnarray*}
	Then write
	
	\begin{eqnarray*}
		\frac{T}{\ell}\text{Var}\left(s_\mathbb{N}^{*2}\right) &=& \frac{T}{\ell}\sum_{i=-\ell}^{\ell}\sum_{k=-\ell}^{\ell}a\left(\frac{i}{\ell}\right)a\left(\frac{k}{\ell}\right)\text{Cov}\left(\frac{1}{T}\sum_{t=1+|i|}^{T}\overline{U}_t\overline{U}_{t-|i|},\frac{1}{T}\sum_{t=1+|k|}^{T}\overline{U}_t\overline{U}_{t-|k|}\right)\\
		&=&\frac{1}{\ell}\sum_{i=-\ell}^{\ell}\sum_{k=-\ell}^{\ell}a\left(\frac{i}{\ell}\right)a\left(\frac{k}{\ell}\right)\left(\frac{1}{T}\sum_{t=1+|i|}^{T}\sum_{s=1+|k|}^{T}E(\overline{U}_t\overline{U}_{t-|i|}\overline{U}_s\overline{U}_{s-|k|}) \right.\\
		&&\left.- \frac{1}{T}\sum_{t=1+|i|}^{T}\sum_{s=1+|k|}^{T} E(\overline{U}_{t}\overline{U}_{t-|i|})E(\overline{U}_{s}\overline{U}_{s-|k|}) \right)\\
		&=&\frac{1}{\ell}\sum_{i=-\ell}^{\ell}\sum_{k=-\ell}^{\ell}a\left(\frac{i}{\ell}\right)a\left(\frac{k}{\ell}\right)\sum_{r=1-T}^{T-1} \phi_T(r,i,k) [\sigma(r)\sigma(r+k-i) + \\ &&\sigma(r-i)\sigma(r+k) + \kappa(k,-r,i-r) ] := I_1 + I_2 + I_3,
	\end{eqnarray*}
	where, by tedious but trivial calculation (e.g., Chapter 9 of \citealp{Anderson}), it is easy to know that $\lim_{T\to\infty}\phi_T(r,i,k)=1$ for every $r,i,k$, $0\leq\phi_T(r,i,k)\leq 1$ and $\phi_T(r,i,k) \geq 1 - \frac{|r|+|i|+|k|}{T}$.
	
	Consider $I_1$. Write
	\begin{eqnarray*}
		I_1 &=& \frac{1}{\ell}\sum_{u=1-T}^{T-1} \sum_{v=u-2\ell}^{u+2\ell}\sum_{s=\max(u,v)-\ell}^{\min(u,v)+\ell}a\left(\frac{u-s}{\ell}\right)a\left(\frac{v-s}{\ell}\right)\phi_T(u,u-s,v-s) \sigma(u)\sigma(v)\\
		&=&\frac{1}{\ell}\sum_{u,v=-m}^{m}\sum_{s=\max(u,v)-\ell}^{\min(u,v)+\ell}a\left(\frac{u-s}{\ell}\right)a\left(\frac{v-s}{\ell}\right)\phi_T(u,u-s,v-s) \sigma(u)\sigma(v) + o(1)\\
		&=&\frac{1}{\ell}\sum_{u,v=-m}^{m}\sum_{s=-\ell}^{\ell}a\left(\frac{s}{\ell}\right)^2\phi_T(u,u-s,v-s) \sigma(u)\sigma(v) + o(1)\\
		&=&\frac{1}{\ell}\sum_{u,v=-m}^{m}\sum_{s=-\ell}^{\ell}a\left(\frac{s}{\ell}\right)^2\sigma(u)\sigma(v) + o(1) \to  s_{\mathbb{N}}^{4}\int_{-1}^{1}a^2(x)\mathrm{d}x
	\end{eqnarray*}
	by $\sum_{i=-\infty}^{\infty}|\sigma(i)|<\infty$ and selecting some $m\to \infty$ and $m/\ell \to 0$. Similarly, we have 
	
	\begin{eqnarray*}
		I_2 \to s_{\mathbb{N}}^{4}\int_{-1}^{1}a^2(x)\mathrm{d}x.
	\end{eqnarray*}
	
	\medskip
	
	To complete the proof, it suffices to show $I_3 \to 0$. In view of the facts that $a(\cdot)$ is finite and $1/\ell\to 0$, we need only to show $\sum_{r,s,t=-\infty}^{\infty}|\kappa(r,s,t)|<\infty$. By construction, we have $\sum_{r,s,t=-\infty}^{\infty}|\kappa(r,s,t)|\leq O(1)\sum_{0\leq r\leq s\leq t < \infty}|\kappa(r,s,t)|$, so focus on $\sum_{0\leq r\leq s\leq t < \infty}|\kappa(r,s,t)| $ below.
	
	\begin{eqnarray*}
		\sum_{0\leq r\leq s\leq t < \infty}|\kappa(r,s,t)| &=& \sum_{0< r < s < t < \infty}|\kappa(r,s,t)| + \sum_{0 = r < s < t < \infty}|\kappa(r,s,t)| \\
		&&+\sum_{0\leq r < s = t < \infty}|\kappa(r,s,t)| + \sum_{0\leq r = s\leq t < \infty}|\kappa(r,s,t)|\\
		&:=& I_{4} + I_{5} +I_{6} + I_{7}.
	\end{eqnarray*}
	
	The most difficult term to deal with is $I_4$. Since 
	
	\begin{eqnarray*}
		E[\overline{U}_0]E\left[\left(\overline{U}_r\overline{U}_s\overline{U}_t\right)^{(t-0,*)}\right] = E\left[\overline{U}_0\left(\overline{U}_r\overline{U}_s\overline{U}_t\right)^{(t-0,*)}\right]=E\left[\overline{U}_0E\left[\left(\overline{U}_r\overline{U}_s\overline{U}_t\right)^{(t-0,*)}\mid \mathscr{F}_0\right] \right] = 0,
	\end{eqnarray*}  
	using Jensen's inequality and H\"older's inequality yields that
	
	\begin{eqnarray*}
		|E\left[\overline{U}_0\overline{U}_r\overline{U}_s\overline{U}_t\right]|&=&\left|E\left[\overline{U}_0E\left[\overline{U}_r\overline{U}_s\overline{U}_t\mid \mathscr{F}_0\right]\right] - E\left[\overline{U}_0E\left[\left(\overline{U}_r\overline{U}_s\overline{U}_t\right)^{(t-0,*)}\mid \mathscr{F}_0\right] \right] \right| \\
		&\leq& \|\overline{U}_0\|_4^3\left(\|\overline{U}_r - \overline{U}_r^{(r-0,*)}\|_4+\|\overline{U}_s-\overline{U}_s^{(s-0,*)}\|_4+\|\overline{U}_t-\overline{U}_t^{(t-0,*)}\|_4\right)\\
		&=&O(1)(\lambda_{r,4}+\lambda_{s,4}+\lambda_{t,4}).
	\end{eqnarray*}
	
	Similarly,  we have
	
	\begin{eqnarray*}
		\left|E\left[\overline{U}_0\overline{U}_r\left(\overline{U}_s\overline{U}_t-(\overline{U}_s\overline{U}_t)^{(t-r,*)}\right)\right]\right|=O(1)(\lambda_{s-r,4}+\lambda_{t-r,4}),
	\end{eqnarray*}
	and 
	
	\begin{eqnarray*}
		\left|E\left[\overline{U}_0\overline{U}_r\overline{U}_s\left(\overline{U}_t- \overline{U}_t^{(t-s,*)}\right)\right]\right|=O(1)\lambda_{t-s,4} .
	\end{eqnarray*}
	Putting the above results together, we have

	\begin{eqnarray*}
		\left|E\left[\overline{U}_0\overline{U}_r \overline{U}_s\overline{U}_t \right]\right|\leq O(1)\min\left(\lambda_{r,4}+\lambda_{s,4}+\lambda_{t,4}, \lambda_{s-r,4}+\lambda_{t-r,4},\lambda_{t-s,4}\right).
	\end{eqnarray*}
	
	Next, define the following three sets:
	
	\begin{eqnarray*}
		\mathcal{S}_{ts} &=& \{r,0:(t-s)\geq \max(s-r,r-0)\},\\
		\mathcal{S}_{sr} &=& \{0,t:(s-r)\geq \max(t-s,r-0)\},\\
		\mathcal{S}_{r0} &=& \{s,t:(r-0)\geq \max(s-r,t-s)\}.
	\end{eqnarray*}
	
	Note that the cardinalities (denoted by $\#$) of these sets are bounded as follows:
	
	\begin{eqnarray*}
		\#\mathcal{S}_{ts} \leq O(1)(t-s), \quad \#\mathcal{S}_{sr} \leq O(1)(s-r), \quad\text{and}\quad \#\mathcal{S}_{r0} \leq O(1)(r-0)^2,
	\end{eqnarray*}
	which further yields that
	
	\begin{eqnarray*}
		&&\sum_{0 < r < s < t<\infty}|v(r,s,t) - \sigma(r)\sigma(s-t)|\\
		&=&\sum_{0 < r < s < t<\infty} \left|E (\overline{U}_0\overline{U}_r\overline{U}_s\overline{U}_t-\overline{U}_0\overline{U}_r(\overline{U}_s\overline{U}_t)^{(t-r,*)})\right|\\
		&\leq &O(1)\left(\sum_{r=1}^{\infty}r^2\lambda_{r,4} + \sum_{s=2}^{\infty}\sum_{r=1}^{s-1}(s-r)\lambda_{s-r,4} + \sum_{t=2}^{\infty}\sum_{s=1}^{t-1}(t-s)\lambda_{t-s,4} \right)=O(1).
	\end{eqnarray*}
	
	In addition, we have
	
	\begin{eqnarray*}
		\sum_{0<r<s<t<\infty}|\sigma(s)\sigma(r-t)|\leq \sum_{s=1}^{\infty}|\sigma(s)| \left(\sum_{t=2}^{\infty}\sum_{r=1}^{t-1}|\sigma(r-t)|\right)\leq\sum_{s=1}^{\infty}|\sigma(s)| \sum_{j=1}^{\infty}j|\sigma(j)| =O(1).
	\end{eqnarray*}
	Similarly, we have $\sum_{0<r<s<t<\infty}|\sigma(t)\sigma(r-s)|=O(1)$. Thus, we can obtain
	
	\begin{eqnarray*}
		I_4\leq \sum_{0 < r < s < t<\infty}\left(|v(r,s,t) - \sigma(r)\sigma(s-t) + |\sigma(s)\sigma(r-t)| + |\sigma(t)\sigma(r-s)|\right)=O(1).
	\end{eqnarray*}
	
	Similarly, we have $|I_j| < \infty$ for $j=5,6,7$. Collecting the above results, the proof of the second result is now completed. 
	
	\hspace*{\fill}{$\blacksquare$}
\bigskip

		\noindent \textbf{Proof of Theorem \ref{Theo2.2}:}
	\medskip
	
\noindent (1). Our goal is to show that
	
	\begin{eqnarray}\label{eq.a.3}
		S_{\mathbb{N}}^*\to_{D^*} N(0,\sigma_u^2),
	\end{eqnarray}
	which in connection with Theorem \ref{Theo2.1} immediately yields the result. In order to do so, we rewrite $S_{\mathbb{N}}^*$ as follows:
	
	\begin{eqnarray}\label{eq.a.4}
		S_{\mathbb{N}}^* = \sum_{j=1}^K \nu_{j}^*  + \sum_{j=1}^K \varpi_{j}^*,
	\end{eqnarray}
	where $\nu_{j}^*  =\sum_{t=B_j+1}^{B_j+r_1}  \frac{U_t^\top 1_{N}\xi_t}{\sqrt{\mathbb{N}}}  $, $ \varpi_{j}^*  =\sum_{t=B_j+r_1+1}^{B_j+r_1+r_2}   \frac{U_t^\top 1_{N}\xi_t}{\sqrt{\mathbb{N}}} $, and $B_j = (j-1)(r_1+r_2)$. Without loss of generality, suppose that $K= T/(r_1+r_2) $ is an integer for simplicity. Otherwise, one needs to include the remaining terms in \eqref{eq.a.4} which are negligible for an obvious reason. In addition, we let 
	
	\begin{eqnarray}\label{eq.a.5}
		(r_1, r_2)\to (\infty,\infty), \quad \left(\frac{r_2}{r_1},\frac{r_1}{T}\right)\to (0,0),\quad r_1\ge \ell ,
	\end{eqnarray}
	so the blocks $\varpi_{j}^*$'s are mutually independent by the construction of $\xi_t$'s. Note that by $\frac{r_2}{r_1}\to 0$ of \eqref{eq.a.5}, it is easy to know that $\frac{Kr_2}{T}\to 0$ and $ \frac{Kr_1}{T}\to 1$.
	
	We now write
	\begin{eqnarray*}
		&&EE^*\left[ \left( \sum_{j=1}^K \varpi_{j}^* \right)^2\right] =  \sum_{j=1}^K EE^*[(\varpi_{j}^*)^2] \nonumber \\
		&\le &\frac{1}{\mathbb{N}}\sum_{j=1}^K\sum_{s= -r_2+1}^{r_2-1}\sum_{t=B_j+r_1+1}^{B_j+r_1+r_2-|s|} a\left(\frac{s}{\ell} \right)   |E[U_t^\top 1_{N} U_{t+s}^\top 1_{N}]| \nonumber \\
		&\le &O(1)\frac{  1}{T}\sum_{j=1}^K \sum_{t=B_j+r_1+1}^{B_j+r_1+r_2} \frac{1}{N}  \sum_{s= -r_2+1}^{r_2-1}  |E[U_t^\top 1_{N} U_{t+s}^\top 1_{N}]| \le O(1) \frac{Kr_2 }{T}=o(1),
	\end{eqnarray*}
	where the second inequality follows from $a(\cdot)$ being bounded on $[-1,1]$, and the third inequality follows from $ \frac{1}{N}  \sum_{s= -r_2+1}^{r_2-1}  |E[U_t^\top 1_{N} U_{t+s}^\top 1_{N}]| =O(1)$ by Proposition \ref{Prop2.1}.  Therefore, the term $\sum_{j=1}^K \varpi_{j}^*$ of \eqref{eq.a.4} is negligible.
	
	\medskip
	
	Next, we employ Lindeberg CLT to establish the asymptotic normality of $ \sum_{j=1}^K \nu_{j}^*$. Note that by the first result of Theorem \ref{Theo2.3}.1, we know that $E^*[(S_{\mathbb{N}}^* )^2] \to_P \sigma_u^2$.  As we have shown that  $\sum_{j=1}^K \varpi_{j}^*$ of \eqref{eq.a.4} is negligible, we conclude that $E^* [ \sum_{j=1}^K \nu_{j}^*  ]^2 \to_P \sigma_u^2 $. That said, we need only to verify that for $\forall \epsilon>0$
	
	\begin{eqnarray}\label{eq.a.6}
		\sum_{j=1}^K E^* \left[(\nu_{j}^*)^2\cdot I\left(|\nu_{j}^*|>\epsilon\right) \right] =o_P(1),
	\end{eqnarray}
	which follows from
	
	\begin{eqnarray*}
		\sum_{j=1}^K E|E^* \left[(\nu_{j}^*)^2\cdot I\left(|\nu_{j}^*|>\epsilon\right) \right]| = \sum_{j=1}^K E \left[(\nu_{j}^*)^2\cdot I\left(|\nu_{j}^*|>\epsilon\right) \right] =o(1).
	\end{eqnarray*}
	Thus, we write
	
	\begin{eqnarray*} 
		\sum_{j=1}^K E[(\nu_{j}^* )^2\cdot I ( |\nu_{j}^*|>\epsilon ) ] &\le &\epsilon^{-2} \sum_{j=1}^K E| \nu_{j}^*|^{4}  =\epsilon^{-2}\frac{1}{T^{2}} \sum_{j=1}^K E\left(\sum_{t=B_j+1}^{B_j+r_1}  \overline{U}_t\xi_t\right)^{4}\nonumber \\
		&=&O(1)\frac{ Kr_1^{2} }{T^{2}} =O(1)\frac{r_1}{T} =o(1),
	\end{eqnarray*}
	where the first inequality follows from Cauchy-Schwarz inequality and Chebyshev's inequality, the second equality follows from the fact $E (\sum_{t=B_j+1}^{B_j+r_1}  \overline{U}_t\xi_t )^{4} = O(r_1^2)$ by Proposition \ref{Prop2.1}. Thus, we can conclude the validity of \eqref{eq.a.6}. Based on the above development, we are ready to conclude that \eqref{eq.a.3} holds.

\medskip
	
\noindent (2). Note that conditional on the original sample $S_{\mathbb{N}}^* = \frac{1}{\sqrt{T}}\sum_{t=1}^T \overline{U}_t \xi_t$ is a weighted sum of $\{\xi_t\}_{t=1}^T$ with fixed and known weights, therefore we mainly employ the time series techniques to prove this Berry-Esseen bound.

We rely on Theorem 1 in \cite{shergin1980convergence}, i.e., suppose $\{x_t\}_{t=1}^T$ is a sequence of $m$-dependent time series with $m=1$ and $E|x_t^3|<\infty$, then
$$
\sup_{w\in\mathbb{R}}\left|\Pr(S_t/\sigma_T \leq w) - \Psi(x)  \right| \leq a \sigma_T^{-3}\sum_{t=1}^{T}E|x_t^3|,
$$
where $S_T = \sum_{t=1}^{T}x_t$, $\sigma_T^2 = E(S_T^2)$ and $a$ is a constant not depending on $T$ and $m$.

Define the following sequence of time series:
\begin{eqnarray*}
\eta_t &=& \overline{U}_{(t-1)(\ell+1)+1} \xi_{(t-1)(\ell+1)+1} + \cdots + \overline{U}_{t(\ell+1)} \xi_{t(\ell+1)}, \quad t = 1,2,...,b,\\
\eta_{b+1}&=& \overline{U}_{b(\ell+1)+1} \xi_{b(\ell+1)+1} +\cdots + \overline{U}_{T} \xi_{T},
\end{eqnarray*}
where $b = \lfloor T/\ell \rfloor$. Then conditional on the original sample, $\{\eta_t\}$ is a sequence of $1$-dependent time series. Hence, by using Theorem 1 in \cite{shergin1980convergence} and $S_{\mathbb{N}}^* = \frac{1}{\sqrt{T}}\sum_{t=1}^{b+1}\eta_t$, we have
$$
\sup_{w \in \mathbb{R}}\left|\text{\normalfont Pr}^*(S_{\mathbb{N}}^* \le w) - \Psi\left(\frac{w}{s_{\mathbb{N}}^*}\right) \right| \leq a s_{\mathbb{N}}^{*-3}\sum_{t=1}^{b+1}E^*|(\eta_t/\sqrt{T})^3|.
$$

To complete the proof, it is sufficient to show $E^*|(\eta_t/\sqrt{T})^3| = O_P((\ell/T)^{3/2})$, which holds since $EE^*|(\eta_t/\sqrt{T})^3| = E|(\eta_t/\sqrt{T})^3| = O((\ell/T)^{3/2})$.

\medskip

\noindent (3). By using the second-order Edgeworth expansion results of \cite{rhee1985edgeworth,rhee1986characteristic} to the $1$-dependent time series $\{\eta_t\}$ constructed in part (2), we have
$$
\sup_{w\in \mathbb{R}}\left| \text{\normalfont Pr}^*(S_{\mathbb{N}}^* \le w) - \Psi\left(\frac{w}{s_{\mathbb{N}}^*}\right) - \frac{1}{6}\kappa_{\mathbb{N}}^{*3}\left(1-\frac{w^2}{s_{\mathbb{N}}^{*2}}\right)\psi \left(\frac{w}{s_{\mathbb{N}}^*}\right)\right| = O_P\left(\frac{\ell}{T}\right)
$$
with $\kappa_{\mathbb{N}}^{*3} = E^*[S_{\mathbb{N}}^{*3}]$. Next, it can be seen that  
$$
\sup_{w \in \mathbb{R}^+}\left| \text{\normalfont Pr}^*(-w \leq S_{\mathbb{N}}^* \le w) -  \left(2\Psi\left(\frac{w}{s_{\mathbb{N}}^*}\right) -1\right) \right| = O_P\left(\frac{\ell}{T}\right)
$$
since $\left(1-\frac{w^2}{s_{\mathbb{N}}^{*2}}\right)\psi \left(\frac{w}{s_{\mathbb{N}}^*}\right)$ is an even function of $w$ and $\Psi\left(-w\right) = 1 - \Psi\left(w\right)$. Therefore, we have
$$
\sup_{w \in \mathbb{R}^+}\left| \text{\normalfont Pr}^*(|S_{\mathbb{N}}^*| \le w) - \Pr(|S_{\mathbb{N}}| \le  w) -  2\left(\Psi\left(\frac{w}{s_{\mathbb{N}}^*}\right) -\Psi\left(\frac{w}{s_{\mathbb{N}}}\right)\right) \right| = O_P\left(\frac{\ell}{T}\right).
$$

\hspace*{\fill}{$\blacksquare$}

\medskip

\noindent \textbf{Proof of Theorem \ref{Theo2.4}:}
	\medskip
	
Before we start our investigation, we show that 

\begin{eqnarray}\label{thetabc}
\sqrt{\mathbb{N}}(\widehat{\theta}_{\text{bc}}  -\theta_0)  \simeq \frac{1}{\sqrt{\mathbb{N}}}\sum_{t=1}^T W_t^\top U_t  +\sqrt{\frac{T}{N}}\mu_{\mathbb{N}, C}.
\end{eqnarray}
Write

\begin{eqnarray*}
\sqrt{\mathbb{N}}(\widehat{\theta}_{\text{bc}}  -\theta_0) & \simeq &2\sqrt{\mathbb{N}}(\widehat{\theta}  -\theta_0)  - \frac{1}{\sqrt{2}}  \sqrt{\frac{\mathbb{N}}{2}}(\widehat{\theta}_{S_1}-\theta_0)-\frac{1}{\sqrt{2}}  \sqrt{\frac{\mathbb{N}}{2}}(\widehat{\theta}_{S_2}-\theta_0)\nonumber \\
&=&\Big\{  \frac{2}{\sqrt{\mathbb{N}}} \sum_{t=1}^TW_t^\top U_t +2\sqrt{\frac{N}{T}}\mu_{\mathbb{N}, B} + 2\sqrt{\frac{T}{N}}\mu_{\mathbb{N}, C} \nonumber  \\
&&-\frac{1}{\sqrt{2}}\left(\sqrt{\frac{2}{\mathbb{N}}}\sum_{i=1}^N\sum_{t\in S_1}  w_{it}u_{it} +  \sqrt{\frac{N}{T/2}} \mu_{\mathbb{N}, B}+ \sqrt{\frac{T/2}{N }} \mu_{\mathbb{N}, C}  \right) \nonumber \\
&&-\frac{1}{\sqrt{2}}\left(\sqrt{\frac{2}{\mathbb{N}}}\sum_{i=1}^N\sum_{t\in S_2} w_{it} u_{it}+ \sqrt{\frac{N}{T/2}} \mu_{\mathbb{N}, B} + \sqrt{\frac{T/2}{N }} \mu_{\mathbb{N}, C}  \right)\Big\}\nonumber  \\
&=& \frac{1}{\sqrt{\mathbb{N}}} \sum_{t=1}^T W_t^\top U_t +\sqrt{\frac{N}{T}}\left(2\mu_{\mathbb{N}, B}  -\mu_{\mathbb{N}, B}-\mu_{\mathbb{N}, B}\right) +\sqrt{\frac{T}{N}}\left(2\mu_{\mathbb{N}, C}   -\frac{1}{2}\mu_{\mathbb{N}, C} -\frac{1}{2}\mu_{\mathbb{N}, C}\right) \nonumber \\
& =& \frac{1}{\sqrt{\mathbb{N}}} \sum_{t=1}^T W_t^\top U_t  +\sqrt{\frac{T}{N}}\mu_{\mathbb{N}, C},
\end{eqnarray*}
where the first step follows from \eqref{Eq.2.10} of the main text, and $w_{it}^\top$ stands for the $i^{th}$ row of $W_t$. Therefore, equation \eqref{thetabc} holds, and we are ready to proceed.

\medskip

(1). First, we note that

\begin{eqnarray}\label{defVNT}
 \frac{\widehat{F}^\top \widehat{F}}{T} &=& \frac{1}{T}\sum_{t=1}^T\widehat{f}_t\widehat{f}_t^\top = \frac{1}{TN^2}\sum_{t=1}^T\widehat{\Gamma}^\top (Y_t- X_t\widehat{\theta})(Y_t- X_t\widehat{\theta})^\top \widehat{\Gamma}=V_{\mathbb{N}},
\end{eqnarray}
where $V_{\mathbb{N}}$ includes the largest $p$ eigenvalues of $ \frac{1}{\mathbb{N}}\sum_{t=1}^T  (Y_t- X_t\widehat{\theta})(Y_t- X_t\widehat{\theta})^\top$ in descending order, the second equality follows from the definition of $\widehat{f}_t$, and the third equality follows from the definition of PCA.  Second, we note that

\begin{eqnarray}\label{defPINT}
\frac{1}{N}\|\widehat{\Gamma}^\top (\widehat{\Gamma} - \Gamma_0 \Pi_{\mathbb{N}}) \| =o_P(1)\quad \text{and}\quad\frac{1}{\sqrt{N}}\|\widehat{\Gamma}\Pi_{\mathbb{N}}^{-1} -\Gamma_0\| =o_P(1),
\end{eqnarray}
where $\Pi_{\mathbb{N}}^{-1} =V_{\mathbb{N}} ( \frac{\Gamma_0^\top\widehat{\Gamma}}{N} )^{-1} (\frac{F^\top F}{T} )^{-1}$, and the two results follow from Proposition A.1 and Lemma A.3 of \cite{Bai} after interchanging $i$ and $t$ dimensions.

In order to prove the first result, it suffices to only consider the following term:

\begin{eqnarray*}
\frac{1}{\mathbb{N}}\sum_{t=1}^T X_t^\top M_{\widehat{\Gamma}} \widehat{\Omega}  \widehat{\Gamma} \left( \frac{\widehat{F}^\top \widehat{F}}{T}\right)^{-1} \widehat{f}_t.
\end{eqnarray*}
'
Thus, we write

\begin{eqnarray*}
&&\frac{1}{NT}\sum_{t=1}^T X_t^\top M_{\widehat{\Gamma}} \widehat{\Omega}  \widehat{\Gamma} \left( \frac{\widehat{F}^\top \widehat{F}}{T}\right)^{-1} \widehat{f}_t\nonumber \\
&=&\frac{1}{NT}\sum_{t=1}^T X_t^\top M_{\widehat{\Gamma}}\frac{1}{T}\sum_{s=1}^T (Y_s-X_s\widehat{\theta})(Y_s-X_s\widehat{\theta})^\top  \cdot \widehat{\Gamma} \cdot V_{\mathbb{N}}^{-1} \cdot \frac{1}{N}\widehat{\Gamma}^\top  (Y_t-X_t\widehat{\theta} )    \nonumber \\
&\simeq &\frac{1}{NT}\sum_{t=1}^T X_t^\top M_{\widehat{\Gamma}}\frac{1}{T}\sum_{s=1}^T U_sU_s^\top \cdot \widehat{\Gamma} \cdot V_{\mathbb{N}}^{-1} \cdot \frac{1}{N}\widehat{\Gamma}^\top \Gamma_0 f_t   \nonumber \\
&\simeq &\frac{1}{NT}\sum_{t=1}^T X_t^\top M_{\widehat{\Gamma}}\frac{1}{T}\sum_{s=1}^T U_sU_s^\top \cdot\widehat{\Gamma}\cdot V_{\mathbb{N}}^{-1}\cdot \Pi_{\mathbb{N}}^{-1} f_t    \nonumber \\
&= &\frac{1}{NT}\sum_{t=1}^T X_t^\top M_{\widehat{\Gamma}}\frac{1}{T}\sum_{s=1}^T U_sU_s^\top \widehat{\Gamma}\Pi_{\mathbb{N}}^{-1} \left( \frac{\Gamma_0^\top \widehat{\Gamma}\Pi_{\mathbb{N}}^{-1}}{N}\right)^{-1}\left( \frac{F^\top F}{T}\right)^{-1} f_t \nonumber \\
&\simeq &\frac{1}{NT}\sum_{t=1}^T X_t^\top M_{\widehat{\Gamma}}\frac{1}{T}\sum_{s=1}^T U_sU_s^\top \Gamma_0 \left( \frac{\Gamma_0^\top\Gamma_0}{N}\right)^{-1}\left( \frac{F^\top F}{T}\right)^{-1} f_t   \nonumber \\
&\to_P & \mu_C ,
\end{eqnarray*}
where the first step follows from \eqref{defVNT} and the definitions of $\widehat{\Omega}$ and $\widehat{f}_t$, the second step follows from expanding $(Y_t-X_t\widehat{\theta} )$ in three places using some tedious (but obvious) development, the third step follows from the first equality of \eqref{defPINT}, and the fifth step follows from the second equality of \eqref{defPINT}.

Therefore, we have shown $\widehat{\mu}_C\to_P \mu_C$, which in connection with \eqref{thetabc} further indicates the first result of this theorem.

\medskip

(2). Next, we consider $\widehat{\theta}^*-\widehat{\theta}$. By design, it is easy to know that
	
	\begin{eqnarray}\label{eq.a.8}
\widehat{\theta}^*-\widehat{\theta}&=& \widehat{\Sigma}_1^{-1}\cdot \frac{1}{\mathbb{N}}\sum_{t=1}^TX_t^\top M_{\widehat{\Gamma}}  \widehat{U}_t  \xi_t\nonumber \\
		&=&\widehat{\Sigma}_1^{-1}\cdot \frac{1}{\mathbb{N}}\sum_{t=1}^TX_t^\top M_{\widehat{\Gamma}}U_t\xi_t+\frac{1}{\mathbb{N}}\sum_{t=1}^TX_t^\top M_{\widehat{\Gamma}}( X_t\theta_0 + \Gamma_0 f_t-X_t \widehat{\theta} - \widehat{\Gamma} \widehat{f}_t)\xi_t.
	\end{eqnarray}
	Note that by bringing  $\widehat{f}_t=\frac{1}{N}\widehat{\Gamma}^\top (Y_t-X_t\widehat{\theta})$ in $X_t\theta_0 + \Gamma_0 f_t-X_t \widehat{\theta} - \widehat{\Gamma} \widehat{f}_t $, we can obtain 
	
	\begin{eqnarray}\label{eq.a.9}
		X_t\theta_0 + \Gamma_0 f_t-X_t \widehat{\theta} - \widehat{\Gamma} \widehat{f}_t = M_{\widehat{\Gamma}} X_t (\theta_0 -\widehat{\theta} ) +M_{\widehat{\Gamma}}\Gamma_0 f_t+P_{\widehat{\Gamma}} U_t.
	\end{eqnarray}
	Bringing \eqref{eq.a.9} in \eqref{eq.a.8}, the term $P_{\widehat{\Gamma}} U_t$ disappear automatically as $M_{\widehat{\Gamma}}P_{\widehat{\Gamma}} U_t=0$.  Thus, we can write further
	
	\begin{eqnarray*}
\widehat{\theta}^*-\widehat{\theta}=\widehat{\Sigma}_1^{-1}\cdot \frac{1}{\mathbb{N}}\sum_{t=1}^TX_t^\top M_{\widehat{\Gamma}}(U_t+\Gamma_0 f_t)\xi_t+\widehat{\Sigma}_1^{-1}\cdot \frac{1}{\mathbb{N}}\sum_{t=1}^TX_t^\top M_{\widehat{\Gamma}}X_t( \theta_0 -\widehat{\theta} )\xi_t,
	\end{eqnarray*}
where the definition of $\widehat{\Sigma}_1$ should be obvious and is thus omitted. Moreover, we know that
	
	\begin{eqnarray*} 
		\frac{1}{\mathbb{N}} \sum_{t,s=1}^{T} (\theta_0 -\widehat{\theta} )^\top  X_t^\top M_{\widehat{\Gamma}}  X_t X_s^\top M_{\widehat{\Gamma}}    X_t (\theta_0 -\widehat{\theta} )  E[\xi_t \xi_s] =o_P(1),
	\end{eqnarray*}
	where the last step is obvious in view of  $\| \theta_0 -\widehat{\theta} \| =O_P\left(\frac{1}{\sqrt{\mathbb{N}}} \right)$ and Assumption \ref{Assumption2}.
	
	Therefore, in order to establish the asymptotic distribution of $\sqrt{\mathbb{N}}(\widehat{\theta}^*-\widehat{\theta})$, we need only to study 
	\begin{eqnarray*}
		\widehat{\Sigma}_1^{-1}\cdot \frac{1}{\mathbb{N}}\sum_{t=1}^TX_t^\top M_{\widehat{\Gamma}}(U_t+\Gamma_0 f_t)\xi_t.
	\end{eqnarray*}
	
	In view of the proof of Theorem 3 of \cite{Bai}, one can decompose $M_{\widehat{\Gamma}}(U_t+\Gamma_0f_t)\xi_t$, so the two bias terms will arise. However, both terms include $\{\xi_t\}$ that is independent of all the other variables and has mean 0, as a consequence these two terms will vanish asymptotically under Assumption \ref{Assumption2}. Therefore, we have
	
	\begin{eqnarray*}
		\widehat{\Sigma}_1^{-1}\cdot \frac{1}{\mathbb{N}}\sum_{t=1}^TX_t^\top M_{\widehat{\Gamma}}(U_t+\Gamma_0 f_t)\xi_t\to_{D^*} N(0,\Sigma_1^{-1}\Sigma_2\Sigma_1^{-1}),
	\end{eqnarray*}
	which in connection with the first result of this theorem completes the proof. \hspace*{\fill}{$\blacksquare$}

		\medskip
		
		\noindent\textbf{Proof of Corollary \ref{Coro2.2}:}
		\medskip
		
		Still, we decompose $S_{\mathbb{N}}^*$ as follows.
		
		\begin{eqnarray*}
			S_{\mathbb{N}}^* = \sum_{j=1}^K \nu_{j}^*  + \sum_{j=1}^K \varpi_{j}^*,
		\end{eqnarray*}
		where  $\nu_{j}^*  =\sum_{t=B_j+1}^{B_j+r_1}  \frac{U_t^\top 1_{N_t}\xi_t}{\sqrt{\mathbb{N}}} $, $ \varpi_{j}^*  =\sum_{t=B_j+r_1+1}^{B_j+r_1+r_2}   \frac{U_t^\top 1_{N_t}\xi_t}{\sqrt{\mathbb{N}}} $, and $B_j = (j-1)(r_1+r_2)$. Without loss of generality we suppose that $K= T/(r_1+r_2) $ is an integer for simplicity In addition, we let 
		
		\begin{eqnarray*}
			(r_1, r_2)\to (\infty,\infty), \quad \left(\frac{r_2}{r_1},\frac{r_1}{T}\right)\to (0,0),\quad r_1\ge \ell ,
		\end{eqnarray*}
		so the blocks $\varpi_{j}^*$'s are mutually independent by the construction of $\xi_t$'s. Note that  
		
		\begin{eqnarray*}
			&&EE^*\left[ \left( \sum_{j=1}^K \varpi_{j}^* \right)^2\right] =  \sum_{j=1}^K EE^*[(\varpi_{j}^*)^2] \le \frac{1}{\mathbb{N}}\sum_{j=1}^K\sum_{s= -r_2+1}^{r_2-1}\sum_{t=B_j+r_1+1}^{B_j+r_1+r_2-|s|} a\left(\frac{s}{\ell} \right)   |E[U_t^\top 1_{N_t} U_{t+s}^\top 1_{N_{t+s}}]| \nonumber \\
			&\le &O(1)\frac{  \overline{N}}{\mathbb{N}}\sum_{j=1}^K \sum_{t=B_j+r_1+1}^{B_j+r_1+r_2} \frac{1}{\overline{N}}  \sum_{s= -r_2+1}^{r_2-1}  |E[U_t^\top 1_{N_t} U_{t+s}^\top 1_{N_{t+s}}]| \le O(1) \frac{Kr_2 }{T}=o(1),
		\end{eqnarray*}
		where the second inequality follows from $a(\cdot)$ being bounded on $[-1,1]$, and the third inequality follows from $ \frac{1}{\overline{N}}  \sum_{s= -r_2+1}^{r_2-1}  |E[U_t^\top 1_{N_t} U_{t+s}^\top 1_{N_{t+s}}]| =O(1)$ by $\frac{\overline{N}T}{\mathbb{N}}\to c\in (0,\infty) $ of Assumption \ref{Assumption4}.  Therefore, the term $\sum_{j=1}^K \varpi_{j}^*$  is negligible.
		
		\medskip
		
		We can then establish the asymptotic normality of $ \sum_{j=1}^K \nu_{j}^*$ in a way almost identical to those presented in the proof of Theorem \ref{Theo2.2}. The proof is now completed.  \hspace*{\fill}{$\blacksquare$}	
		
				\medskip
				
				\noindent \textbf{Proof of Corollary \ref{Coro1}:}
		\medskip
		
		First, note that it is easy to show that
		
\begin{eqnarray*}
\sqrt{\mathbb{N}}( \widehat{\theta} -\theta_0)  = \left(\frac{1}{\mathbb{N}}\sum_{t=1}^T X_t^{\dag, \top} X_t^\dag \right)^{-1}\frac{1}{\sqrt{\mathbb{N}}}\sum_{t=1}^T X_t^{\dag, \top} U_t +o_P(1),
\end{eqnarray*}
where the equality follows from Assumption \ref{Assumption1}. 
		
Second, we note that
		
\begin{eqnarray*}
	\sqrt{\mathbb{N}}(\widetilde{\theta} -\widehat{\theta}) &=& \left(\frac{1}{\mathbb{N}}\sum_{t=1}^T X_t^{\dag, \top} X_t^\dag \right)^{-1}\frac{1}{\sqrt{\mathbb{N}}}\sum_{t=1}^T X_t^{\dag, \top} U_t  \xi_t  \\
	& & - \left(\frac{1}{\mathbb{N}}\sum_{t=1}^T X_t^{\dag, \top} X_t^\dag \right)^{-1}\frac{1}{T\sqrt{\mathbb{N}}}\sum_{t=1}^T \sum_{s=1}^TX_t^{\dag, \top} U_s \xi_t \nonumber \\
	&&+ \left(\frac{1}{\mathbb{N}}\sum_{t=1}^T X_t^{\dag, \top} X_t^\dag \right)^{-1}\frac{1}{\mathbb{N}}\sum_{t=1}^T X_t^{\dag, \top}X_t^{\dag }\xi_t \cdot \sqrt{\mathbb{N}} (\theta_0- \widehat{\theta})\nonumber \\
	&=& \left(\frac{1}{\mathbb{N}}\sum_{t=1}^T X_t^{\dag, \top} X_t^\dag \right)^{-1}\frac{1}{\sqrt{\mathbb{N}}}\sum_{t=1}^T X_t^{\dag, \top} U_t  \xi_t + o_P(1),
\end{eqnarray*}
where the second equality follows from the conditions in the body of this corollary and Assumptions \ref{Assumption1} and \ref{Assumption2} with some obvious reasons.
		
		Using a procedure similar to that of Theorem \ref{Theo2.2}, it is easy to show that the result follows. Thus, the details are omitted. \hspace*{\fill}{$\blacksquare$}
		\medskip

	{\small
		\noindent \section*{Appendix B}	

		\renewcommand{\theequation}{B.\arabic{equation}}
		\renewcommand{\thesubsection}{B.\arabic{subsection}}
		\renewcommand{\thefigure}{B.\arabic{figure}}
		\renewcommand{\thelemma}{B.\arabic{lemma}}
		\renewcommand{\theremark}{B.\arabic{remark}}
		\renewcommand{\thecorollary}{B.\arabic{corollary}}
		\renewcommand{\theassumption}{B.\arabic{assumption}}
		
		\setcounter{equation}{0}
		\setcounter{lemma}{0}
		\setcounter{subsection}{0}
		\setcounter{table}{0}
		\setcounter{figure}{0}
		\setcounter{remark}{0}
		\setcounter{corollary}{0}
		\setcounter{assumption}{0}

		In this appendix, we first   introduce a few definitions in Appendix \ref{SecB.1} to facilitate development of the preliminary lemmas. Finally, we summarize the preliminary lemmas in Appendix \ref{SecB.2} and give their proofs in Appendix \ref{SecB.3}.

		\subsection{Notation and Definitions}\label{SecB.1}
		
		We now introduce some other notation, which will be repeatedly used in the following development. First, we would like to emphasize that from this point and onwards, we let
		
		\begin{eqnarray}\label{eq.b.1}
			U_t \equiv U_{t\gamma}=g_\gamma(\varepsilon_t,\ldots, \varepsilon_{t-\gamma}) 
		\end{eqnarray}
		for notational simplicity, in which $\gamma \equiv \gamma_T\to \infty$ and $\gamma/T\to 0$. $\overline{U}_t$ is then defined accordingly. Also, without loss of generality, let $T \equiv 2 n \gamma$ denote the integer part, in which $n$ stands for the number of blocks. Otherwise, we have to take into account the remaining terms, which are negligible for an obvious reason.
		
		Denote the following $\sigma$-field:
		
		\begin{eqnarray}\label{eq.b.2}
			\mathscr{F}_{\gamma} =\sigma(\underbrace{\varepsilon_{-\gamma+1},\ldots, \varepsilon_0}_{\text{$1^{st}$ block}}, \underbrace{\varepsilon_{\gamma+1},\ldots \varepsilon_{2\gamma}}_{\text{$2^{nd}$ block}},\ldots, \underbrace{\varepsilon_{(2n-1)\gamma +1},\ldots, \varepsilon_{2n\gamma}}_{\text{$(n+1)^{th}$ block}}),
		\end{eqnarray}
		and let $E_{\mathscr{F}_{\gamma}}[\cdot] = E[\cdot \mid \mathscr{F}_{\gamma}]$ and $\text{Pr}_{\mathscr{F}_{\gamma}}(\cdot) = \Pr(\cdot \mid \mathscr{F}_{\gamma})$ respectively be the conditional expectation and the conditional probability induced by $\mathscr{F}_{\gamma}$. Also, with respect to $\mathscr{F}_t$ of Section \ref{SecA.1}, we define
		
		\begin{eqnarray*}
			\mathscr{F}_{t}^* = \sigma(\varepsilon_t,\ldots,\varepsilon_{1},\varepsilon_{0}^\prime,\varepsilon_{-1}^\prime,\ldots).
		\end{eqnarray*}
		
		\medskip
		
		\noindent \textbf{Decomposition of $S_{\mathbb{N}}$:}
		\medskip
		
		We are now ready to decompose $S_{\mathbb{N}}$. First let
		
		\begin{eqnarray*}
			S_{\mathbb{N}} = \frac{1}{\sqrt{n}} \sum_{j=1}^n  (\overline{S}_{j|\mathbb{N}}+ \widetilde{S}_{j|\mathbb{N}}),
		\end{eqnarray*}
		where $\overline{S}_{j|\mathbb{N}} = \frac{1}{\sqrt{2\gamma}} \sum_{t=(2j-2)\gamma+1}^{(2j-1)\gamma} \overline{U}_t$ and $ \widetilde{S}_{j|\mathbb{N}} = \frac{1}{\sqrt{2\gamma}} \sum_{t=(2j-1)\gamma+1}^{2j\gamma} \overline{U}_t$. By design, $\{\overline{S}_{j|\mathbb{N}}\}_{j=1}^n$ and $\{\widetilde{S}_{j|\mathbb{N}}\}_{j=1}^n$ are two sequences of independent variables, respectively.
		
		Using \eqref{eq.b.2}, we can also decompose $S_{\mathbb{N}}  $ into the following two parts:
		
		\begin{eqnarray*}
			S_{\mathbb{N}} = \overline{S}_{\mathbb{N}|\gamma} + \widetilde{S}_{\mathbb{N}|\gamma},
		\end{eqnarray*}
		where $\overline{S}_{\mathbb{N}|\gamma} = \frac{1}{\sqrt{T}} \sum_{t=1}^T (\overline{U}_t - E [\overline{U}_t \mid \mathscr{F}_{\gamma}])$ and $\widetilde{S}_{\mathbb{N}|\gamma} = \frac{1}{\sqrt{T}} \sum_{t=1}^T E[\overline{U}_t\mid \mathscr{F}_{\gamma}] $.
		
		Below, we further decompose $\overline{S}_{\mathbb{N}|\gamma}$ and $\widetilde{S}_{\mathbb{N}|\gamma}$. Then $\overline{S}_{\mathbb{N}|\gamma}$ can be written as follows:
		
		\begin{eqnarray*}
			\sqrt{T}\overline{S}_{\mathbb{N}|\gamma} &=&\sqrt{2\gamma}\sum_{j=1}^n \overline{V}_j ,\quad \overline{V}_j = \frac{1}{\sqrt{2\gamma}}(\overline{V}_{j, \text{odd}} + \overline{V}_{j, \text{even}}),
		\end{eqnarray*}
		where $\overline{V}_{j, \text{odd}}= \sum_{t=(2j-2)\gamma+1}^{(2j-1)\gamma}(\overline{U}_t - E [\overline{U}_t \mid \mathscr{F}_{\gamma}])$ and $\overline{V}_{j, \text{even}} = \sum_{t=(2j-1)\gamma+1}^{2j\gamma}(\overline{U}_t - E [\overline{U}_t \mid \mathscr{F}_{\gamma}])$. By construction, the blocks $\overline{V}_j$ with $j\in [n]$ are independent variables under the conditional probability measure $\text{Pr}_{\mathscr{F}_{\gamma}}(\cdot)$.
		
		Let $\widetilde{V}_0 = \frac{1}{\sqrt{2\gamma}} \sum_{t=1}^{\gamma}E[\overline{U}_t\mid \mathscr{F}_{\gamma}]$, $\widetilde{V}_n = \frac{1}{\sqrt{2\gamma}} \sum_{t=(2n-1)\gamma+1}^{2n\gamma}E[\overline{U}_t\mid \mathscr{F}_{\gamma}]$, and $$\widetilde{V}_j =  \frac{1}{\sqrt{2\gamma}} \sum_{t=(2j-1)\gamma+1}^{(2j+1)\gamma}E[\overline{U}_t\mid \mathscr{F}_{\gamma}]$$ for $  j\in [n-1]$.  Note that  $\{\widetilde{V}_j\}_{j=0}^{n}$ is a sequence of independent variables under the probability measure $\text{Pr}(\cdot)$. We then decompose $\widetilde{S}_{\mathbb{N}|\gamma}$ as follows:
		\begin{eqnarray*}
			\widetilde{S}_{\mathbb{N}|\gamma} = \frac{1}{\sqrt{n}}\sum_{j=0}^n \widetilde{V}_j.
		\end{eqnarray*}
		
		Accordingly, we define the following coupled variables for $j\in [n]$:
		
		\begin{eqnarray*}
			V_{j,\text{odd}}^* =\frac{1}{\sqrt{2\gamma}} \sum_{t=(2j-2)\gamma+1}^{(2j-1)\gamma}\overline{U}_t^{(t-(2j-2)\gamma,*)} \quad\text{and}\quad V_{j,\text{even}}^* = \frac{1}{\sqrt{2\gamma}}\sum_{t=(2j-1)\gamma+1}^{2j\gamma}\overline{U}_t^{(t-(2j-1)\gamma,*)},
		\end{eqnarray*}
		where by construction $V_{j,\text{odd}}^*$ is independent of $\mathscr{F}_{\gamma}$, and $V_{j,\text{even}}^*$ is independent of $V_{j,\text{odd}}^*$.

		\medskip

		\noindent \textbf{Moments:}
		\medskip
		
		We define the following second moments: 
		
		\begin{eqnarray*}
			&&\overline{\sigma}_{j|\gamma}^{2} = E_{\mathscr{F}_{\gamma}}|\overline{V}_j|^2,\quad \overline{\sigma}_{j}^{2} = E|\overline{V}_j|^2,\quad \widetilde{\sigma}_{j}^{2} = E|\widetilde{V}_j|^2.
		\end{eqnarray*}
		Similarly, we let $\overline{\kappa}_{j|\gamma}^{3}$, $\overline{\kappa}_{j}^{3} $, and $ \widetilde{\kappa}_{j}^{3}$ be the corresponding quantities for the third moments, and let $\overline{\tau}_{j|\gamma}^{4} $, $\overline{\tau}_{j}^{4}$, and $ \widetilde{\tau}_{j}^{4}$ be the fourth moments.
		
		By construction, we have $  E[\overline{S}_{j|\mathbb{N}}^{2}]=E[\widetilde{S}_{j|\mathbb{N}}^{2}] := \widehat{\sigma}_{j}^{2} $. We let further $\sigma_{|\gamma}^{2}= E_{\mathscr{F}_\gamma}[\overline{S}_{\mathbb{N}|\gamma}^{2}]$, and $ \sigma^{2} = E[\sigma_{|\gamma}^{2}]$. Simple algebra shows that
		
		\begin{eqnarray*}
			\sigma_{|\gamma}^{2} = \frac{2\gamma}{T}\sum_{j=1}^{n}\overline{\sigma}_{j|\gamma}^{2}\quad \text{and}\quad
			\sigma^{2} = \frac{2\gamma}{T}\sum_{j=1}^{n}\overline{\sigma}_{j}^{2}.
		\end{eqnarray*}
		
		\medskip
		
		\noindent \textbf{Conditional approximations and distributions:}
		\medskip
		
		Let $F(\cdot)$ be any continuous distribution function and $\{\overline{Z}_j\}_{j\in [n]}$ be i.i.d. random variables and distributed according to $F(\cdot)$ such that
		
		\begin{eqnarray*}
			&&E(\overline{Z}_j) = 0,\quad E(\overline{Z}_j^2)- \overline{\sigma}_{j}^{2} = O(\gamma^{-1}),\quad E(\overline{Z}_j^3) - \overline{\kappa}_{j}^{3} = O(\gamma^{-1}),\\
			&&E(\overline{Z}_j^4) - \overline{\tau}_{j}^{4} = O(\gamma^{-1}).
		\end{eqnarray*}
		The existence of $\overline{Z}_j$ is guaranteed by Lemmas \ref{L3} and \ref{L5} below. Let $\{\widetilde{Z}_j\}_{j\in [n]}$ be an independent copy of $\{\overline{Z}_j\}_{j\in [n]}$.
		
		Define 
		\begin{eqnarray*}
			&&\Delta_{j,\gamma}(w) = \text{Pr}(V_{j,\text{odd}}^*>w) - \text{Pr}(\overline{Z}_j>w),\nonumber \\
			&&\mathscr{U}_{T} = \int_{-c\sqrt{T}}^{c\sqrt{T}} |E(e^{ixS_{\mathbb{N}}}) - E(e^{ix(\overline{Z}+\widetilde{Z})}) | \frac{1}{|x|}\mathrm{d}x,
		\end{eqnarray*}
		where $c$ is a sufficiently large constant,  $ \overline{Z} = n^{-1/2}\sum_{j=1}^{n}\overline{Z}_j$, and $ \widetilde{Z} = n^{-1/2}\sum_{j=1}^{n}\widetilde{Z}_j$.
		
		\medskip
		
		\noindent \textbf{Recursion step:}
		\medskip
		
		For $a > 0$ and $b \in \mathbb{N}$ even, let $G_{a,b}$ be a real-valued random variable with density function 
		
		\begin{eqnarray*}
			g_{a,b}(x) = c_b a \left|\frac{\sin ax}{ax} \right|^b
		\end{eqnarray*}
		for some $c_b > 0 $ only depending on $b$. It is well-known that for even $b$ the Fourier transform $\widehat{g}_{a,b}$ satisfies 
		$\widehat{g}_{a,b}(t) = 2\pi c_b u^{*b}[-a,a](t)$ if $|t|\leq ab$ and $\widehat{g}_{a,b}(t) = 0$ otherwise, where $u^{*b}[-a,a]$ denotes the $b$-fold convolution of the density of the uniform distribution on $[-a,a]$, that is $u[-a,a](t) = \frac{1}{2a}I(-a\leq t\leq a)$. For $b \geq 6$, let $\{H_t\}_{t=1}^T$ be a sequence of i.i.d. random variables with $H_t =_D G_{a,b}$ and independent of $\{\overline{U}_t\}$. Define
		
		\begin{eqnarray*}
			X_t^{\diamond} &=& \overline{U}_t+ H_t - H_{t-1},\nonumber \\
			S_{\mathbb{N}}^{\diamond} &=&\frac{1}{\sqrt{T}}\sum_{t=1}^{T}X_t^{\diamond} = \frac{1}{\sqrt{T}}\sum_{t=1}^{T}\overline{U}_t + \frac{1}{\sqrt{T}}H_T - \frac{1}{\sqrt{T}}H_0,\nonumber \\
			\overline{S}_{j|\mathbb{N}}^\diamond &=& \frac{1}{\sqrt{2\gamma}} \sum_{t=(2j-2)\gamma+1}^{(2j-1)\gamma} \overline{U}_t + \frac{1}{\sqrt{2\gamma}}H_{(2j-1)\gamma} - \frac{1}{\sqrt{2\gamma}}H_{(2j-2)\gamma},\nonumber \\
			\widetilde{S}_{j|\mathbb{N}}^\diamond &=& \frac{1}{\sqrt{2\gamma}} \sum_{t=(2j-1)\gamma+1}^{2j\gamma} \overline{U}_t + \frac{1}{\sqrt{2\gamma}}H_{2j\gamma} - \frac{1}{\sqrt{2\gamma}}H_{(2j-1)\gamma}.
		\end{eqnarray*}
		
		Accordingly, we have $s_{\mathbb{N}}^{\diamond}$, $\kappa_{\mathbb{N}}^{\diamond}$, $\left(\mathscr{T}_{a}^{b}\right)^\diamond$, $\mathscr{U}_{T}^\diamond$, $\Psi_{\mathbb{N}}^{\diamond}$, $\overline{Z}_j^{\diamond}$, $\widetilde{Z}_j^{\diamond}$, $\overline{Z}^{\diamond}$, $\widetilde{Z}^{\diamond}$ and the difference
		
		\begin{eqnarray*}
			\Delta_T^{\diamond}(w) = \mathrm{Pr}\left(S_{\mathbb{N}}^{\diamond}\leq w \right) - \Psi_{\mathbb{N}}^{\diamond}(w).
		\end{eqnarray*}
		By the formulae of $g_{a,b}$ and $\widehat{g}_{a,b}$, we have $E(H_t) = 0$ and $E|H_t|^4 < \infty$. Moreover, by independence we have
		
		\begin{eqnarray*}
			E(e^{ix S_{\mathbb{N}}^{\diamond}}) &=&  E(e^{ix S_{\mathbb{N}}})\cdot E(e^{ix \frac{H_T}{\sqrt{T}}  })\cdot E(e^{-ix \frac{H_0}{\sqrt{T}}  }),
		\end{eqnarray*}
		which, in connection with the definitions of $g_{a,b}(x)$ and $\widehat{g}_{a,b}(x)$, yields
		
		\begin{eqnarray*}
			E(e^{ix S_{\mathbb{N}}^{\diamond}}) = 0 \text{ for } |x| > \sqrt{T}|ab|.
		\end{eqnarray*}
		
		For $ j \in [n]$, define 
		
		\begin{eqnarray*}
			\sqrt{2\gamma}V_{j,\mathrm{odd}}^{\diamond} &=& \sqrt{2\gamma} V_{j,\mathrm{odd}}^* + H_{(2j-1)\gamma+1} - H_{(2j-2)\gamma+1},\nonumber \\
			\sqrt{2\gamma} V_{j,\mathrm{even}}^{\diamond} &=& \sqrt{2\gamma} V_{j,\mathrm{even}}^* + H_{2j\gamma+1} - H_{(2j-1)\gamma+1},\\
			\Delta_{j,\gamma}^{\diamond}(w) &=& \text{Pr}(V_{j,\text{odd}}^{\diamond}>w) - \text{Pr}(\overline{Z}_j^{\diamond}>w).
		\end{eqnarray*}
		
		By definition, we have $\overline{S}_{j|\mathbb{N}}^\diamond =_D V_{j,\mathrm{odd}}^{\diamond}=_D\widetilde{S}_{j|\mathbb{N}}^\diamond =_D V_{j,\mathrm{even}}^{\diamond}$.  
		\medskip
		
		We now provide the following preliminary lemmas that have already been used in the proofs of Appendix A.
		
		\subsection{Preliminary Lemmas}\label{SecB.2}
		
		In order to prove the main results in Lemma B.8, we will need to introduce Lemmas B.1-B.7 and then given their proofs in this appendix.
		
		\begin{lemma}\label{L0}
			For $a_1 > 0$, $a_2,w \in \mathbb{R}$, let $G(a_1,a_2,w) = \Psi\left(\frac{w}{a_1}\right)+\frac{1}{6}a_2\left(1-\frac{w^2}{a_1^2}\right)\psi\left(\frac{w}{a_1}\right)$. Suppose that $\max_{1\leq i\leq2}|a_i-b_i|\leq y$. Then $\sup_{w \in \mathbb{R}}(w^2+1)|G(a_1,a_2,w)-G(b_1,b_2,w)|\leq O(1)y$.
		\end{lemma}

		\begin{lemma}\label{L1}
			Under Assumption \ref{Assumption1}, the following results hold:
			\begin{enumerate}
				
				\item $\|\overline{V}_{j,\normalfont\text{even}}\|_{\delta} = O(1)$ for each $j\in [n]$;
				
				\item $\|\overline{V}_j -V_{j,\normalfont\text{odd}}^*\|_{\delta} = O(\gamma^{-1/2})$ for each $j\in[n]$;
				
				\item $E(S_{\mathbb{N}}^{3}) = O\left(\frac{1}{\sqrt{T}}\right)$;
				
				\item $E(S_{\mathbb{N}}^{4}) - 3[E(S_{\mathbb{N}}^{2})]^2 = O\left(\frac{1}{T}\right)$.
			\end{enumerate}
		\end{lemma}

		\begin{lemma}\label{L2}
			Let Assumption \ref{Assumption1} hold. Suppose that $f(\cdot)$ is a three times differentiable function with $|f^{(s)}|\leq M_f<\infty$ for $s=0,1,\ldots,3$. Then the following results hold:
			
			\begin{enumerate}
				\item  $\|E_{\mathscr{F}_\gamma}(f(\overline{V}_j) - f(V_{j, \normalfont\text{odd}}^*)\|_1 = O(M_f\gamma^{-1})$ for each $j\in [n]$;
				
				\item  $\|f(\widetilde{V}_j) - f(V_{j, \normalfont\text{even}}^*) \|_1=O(M_f\gamma^{-1})$ for each $j\in [n]$.
			\end{enumerate}
			
			Let further $f(\cdot)$ be a fourth-degree polynomial with coefficients bounded by $M_f$. Then the following results hold:
			\begin{enumerate}
				\item[3.] $\|E_{\mathscr{F}_{\gamma}}(f(\overline{V}_j) - f(V_{j, \normalfont\text{odd}}^*) )\|_1=O(M_f\gamma^{-1})$ for each $j\in [n]$;
				
				\item[4.] $\|f(\widetilde{V}_j) - f(V_{j, \normalfont\text{even}}^*)\|_1=O(M_f\gamma^{-1})$ for each $j\in [n]$.
			\end{enumerate}
		\end{lemma}
		
		\begin{lemma}\label{L3}
			Under Assumption \ref{Assumption1}, the following results of the sample moments hold:
			\begin{enumerate}
				\item $\|\overline{\sigma}_{j|\gamma}^{2} - \overline{\sigma}_{j}^{2} \|_{\delta/2} = \|\overline{\sigma}_{j|\gamma}^{2} - \widehat{\sigma}_{j}^{2} \|_{\delta/2}+O(\gamma^{-1})=O(\gamma^{-1})$ for each $j\in [n]$;
				
				\item $\|\overline{\sigma}_{j}^{2} - \widehat{\sigma}_{j}^{2}\|_{\delta/2} = O(\gamma^{-1})$ for each $j\in [n]$;
				
				\item $\|\sigma_{|\gamma}^{2} - \sigma^{2}\|_{\delta/2} = O(T^{-1}n^{1/2})$;
				
				\item  $\|\overline{\kappa}_{j|\gamma}^{3} - E(\overline{S}_{j|\mathbb{N}}^{3})\|_1 = O(\gamma^{-1})$ and $\|\overline{\kappa}_{j|\gamma}^{3} - \overline{\kappa}_{j}^{3}\|_1 = O(\gamma^{-1})$  for each $j\in[n]$;
				
				\item $E(S_{\mathbb{N}}^{3}) - E(\overline{S}_{\mathbb{N}|\gamma}^{3}) - E(\widetilde{S}_{\mathbb{N}|\gamma}^{3}) = O(T^{-1}n^{1/2})$ and $E(\overline{S}_{\mathbb{N}|\gamma}^{3})+E(\widetilde{S}_{\mathbb{N}|\gamma}^{3})- \left(\frac{2\gamma}{T}\right)^{3/2}\sum_{j=1}^{n}(\overline{\kappa}_{j}^{3} + \widetilde{\kappa}_{j}^{3})=O(T^{-1}n^{1/2})$;
				
				\item $\|\overline{\tau}_{j|\gamma}^{4} - E(\overline{S}_{j|\mathbb{N}}^{4}) \|_1 = O(\gamma^{-1})$ and $\|\overline{\tau}_{j|\gamma}^{4} - \overline{\tau}_{j}^{4}\|_1 =  O(\gamma^{-1})$  for each $j\in[n]$;
				
				\item $\widetilde{\sigma}_{j}^{2}-\overline{\sigma}_{j}^{2} = O(\gamma^{-1})$  for each $j\in[n]$;
				
				\item $\widetilde{\kappa}_{j}^{3}-\overline{\kappa}_{j}^{3} = O(\gamma^{-1})$ for each $j\in[n]$;
				
				\item $\widetilde{\tau}_{j}^{4} - \overline{\tau}_{j}^{4} = O(\gamma^{-1})$ for each $j\in[n]$.
			\end{enumerate}
		\end{lemma}

		\begin{lemma}\label{L4}
			Under Assumption \ref{Assumption1}, for all $x > M \sqrt{T \log T}$ with $M$ being a large constant, we have $\Pr(\sqrt{T}S_{\mathbb{N}}\geq x) = O(T x^{-4})$.
		\end{lemma}

		\begin{lemma}\label{L5}
			Let Assumption \ref{Assumption1} hold. Then $F(\cdot)$ exists and can be chosen such that
			\begin{enumerate}
				\item $\sup_{w\in\mathbb{R}}|F(w) - \Psi_{\gamma,j}(w)|= \sup_{w\in\mathbb{R}}|\mathrm{Pr}(\overline{Z}_j\leq w) - \Psi_{\gamma,j}(w)| =O(\gamma^{-1})$, where
				$$
				\Psi_{\gamma,j}(w) := \Psi\left(\frac{w}{\widehat{\sigma}_j}\right) + \frac{1}{6}E(\overline{S}_{j|\mathbb{N}}^{3})\left(1 - \frac{w^2}{\widehat{\sigma}_j^{2}} \right)\psi\left(\frac{w}{\widehat{\sigma}_j}\right);
				$$
				
				\item $ \sup_{w\in\mathbb{R}}(w^2+1)|\mathrm{Pr}(\overline{Z}+\widetilde{Z}\leq w) - \Psi_{\mathbb{N}}(w)|=O(\gamma^{-1})$, where 
				
				\begin{eqnarray*}
					\Psi_{\mathbb{N}}(w) := \Psi\left(\frac{w}{s_{\mathbb{N}}}\right)+\frac{1}{6}\kappa_{\mathbb{N}}^{3}\left(1-\frac{w^2}{s_{\mathbb{N}}^{2}}\right)\psi \left(\frac{w}{s_{\mathbb{N}}}\right);
				\end{eqnarray*}
				
				\item $E\left(|\overline{Z}_j|^3 I(|\overline{Z}_j|\geq \tau_T)\right) =O(\gamma^{-2})$, where $\tau_T \geq c_\tau\sqrt{\log T}$, and $c_\tau > 0$ is sufficiently large.
			\end{enumerate}
		\end{lemma}

		\begin{lemma}\label{L6}
			Let Assumption \ref{Assumption1} hold. Let  $f$ be a smooth function such that $\sup_{x\in \mathbb{R}}|f^{(s)}(x) |\leq 1$ for $s=0,1,\ldots,8$, and $x_n = x/\sqrt{n}$. Then for $\tau_T \geq c_\tau\sqrt{\log T}$ with $c_\tau > 0$ being sufficiently large, we have
			\begin{enumerate}
				\item $\|E_{\mathscr{F}_\gamma}(f(x_n\overline{V}_j^{\diamond})) - E(f(x_n\overline{Z}_j^{\diamond}))\|_1 \\ =(|x_n|^2 + \tau_T^5|x_n|^5)O(\gamma^{-1})+\sup_{x \in \mathbb{R}}|\Delta_{j,\gamma}^{\diamond}(x)|O(\tau_T^5|x_n|^5+\tau_T^6|x_n|^6)$;
				
				\item $\|E_{\mathscr{F}_\gamma}(f(x_n\overline{V}_j)) - E(f(x_n\overline{Z}_j))\|_1\\=(|x_n|^2 + \tau_T^5|x_n|^5)O(\gamma^{-1})+\sup_{x \in \mathbb{R}}|\Delta_{j,\gamma}^{\diamond}(x)|O(\tau_T^5|x_n|^5+\tau_T^6|x_n|^6)$;
				
				\item $\|E_{\mathscr{F}_\gamma}(f(x_n\widetilde{V}_j^{\diamond})) - E(f(x_n\widetilde{Z}_j^{\diamond}))\|_1 \\ =(|x_n|^2 + \tau_T^5|x_n|^5)O(\gamma^{-1})+\sup_{x \in \mathbb{R}}|\Delta_{j,\gamma}^{\diamond}(x)|O(\tau_T^5|x_n|^5+\tau_T^6|x_n|^6)$;
				
				\item $\|E_{\mathscr{F}_\gamma}(f(x_n\widetilde{V}_j)) - E(f(x_n\widetilde{Z}_j))\|_1 \\ = (|x_n|^2 + \tau_T^5|x_n|^5)O(\gamma^{-1})+\sup_{x \in \mathbb{R}}|\Delta_{j,\gamma}^{\diamond}(x)|O(\tau_T^5|x_n|^+\tau_T^6|x_n|^6)$.
			\end{enumerate}
		\end{lemma}

		\begin{lemma}\label{L7}
			Under Assumption \ref{Assumption1}, for $n \asymp (\log T)^m$ with $m\geq 5$,
			\begin{enumerate}
				
				\item $\sup_{w\in \mathbb{R}}|\Delta_T^{\diamond}(w)|= O(n/T)$;
				
				\item $\mathscr{U}_{T} = O(T^{-1} (\log T)^5) $ with $\mathscr{U}_{T} = \int_{-c\sqrt{T}}^{c\sqrt{T}} |E(e^{ixS_{\mathbb{N}}}) - E(e^{ix(\overline{Z}+\widetilde{Z})}) | \frac{1}{|x|}\mathrm{d}x$.
			\end{enumerate}
		\end{lemma}

		\subsection{Proofs of Preliminary Lemmas}\label{SecB.3}
		
		\noindent \textbf{Proof of Lemma \ref{L0}:}
		
		This lemma follows directly from Taylor expansions of CDF and PDF of normal distributions.
		\medskip
		
		\noindent \textbf{Proof of Lemma \ref{L1}:}
		
		\noindent (1). Without loss of generality, we assume $j=1$ and write
		
		\begin{eqnarray*}
			\| \overline{V}_{1,\text{even}} \|_{\delta} &\leq&\sum_{t=\gamma+1}^{2\gamma} \left\|\overline{U}_t - E[\overline{U}_t \mid \mathscr{F}_{\gamma}]\right\|_{\delta}=\sum_{t=\gamma+1}^{2\gamma} \left\|E[ \overline{U}_t  -\overline{U}_t^{(t-\gamma,*)} \mid \mathscr{F}_{\gamma}, \mathscr{F}_{t}^*]\right\|_{\delta} \\
			&\leq&\sum_{t=\gamma+1}^{2\gamma} \|\overline{U}_t -\overline{U}_t^{(t-\gamma, *)} \|_\delta=\sum_{t=1}^{\gamma} \| \overline{U}_t-\overline{U}_t^*\|_\delta=  \sum_{t=1}^{\gamma}\lambda_{t,\delta}^U=O(1),
		\end{eqnarray*}
		where the first inequality follows from the triangle inequality, the equality follows from $E[\overline{U}_t \mid \mathscr{F}_{\gamma}] =_DE[\overline{U}_t^{(t-\gamma,*)}  \mid \mathscr{F}_{\gamma}, \mathscr{F}_{t}^*]$,  the second inequality follows from the Jensen's inequality, and the second equality follows from the definitions of $\overline{U}_t$ and $\overline{U}_t^{(m,*)}$. 
		
		\medskip
		
		\noindent (2). Focus on $j=1$ without loss of generality. Write
		\begin{eqnarray*}
			\sqrt{2\gamma}\|\overline{V}_1 -V_{1,\text{odd}}^*\|_{\delta}& \leq & \sqrt{2\gamma}\|\overline{V}_1 -V_{1,\text{odd}}^*\|_{\delta} = \|\overline{V}_{1,\text{odd}} -V_{1,\text{odd}}^*+\overline{V}_{1,\text{even}}\|_{\delta} \\
			&\le& \sum_{t=1}^{\gamma} \|\overline{U}_t^{(t,*)}-\overline{U}_t\|_{\delta} + \sum_{t=1}^{\gamma}\|E[\overline{U}_t \mid \mathscr{F}_{\gamma}]\|_{\delta} + \|\overline{V}_{1,\text{even}}\|_{\delta}\\
			&=& \sum_{t=1}^{\gamma} \|\overline{U}_t^*-\overline{U}_t\|_{\delta} + \sum_{t=1}^{\gamma}\|E[\overline{U}_t\mid\mathscr{F}_{\gamma}]\|_{\delta} + O(1) \\
			&=& \sum_{t=1}^{\gamma}\|E[\overline{U}_t\mid \mathscr{F}_{\gamma}]\|_{\delta}  + O(1) =O(1),
		\end{eqnarray*}
		where the second equality follows from the first result of this lemma, the third equality follows from the definitions of $\overline{U}_t$ and $\overline{U}_t^{(t,*)}$, and the fourth equality follows from
		
		\begin{eqnarray*}
			\sum_{t=1}^{\gamma}\| E[\overline{U}_t\mid \mathscr{F}_{\gamma}] \|_{\delta} = \sum_{t=1}^{\gamma} \| E[\overline{U}_t - \overline{U}_t^* \mid \mathscr{F}_{\gamma}] \|_{\delta}
			\le \sum_{t=1}^{\gamma}\|\overline{U}_t-\overline{U}_t^*\|_{\delta} = \sum_{t=1}^{\gamma} \lambda_{t,\delta}^U =O(1).
		\end{eqnarray*}
		
		\medskip
		
		\noindent (3). For $k\leq s \leq t$, we write
		
		\begin{eqnarray*}
			|E (\overline{U}_k\overline{U}_s\overline{U}_t)| &=& |E [\overline{U}_k \cdot E (\overline{U}_s\overline{U}_t \mid \mathscr{F}_k)]|=|E [\overline{U}_k\cdot E (\overline{U}_s\overline{U}_t - \overline{U}_s^{(s-k,*)}\overline{U}_t^{(t-k,*)} \mid  \mathscr{F}_k)]|\nonumber \\
			&\le &\|\overline{U}_k\|_3 \cdot\left\{E [ | E (\overline{U}_s\overline{U}_t - \overline{U}_s^{(s-k,*)}\overline{U}_t^{(t-k,*)} \,|\,  \mathscr{F}_k) |^{\frac{3}{2}} ]\right\}^{\frac{2}{3}}\nonumber \\
			&\le &\|\overline{U}_k\|_3\cdot \| E (\overline{U}_s\overline{U}_t - \overline{U}_s^{(s-k,*)}\overline{U}_t^{(t-k,*)} \,|\,  \mathscr{F}_k)\|_3\nonumber \\
			&\le &\|\overline{U}_k\|_3\cdot \|  \overline{U}_s\overline{U}_t - \overline{U}_s^{(s-k,*)}\overline{U}_t^{(t-k,*)}\|_3\nonumber \\
			&\le &\|\overline{U}_k\|_3\cdot (\| \overline{U}_s\|_3 \|   \overline{U}_t - \overline{U}_t^{(t-k,*)}\|_3 +\|\overline{U}_s- \overline{U}_s^{(s-k,*)} \|_3\|\overline{U}_t^{(t-k,*)} \|_3)\nonumber \\
			&= &\|\overline{U}_k\|_3 \cdot (\| \overline{U}_s\|_3 \|   \overline{U}_t - \overline{U}_t^{(t-k,*)}\|_3 +\|\overline{U}_s- \overline{U}_s^{(s-k,*)} \|_3\|\overline{U}_t  \|_3)\nonumber \\
			&=& \|\overline{U}_0\|_3^2 \cdot (\lambda_{t-k,3}^U + \lambda_{s-k,3}^U),
		\end{eqnarray*}
		where the second equality follows from $E (\overline{U}_k) = 0$ and $E (\overline{U}_s\overline{U}_t)= E (\overline{U}_s^{(s-k,*)}\overline{U}_t^{(t-k,*)} \mid\mathscr{F}_k)$, the first inequality follows from the H\"older inequality, the second inequality follows from the moments monotonicity, the third inequality follows from the Jensen's inequality, and the third and fourth equalities follow from the definition of $\overline{U}_t$.
		
		Similarly, we have $|E (\overline{U}_k\overline{U}_s\overline{U}_t)|\leq\|\overline{U}_0\|_3^2 \lambda_{t-s,3}^U$. Thus, we can write
		
		\begin{eqnarray}\label{eq.b.3}
			E(S_{\mathbb{N}}^{3})&\leq& O(1) \frac{1}{T^{3/2}}\sum_{1\leq k\leq s\leq t\leq T} |E (\overline{U}_k\overline{U}_s\overline{U}_t)| \leq O(1)\frac{1}{T^{3/2}}\sum_{k=1}^{T}\sum_{s=k}^{k+\gamma} \sum_{t=s}^{s+\gamma}   |E (\overline{U}_k\overline{U}_s\overline{U}_t)|\nonumber \\
			&\leq&O(1)\frac{1}{T^{3/2}}\sum_{k=1}^{T}\sum_{s=k}^{k+\gamma} \left(\sum_{t=s}^{s+s-k-1}   |E (\overline{U}_k\overline{U}_s\overline{U}_t)| + \sum_{t=2s-k}^{s+\gamma}   |E (\overline{U}_k\overline{U}_s\overline{U}_t)|\right)\nonumber \\
			&\leq&O(1)\frac{1}{T^{3/2}}\sum_{k=1}^{T}\sum_{s=k}^{k+\gamma} \left(\sum_{t=s}^{2s-k-1}  \lambda_{s-k,3}^U + \sum_{t=2s-k}^{s+\gamma}  \lambda_{t-s,3}^U\right)\nonumber \\
			&=&O(1) \frac{1}{T^{1/2}}\sum_{j=1}^{\gamma}j\lambda_{j,3}^U=O\left(\frac{1}{\sqrt{T}} \right).
		\end{eqnarray}
		
		\medskip
		
		\noindent (4). Expanding $E(S_{\mathbb{N}}^{4})$, we have
		
		\begin{eqnarray*}
			E(S_{\mathbb{N}}^{4})&=&\frac{1}{T^2}\sum_{t=1}^{T}E(\overline{U}_t^4) + \frac{4}{T^2}\sum_{1\leq t < s\leq T}E (\overline{U}_t^3\overline{U}_s) + \frac{4}{T^2}\sum_{1\leq t < s\leq T}E (\overline{U}_t\overline{U}_s^3)\\
			&& + \frac{12}{T^2}\sum_{1\leq t < s < k\leq T}E (\overline{U}_t^2\overline{U}_s\overline{U}_k) + \frac{12}{T^2}\sum_{1\leq t < s < k\leq T}E (\overline{U}_t\overline{U}_s^2\overline{U}_k)\\
			&& + \frac{12}{T^2}\sum_{1\leq t < s < k\leq T}E (\overline{U}_t\overline{U}_s\overline{U}_k^2) + \frac{6}{T^2}\sum_{1\leq t < s \leq T}E (\overline{U}_t^2\overline{U}_s^2)\\
			&& + \frac{24}{T^2}\sum_{1\leq t < s < k < l\leq T}E (\overline{U}_t\overline{U}_s\overline{U}_k\overline{U}_l)\\
			&:=&I_{T,1}+I_{T,2}+I_{T,3}+I_{T,4}+I_{T,5}+I_{T,6}+I_{T,7}+I_{T,8}.
		\end{eqnarray*}
		
		The most difficult term to be dealing with is $I_{T,8}$ that we handle it first. Using similar arguments to the proof of the third result of this lemma, we have for $1\leq t<s<k<l\leq T$
		
		\begin{eqnarray*}
			\|\overline{U}_t\overline{U}_s\overline{U}_k\overline{U}_l  \|_1 \leq O(1) \min(\lambda_{s-t,4}^U,\lambda_{l-s,4}^U+\lambda_{k-s,4}^U,\lambda_{l-k,4}^U).
		\end{eqnarray*}
		
		Then using a procedure similar to \eqref{eq.b.3} we can obtain 
		
		\begin{eqnarray*}
			E|I_{T,8} | &\le &O(1)\frac{1}{T^2}\sum_{1\leq t < s < k < l\leq T}|E (\overline{U}_t\overline{U}_s\overline{U}_k\overline{U}_l)| \le  O(1)\frac{1}{T} \sum_{j=1}^{\gamma}j^2\lambda_{j,4}^U=O\left(\frac{1}{T}\right).
		\end{eqnarray*}
		Similarly, we have $E|I_{T,j}| = O\left(\frac{1}{T}\right)$ for $j=1,2,3$ and
		
		\begin{eqnarray*}
			I_{T,4} + I_{T,5} +I_{T,6} - 18T^{-2}\sum_{t,s,v=1}^{T} E (\overline{U}_t^2)E (\overline{U}_s\overline{U}_v) = O\left(\frac{1}{T}\right).
		\end{eqnarray*}
		
		For $I_{T,7}$, we have 
		\begin{eqnarray*}
			I_{T,7} =3\left(\frac{1}{T}\sum_{t=1}^{T}E(\overline{U}_t^2)\right)^2 + O\left(\frac{1}{T}\right).
		\end{eqnarray*}
		
		Putting the above results together and in view of the expansion of $3[E(S_{\mathbb{N}}^{2})]^2$, the result follows immediately. The proof is now completed. \hspace*{\fill}{$\blacksquare$}
		\medskip
		
		\noindent \textbf{Proof of Lemma \ref{L2}:}
		
		\noindent (1). Without loss of generality, let $j =1$ and write
		\begin{eqnarray*}
			&&\|E_{\mathscr{F}_\gamma}(f(\overline{V}_1) - f(V_{1,\text{odd}}^*) ) - E_{\mathscr{F}_\gamma}(f^{(1)}(V_{1,\text{odd}}^*)(\overline{V}_1 - V_{1,\text{odd}}^*) )\|_1 \\
			&\le & M_f \|\overline{V}_1 - V_{1,\text{odd}}^* \|_2^{2}=O(M_f\gamma^{-1}),
		\end{eqnarray*}
		where the first inequality follows from Taylor expansion, and the equality follows from Lemma \ref{L1}.2. In what follows, we show
		\begin{eqnarray*}
			\|E_{\mathscr{F}_\gamma}(f^{(1)}(V_{1,\text{odd}}^*)(\overline{V}_1 - V_{1,\text{odd}}^*))\|_1 =O(M_f\gamma^{-1}),
		\end{eqnarray*}
		and the result then follows. Note further that we can decompose $f^{(1)}(V_{1,\text{odd}}^*)(\overline{V}_1 - V_{1,\text{odd}}^*)$ as follows:
		
		\begin{eqnarray}\label{eq.b.4}
			&&f^{(1)}(V_{1,\text{odd}}^*)(\overline{V}_1 -V_{1,\text{odd}}^*) \nonumber \\
			&=& f^{(1)}(V_{1,\text{odd}}^*)(2\gamma)^{-1/2}\overline{V}_{1, \text{even}} + f^{(1)}(V_{1,\text{odd}})( (2\gamma)^{-1/2}\overline{V}_{1, \text{odd}}  -V_{1,\text{odd}}^*).
		\end{eqnarray}
		
		Thus, we need only to examine the two terms on the right hand side of \eqref{eq.b.4}.
		
		\medskip
		
		First, consider $ f^{(1)}(V_{1,\text{odd}}^*) \overline{V}_{1, \text{even}}$. Define for $0< m< \gamma$
		
		\begin{eqnarray*}
			V_{1,\text{odd}}^{(>m,*)}&=&\frac{1}{\sqrt{2\gamma}}\sum_{t=\gamma-m}^{\gamma}\overline{U}_t^* \quad \text{and}\quad V_{1,\text{odd}}^{(\leq m,*)}=\frac{1}{\sqrt{2\gamma}}\sum_{t=1}^{\gamma-m-1}\overline{U}_t^*.
		\end{eqnarray*}
		Then write
		
		\begin{eqnarray*}
			&&\|(f^{(1)}(V_{1,\text{odd}}^*)-f^{(1)}(V_{1,\text{odd}}^{(\leq m,*)})-f^{(2)}(V_{1,\text{odd}}^{(\leq m,*)})V_{1,\text{odd}}^{(> m,*)} )  \overline{V}_{1, \text{even}}\|_1\\
			&\leq& O(1)M_f \| \overline{V}_{1, \text{even}} \|_3 \|V_{1,\text{odd}}^{(> m,*)}  \|_3^{2}=O(\gamma^{-1} m),
		\end{eqnarray*}
		where the inequality follows from Taylor expansion and H\"older inequality, and the equality follows from $\| \overline{V}_{1, \text{even}} \|_3=O(1)$ by Lemma \ref{L1}.1, and  $\|V_{1,\text{odd}}^{(> m,*)}  \|_3^{2} =O(\gamma^{-1} m)$ by Proposition \ref{Prop2.1}.
		
		In order to investigate $ f^{(1)}(V_{1,\text{odd}}^{(\leq m,*)})  \overline{V}_{1, \text{even}}$, we define
		
		\begin{eqnarray*}
			V_{1,\text{even}}^{(m,**)}&=& \sum_{t=\gamma+1}^{2\gamma}(\overline{U}_t-E[\overline{U}_t\mid \mathscr{F}_{\gamma}])^{(t-\gamma+m,**)},
		\end{eqnarray*}
		and note that for $\gamma < t \leq 2\gamma$ and $0< m<\gamma$, we have  
		
		\begin{eqnarray}\label{eq.b.5}
			(E [\overline{U}_t\mid \mathscr{F}_{\gamma}])^{(t-\gamma+m,**)} = E [\overline{U}_t\mid \mathscr{F}_{\gamma}]= E[\overline{U}_t^{(t-\gamma+m,**)}\mid \mathscr{F}_{\gamma}].
		\end{eqnarray}
		Then write
		
		\begin{eqnarray*}
			&& \| \overline{V}_{1,\text{even}} - V_{1,\text{even}}^{(m,**)} \|_{\delta} \nonumber \\
			&=&\left\|\sum_{t=\gamma+1}^{2\gamma}(\overline{U}_t - E[\overline{U}_t \mid \mathscr{F}_{\gamma}]) - \sum_{t=\gamma+1}^{2\gamma}(\overline{U}_t - E[\overline{U}_t\mid\mathscr{F}_{\gamma}])^{(t-\gamma+m,**)}\right\|_{\delta}\nonumber \\
			&\le &\left\|  \sum_{t=\gamma+1}^{2\gamma}(\overline{U}_t-  \overline{U}_t^{(t-\gamma+m,**)})   \right\|_{\delta} \le O(1) \sum_{t=\gamma+1}^{2\gamma} \|\overline{U}_t  - \overline{U}_t^{(t-\gamma+m,**)} \|_\delta \nonumber\\
			&=& O(1)\sum_{t= 1}^{\gamma-m} \|\overline{U}_{\gamma+t}  - \overline{U}_{t+\gamma}^{(t+m,**)} \|_\delta \le O(1) \sum_{t=m}^{\gamma} \frac{t^2}{m^2} \|\overline{U}_t  - \overline{U}_t^* \|_\delta =O(m^{-2}),
		\end{eqnarray*}
		where the first inequality follows from \eqref{eq.b.5} and the triangle inequality, the third inequality follows from the definition of $\overline{U}_t$ and the fact that $\frac{t}{m}>1$, and the third equality follows from Assumption \ref{Assumption1}. Thus, we immediately obtain that
		
		\begin{eqnarray}\label{eq.b.6}
			\| f^{(1)}(V_{1,\text{odd}}^{(\leq m,*)})  (\overline{V}_{1, \text{even}} - V_{1, \text{even}}^{(m,**)}) \|_1 \leq O(1)M_f \| \overline{V}_{1, \text{even}} - V_{1, \text{even}}^{(m,**)} \|_1 = O(M_fm^{-2}).
		\end{eqnarray}
		
		Since $V_{1,\text{odd}}^{(\leq m,*)}$ and $V_{1, \text{even}}^{(m,**)}$ are independent under the probability measure $\Pr_{\mathscr{F}_\gamma}$, we have
		
		\begin{eqnarray}\label{eq.b.7}
			E_{\mathscr{F}_\gamma}[f^{(1)}(V_{1,\text{odd}}^{(\leq m,*)}) \cdot V_{1, \text{even}}^{(m,**)}]=0.
		\end{eqnarray}
		Hence, by \eqref{eq.b.6} and \eqref{eq.b.7}
		
		\begin{eqnarray*}
			\|E_{\mathscr{F}_\gamma}[f^{(1)}(V_{1,\text{odd}}^{(\leq m,*)}) \cdot \overline{V}_{1, \text{even}} ] \|_1 = O(M_fm^{-2}).
		\end{eqnarray*} 
		
		\medskip
		
		Since $V_{1,\text{odd}}^{(\leq m,*)}$ and $V_{1,\text{odd}}^{(> m,**)}   V_{1, \text{even}}^{(m,**)}$ are independent under the probability measure $\Pr_{\mathscr{F}_{\gamma}}$, we obtain that
		
		\begin{eqnarray*}
			&&\sqrt{2\gamma} \|E_{\mathscr{F}_{\gamma}}[f^{(2)}(V_{1,\text{odd}}^{(\leq m,*)})V_{1,\text{odd}}^{(> m,**)}   V_{1, \text{even}}^{(m,**)}]\|_1\\
			&\leq &O(1)M_f\sqrt{2\gamma}\|E_{\mathscr{F}_{\gamma}}[V_{1,\text{odd}}^{(> m,**)}   V_{1, \text{even}}^{(m,**)}]\|_1\\
			&\le &O(1)M_f \left\|\sum_{t=\gamma-m}^{\gamma}\sum_{k=\gamma+1}^{2\gamma}E_{\mathscr{F}_{\gamma}}[\overline{U}_t^{(t-\gamma + m,**)} \overline{U}_k^{(k-\gamma + m,**)} ] \right\|_1.
		\end{eqnarray*}
		For $\gamma - m\leq t \leq \gamma$, $\overline{U}_t^{(t-\gamma+m,**)}$ is independent of $\mathscr{F}_{\gamma}$, so we have 
		
		\begin{eqnarray}\label{eq.b.8}
			E_{\mathscr{F}_{\gamma}} [\overline{U}_t^{(t-\gamma+m,**)} ] = E(\overline{U}_t) = 0.
		\end{eqnarray}
		
		Also, by the conditional independence of $\overline{U}_t^{(t-\gamma+m,**)}$ and $\overline{U}_k^{(k-t,**)}$ for $\gamma-m\leq t\leq \gamma$, $\gamma+1\leq k\leq 2\gamma$, we have
		
		\begin{eqnarray}\label{eq.b.9}
			&&\sum_{t=\gamma-m}^{\gamma}\sum_{k=\gamma+1}^{2\gamma}E_{\mathscr{F}_\gamma}[\overline{U}_t^{(t-\gamma+m,**)} \overline{U}_k^{(k-t,**)}]\nonumber \\
			&=& \sum_{t=\gamma-m}^{\gamma}\sum_{k=\gamma+1}^{2\gamma}E_{\mathscr{F}_\gamma}[\overline{U}_t^{(t-\gamma+m,**)}] E_{\mathscr{F}_\gamma}[\overline{U}_k^{(k-t,**)}]= 0,
		\end{eqnarray}
		where the last equality follows from \eqref{eq.b.8}. Then we write 
		
		\begin{eqnarray*}
			&&\sqrt{2\gamma} \|E_{\mathscr{F}_{\gamma}}[ V_{1,\text{odd}}^{(> m,**)}   V_{1, \text{even}}^{(m,**)} ]\|_1\\
			&=& \left\|\sum_{t=\gamma-m}^{\gamma}\sum_{k=\gamma+1}^{2\gamma}E_{\mathscr{F}_\gamma}[\overline{U}_t^{(t-\gamma+m,**)}]\cdot (\overline{U}_k^{(k-\gamma+m,**)}-\overline{U}_k^{(k-t,**)}) \right\|_1 \\
			&\leq& \sum_{t=\gamma-m}^{\gamma}\sum_{k=\gamma+1}^{2\gamma}\|\overline{U}_t\|_2   \|\overline{U}_k^{(k-\gamma+m,**)}-\overline{U}_k^{(k-t,**)} \|_2 \\
			&=&\sum_{t=0}^{m}\sum_{k=1}^{\gamma}\|\overline{U}_t\|_2  \|\overline{U}_k-\overline{U}_k^{(k+t,*)} \|_2 \leq O(1)\sum_{k=1}^{\gamma}k \|\overline{U}_k-\overline{U}_k^{*} \|_2 ,
		\end{eqnarray*}
		where the first equality follows from \eqref{eq.b.9}, and then the first inequality follows from Cauchy-Schwarz inequality. Finally, we can conclude that
		
		\begin{eqnarray*}
			\|E_{\mathscr{F}_{\gamma}}[f^{(2)}(V_{1,\text{odd}}^{(\leq m,*)})V_{1,\text{odd}}^{(> m,*)}   \overline{V}_{1, \text{even}} ]\|_1= M_fO(\gamma^{-1/2}+m^{-2}) = M_fO(\gamma^{-1/2}),
		\end{eqnarray*}
		where the last equality follows from letting $m=\gamma^{1/3}$. 
		
		\medskip
		
		Second, using a similar (but simpler) strategy, we can show that
		
		\begin{eqnarray*}
			\|f^{(1)}(V_{1,\text{odd}}^*)( (2\gamma)^{-1/2}\overline{V}_{1, \text{odd}}  -V_{1,\text{odd}}^*)\|_1 = O(M_f\gamma^{-1}).
		\end{eqnarray*}
		
		Collecting the above results, the proof of the first result is now completed.

		\medskip
		
		\noindent (2). Again, without loss of generality, let $j=1$, and the proof of the second result can be done in a way similar to that for the first result. We omit the details herewith.
		
		\medskip 
		
		\noindent (3)-(4). The proofs of these two results are much similar to those for the first two results of this lemma. The only difference is that we use H\"older inequality instead of the bounded derivatives whenever necessary. \hspace*{\fill}{$\blacksquare$}

		\bigskip

		\noindent \textbf{Proof of Lemma \ref{L3}:}
		\medskip
		
		\noindent (1)-(2). Without loss of generality, let $j=1$. We first establish that $\|\overline{\sigma}_{j|\gamma}^{2} - \widehat{\sigma}_{j}^{2} \|_{\delta/2} = O_P(1/\gamma)$. Write
		
		\begin{eqnarray*}
			2\gamma(\overline{\sigma}_{j|\gamma}^{2} - \widehat{\sigma}_{j}^{2})&=& E_{\mathscr{F}_\gamma}\left([\overline{V}_{1,\text{odd}}+\overline{V}_{1,\text{even}}]^2\right) - 2\gamma \widehat{\sigma}_{j}^{2}\nonumber\\
			&=& E_{\mathscr{F}_\gamma}\left([\sum_{k=1}^{\gamma}(\overline{U}_k^*+(\overline{U}_k-\overline{U}_k^*) - E[\overline{U}_k\mid \mathscr{F}_\gamma])+\overline{V}_{1,\text{even}}]^2\right) - 2\gamma \widehat{\sigma}_{j}^{2}.
		\end{eqnarray*}
		
		By squaring out the first term, we have a sum of squared terms and a sum of interaction terms. Here, we treat the interaction terms first:
		
		\begin{eqnarray*}
			&&2\sum_{t=1}^{\gamma}\sum_{s=1}^{\gamma}E_{\mathscr{F}_\gamma}\left[\overline{U}_t^*(\overline{U}_s-\overline{U}_s^*)-\overline{U}_t^*E[\overline{U}_s\mid \mathscr{F}_\gamma]-E[\overline{U}_t\mid\mathscr{F}_\gamma](\overline{U}_s-\overline{U}_s^*) \right]\\
			&&+2\sum_{t=1}^{\gamma}E_{\mathscr{F}_\gamma}\left[\overline{V}_{1,\text{even}}\overline{U}_t^* +\overline{V}_{1,\text{even}}(\overline{U}_t-\overline{U}_t^*)- \overline{V}_{1,\text{even}}E[\overline{U}_t\mid \mathscr{F}_\gamma]\right]\\
			&:=&2I_{\gamma,1}-2I_{\gamma,2}-2I_{\gamma,3}+2I_{\gamma,4}+2I_{\gamma,5}-2I_{\gamma,6}.
		\end{eqnarray*}
		
		Consider $I_{\gamma,1}$, and write
		
		\begin{eqnarray*}
			I_{\gamma,1}&=&\sum_{s=1}^{\gamma}\sum_{t=s}^{\gamma}E\left[\overline{U}_t^*(\overline{U}_s-\overline{U}_s^*)\mid \mathscr{F}_\gamma\right]+\sum_{s=1}^{\gamma}\sum_{t=1}^{s-1}E \left[\overline{U}_t^*(\overline{U}_s-\overline{U}_s^*)\mid \mathscr{F}_\gamma\right]\\
			&=&\sum_{s=1}^{\gamma}\sum_{t=s}^{\gamma}E \left[(\overline{U}_s-\overline{U}_s^*)E(\overline{U}_t^*\mid \mathscr{F}_s,\mathscr{F}_s^*)\mid \mathscr{F}_\gamma\right] +\sum_{s=1}^{\gamma}\sum_{t=1}^{s-1}E \left[\overline{U}_t^*(\overline{U}_s-\overline{U}_s^*)\mid \mathscr{F}_\gamma\right].
		\end{eqnarray*}
		
		Note that for $s\le t\le \gamma$
		
		\begin{eqnarray*}
			E_{\mathscr{F}_\gamma}(\overline{U}_t^*\mid \mathscr{F}_s,\mathscr{F}_s^*)=_D E (\overline{U}_t\mid\mathscr{F}_s)=E (\overline{U}_t-\overline{U}_t^{(t-s,*)}\mid \mathscr{F}_s),
		\end{eqnarray*}
		where the equality follows from the fact that $E ( \overline{U}_t^{(t-s,*)}\mid\mathscr{F}_s)=0$. Thus, by Cauchy-Schwarz inequality and Jensen's inequality, we have
		
		\begin{eqnarray*}
			\|I_{\gamma,1} \|_{\delta/2}&\leq&O(1)\sum_{s=1}^{\gamma}\sum_{t=s}^{\gamma}\|\overline{U}_s-\overline{U}_s^*\|_\delta \|\overline{U}_t-\overline{U}_t^{(t-s,*)}\|_\delta\\
			&&+O(1)\sum_{s=1}^{\gamma}\sum_{t=1}^{s-1}\|\overline{U}_s-\overline{U}_s^*\|_\delta \|\overline{U}_t^*\|_\delta  =O(1),
		\end{eqnarray*}
		where the equality follows from Assumption \ref{Assumption1}.
		
		Consider $I_{\gamma,2}$. Since $E[\overline{U}_t^*\mid \mathscr{F}_\gamma] = E [\overline{U}_t] = 0$ for $1\leq t\leq \gamma$, we have $I_{\gamma,2}=0$.
		
		Consider $I_{\gamma,3}$. By Jensen's inequality, we have 
		
		\begin{eqnarray*}
			\|E[\overline{U}_t\, | \, \mathscr{F}_\gamma]\|_\delta=\|E[\overline{U}_t-\overline{U}_t^*\mid\mathscr{F}_\gamma]\|_\delta\leq \|\overline{U}_t-\overline{U}_t^*\|_\delta.
		\end{eqnarray*}
		Thus, we can obtain that
		
		\begin{eqnarray*}
			\|I_{\gamma,3}\|_{\delta/2}\leq O(1)\left(\sum_{t=1}^{\gamma}\|\overline{U}_t-\overline{U}_t^*\|_\delta\right)^2=O(1).
		\end{eqnarray*}
		
		Consider $I_{\gamma,4}$. Note that $\overline{U}_t$ and $\overline{U}_s^{(s-t,*)}$ are independent for $1\leq t\leq \gamma$ and $\gamma+1\leq s\leq 2\gamma$, and $E[\overline{U}_t^* \mid\mathscr{F}_\gamma] = 0$. Thus we can write
		\begin{eqnarray*}
			\sum_{t=1}^{\gamma}E_{\mathscr{F}_\gamma}\left[\overline{V}_{1,\text{even}}\overline{U}_t^*\right] =\sum_{t=1}^{\gamma}\sum_{s=\gamma+1}^{2\gamma}E\left[(\overline{U}_s-\overline{U}_s^{(s-t,*)})\overline{U}_t^*\mid \mathscr{F}_\gamma\right].
		\end{eqnarray*}
		
		By Cauchy-Schwarz inequality and Jensen's inequality, 
		\begin{eqnarray*}
			\|I_{\gamma,4}\|_{\delta/2}&\leq&O(1)\sum_{t=1}^{\gamma}\sum_{s=\gamma+1}^{2\gamma} \|\overline{U}_t^* \|_\delta  \|\overline{U}_s-\overline{U}_s^{(s-t,*)}\|_\delta \le O(1)\sum_{t=1}^{\infty}t \|\overline{U}_t-\overline{U}_t^*\|_\delta<\infty.
		\end{eqnarray*}
		
		Consider $I_{\gamma,5}$. By using Cauchy-Schwarz inequality, Jensen's inequality and Lemma \ref{L1}.1, we have
		
		\begin{eqnarray*}
			\|I_{\gamma,5}\|_{\delta/2}\leq O(1)\sum_{t=1}^{\gamma}\|\overline{U}_t-\overline{U}_t^*\|_\delta =O(1).
		\end{eqnarray*}
		Similarly, we have $I_{\gamma,6} = O(1)$. 
		
		\medskip
		
		We next deal with the squared terms:
		
		\begin{eqnarray*}
			&&\sum_{t=1}^{\gamma}\sum_{s=1}^{\gamma}E_{\mathscr{F}_\gamma}\left[\overline{U}_t^*\overline{U}_s^*+(\overline{U}_t-\overline{U}_t^*)(\overline{U}_s-\overline{U}_s^*)\right]\\
			&& -\sum_{t=1}^{\gamma}\sum_{s=1}^{\gamma} E_{\mathscr{F}_\gamma}(\overline{U}_t)E_{\mathscr{F}_\gamma}(\overline{U}_s)+E_{\mathscr{F}_\gamma}[\overline{V}_{1,\text{even}}^2]\\
			&:=&I_{\gamma,7}+I_{\gamma,8}+I_{\gamma,9}+I_{\gamma,10}.
		\end{eqnarray*}
		
		Consider $I_{\gamma,7}$. For $1\leq t,s\leq \gamma$, since $E_{\mathscr{F}_\gamma} [\overline{U}_t^*\overline{U}_s^* ] = E [\overline{U}_t^*\overline{U}_s^* ] = E [\overline{U}_t\overline{U}_s ]$, we have $I_{\gamma,7} = 2\gamma \widehat{\sigma}_{1}^{2}$. Similar to the above development, we can show that $\|I_{\gamma,8}\|_{\delta/2}=O(1)$, $\|I_{\gamma,9}\|_{\delta/2}=O(1)$ and $\|I_{\gamma,10}\|_{\delta/2}=O(1)$. Hence, we have proved that $\|\overline{\sigma}_{j|\gamma}^{2} - \widehat{\sigma}_{j}^{2} \|_{\delta/2} = O(\gamma^{-1})$. In addition, from the above arguments, we can easily establish $|\overline{\sigma}_{j}^{2} - \widehat{\sigma}_{j}^{2}| = O(\gamma^{-1})$. Hence, the first two results follow.
		
		\medskip
		
		\noindent (3). Let $\mathcal{I}=\{1,3,5,\ldots\}$ and $\mathcal{J}=\{2,4,6,\ldots\}$ such that $\mathcal{I}\cup \mathcal{J}=\{1,2,\ldots,n\}$, and write
		
		\begin{eqnarray*}
			\|\sum_{j=1}^{n}(\overline{\sigma}_{j|\gamma}^{2}-\overline{\sigma}_{j}^{2})\|_{\delta/2}\leq \|\sum_{j\in\mathcal{I}}(\overline{\sigma}_{j|\gamma}^{2}-\overline{\sigma}_{j}^{2})\|_{\delta/2} + \|\sum_{j\in\mathcal{J}}(\overline{\sigma}_{j|\gamma}^{2}-\overline{\sigma}_{j}^{2})\|_{\delta/2}.
		\end{eqnarray*}
		Note that $\{(\overline{\sigma}_{j|\gamma}^{2}-\overline{\sigma}_{j}^{2})\}_{j\in\mathcal{I}}$ is a sequence of independent random variables under the probability measure $\Pr$, and the same is true for $\{(\overline{\sigma}_{j|\gamma}^{2}-\overline{\sigma}_{j}^{2})\}_{j\in\mathcal{J}}$. By Proposition \ref{Prop2.1}, we have
		
		\begin{eqnarray*}
			\|\sum_{j=1}^{n}(\overline{\sigma}_{j|\gamma}^{2}-\overline{\sigma}_{j}^{2})\|_{\delta/2}\leq n^{1/2}\|\overline{\sigma}_{j|\gamma}^{2}-\overline{\sigma}_{j}^{2} \|_{\delta/2}.
		\end{eqnarray*}
		
		Hence, the result follows from the first result of this lemma.
		
		\medskip
		
		\noindent (4). Note that $E_{\mathscr{F}_\gamma}(V_{j, \normalfont\text{odd}}^{*3})=E(V_{j, \normalfont\text{odd}}^{*3})=E(\overline{S}_{j|\mathbb{N}}^{3})$. By Lemma \ref{L2}.3, we have
		\begin{eqnarray*}
			\|\overline{\kappa}_{j|\gamma}^{3} - E(\overline{S}_{j|\mathbb{N}}^{3})\|_1 = O(\gamma^{-1}).
		\end{eqnarray*}
		
		Similarly, we have $\|\overline{\kappa}_{j|\gamma}^{3} - \overline{\kappa}_{j}^{3}\|_1 = O(\gamma^{-1})$.
		
		\medskip
		
		\noindent (5). We first prove the first equality. Since $E_{\mathscr{F}_\gamma}[\overline{S}_{\mathbb{N}|\gamma}] = 0$, using the iterated law of expectation, we have
		
		\begin{eqnarray*}
			E(\overline{S}_{\mathbb{N}}^{3}) = E(\overline{S}_{\mathbb{N}|\gamma}^{3}) + E(\widetilde{S}_{\mathbb{N}|\gamma}^{3}) + 3E(\widetilde{S}_{\mathbb{N}|\gamma}E_{\mathscr{F}_\gamma}(\overline{S}_{\mathbb{N}|\gamma}^{2}) ).
		\end{eqnarray*}
		
		It then suffices to show $E(\widetilde{S}_{\mathbb{N}|\gamma}E_{\mathscr{F}_\gamma}(\overline{S}_{\mathbb{N}|\gamma}^{2}))=O(T^{-1}n^{1/2})$. Since $E[\widetilde{S}_{\mathbb{N}|\gamma}] = 0$, we have 
		\begin{eqnarray*}
			|E(\widetilde{S}_{\mathbb{N}|\gamma}E_{\mathscr{F}_\gamma}(\overline{S}_{\mathbb{N}|\gamma}^{2}) )|=|E(\widetilde{S}_{\mathbb{N}|\gamma}[E_{\mathscr{F}_\gamma}(\overline{S}_{\mathbb{N}|\gamma}^{2})-E(\overline{S}_{\mathbb{N}|\gamma}^{2})])|.
		\end{eqnarray*}
		
		By H\"older's inequality and the third result of this lemma, we have $|E(\widetilde{S}_{\mathbb{N}|\gamma}E_{\mathscr{F}_\gamma}(\overline{S}_{\mathbb{N}|\gamma}^{2}) )|=O(T^{-1}n^{1/2})$.
		
		We next prove the second equality. Since $\{V_{j,\text{odd}}^*\}_{j=1}^n$ and $\{V_{j,\text{even}}^*\}_{j=1}^n$ are receptively $\Pr_{\mathscr{F}_\gamma}$-independent and $\Pr$-independent, we have the second equality.
		
		\medskip
		
		\noindent (6)-(9). The proofs are very much similar to those of the first four results of this lemma, so we omit the details. \hspace*{\fill}{$\blacksquare$}

		\bigskip

		\noindent \textbf{Proof of Lemma \ref{L4}:}
		\medskip

		We define a few variables to facilitate development. Define $\overline{U}_{i,m} = E[\overline{U}_i\mid\varepsilon_{i},\ldots,\varepsilon_{i-m}]$, $S_{t} = \sum_{i=1}^{t}\overline{U}_i$ and $S_{t,m} = \sum_{i=1}^{t}\overline{U}_{i,m}$. Let $x_m$, $m=1,\ldots,\gamma$ be a positive sequence such that $\sum_{m=1}^\gamma x_m\leq 1$. Hence, we can rewrite $\overline{U}_t$ as 
		
		\begin{eqnarray*}
			\overline{U}_t = \overline{U}_t - \overline{U}_{t,\gamma} + \sum_{m=1}^{\gamma}(\overline{U}_{t,m} - \overline{U}_{t,m-1}) +  \overline{U}_{t,0}.
		\end{eqnarray*}
		Define $X_{t,m} = \sum_{i=1}^{t}(\overline{U}_{i,m} -\overline{U}_{i,m-1})$ and thus $S_{t,\gamma} - S_{t,0} = \sum_{i=1}^{t}(\overline{U}_{i,\gamma}-\overline{U}_{i,0})=\sum_{m=1}^{\gamma} X_{t,m}$. Let $\overline{X}_{T,m} = \max_{1\leq t\leq T}|X_{t,m}|$. For each $1\leq m\leq \gamma$, let $Y_{t,m} = \sum_{i = 1 + (t-1)m}^{\min\{tm,T\}}(\overline{U}_{i,m}-\overline{U}_{i,m-1})$, in which $1\leq t\leq l$ and $ l = \lfloor T/m \rfloor+1$. Define $\lfloor t \rfloor_{m}:= \lfloor t/m \rfloor m$. 
		
		We are now ready to start the proof:
		\begin{eqnarray*}
			&&\text{Pr}\left(\overline{X}_{T,m}\geq 3x_j x\right) \leq \text{Pr}\left(\max_{1\leq t\leq T}|X_{\lfloor t \rfloor_{m},m}|\geq 2x_j x\right) +\text{Pr}\left(\max_{1\leq t\leq T}|X_{\lfloor t \rfloor_{m},m}-X_{t,m}|\geq x_j x\right)\\
			&\leq&\text{Pr}\left(\max_{1\leq s\leq l} \left| \sum_{t=1}^{s}(1 + (-1)^t)/2\times Y_{t,m}\right|\geq x_j x \right) +\text{Pr}\left(\max_{1\leq s\leq l}\left| \sum_{t=1}^{s}(1 - (-1)^t)/2\times Y_{t,m} \right|\geq x_j x \right)\\
			&&+ \sum_{t=1}^{l}\text{Pr}\left(\max_{1 + (t-1)m\leq j\leq \min\{tm,T\}}\left|\sum_{i=1 + (t-1)m}^{j}(\overline{U}_{i,m}-\overline{U}_{i,m-1})\right|\geq x_j x\right)\\
			&:=&I_{m,1} + I_{m,2} + I_{m,3}.
		\end{eqnarray*}
		Since $Y_{2,m},Y_{4,m},\ldots$ are independent, by the classical Nagaev inequality for independent random variables (Corollary 1.8 of \citealp{nagaev1979large}), we have
		
		\begin{eqnarray*}
			I_{m,1}&\leq& \left(1 + \frac{2}{4}\right)^{4} \frac{\sum_{s=1}^{l}E\left(Y_{s,m}^4\right)}{x^4} + 2\exp\left(-\frac{2x^2}{e^4(4+2)^2\sum_{s=1}^{l}E\left(Y_{s,m}^2\right)} \right).
		\end{eqnarray*}
		In addition, since $\{\overline{U}_{k,m}-\overline{U}_{k,m-1}\}_{k=1}^T$ are martingale differences with respect to $\sigma(\varepsilon_{k-m},\varepsilon_{k-m-1},\ldots)$, by Burkholder's inequality, we have
		\begin{eqnarray*}
			\left[E\left(Y_{s,m}^4 \right)\right]^{1/4}&\leq& O(1)\left(E\left[\sum_{i = 1 + (t-1)m}^{\min\{tm,T\}}|\overline{U}_{i,m}-\overline{U}_{i,m-1}|^2\right]^2 \right)^{1/4}\\
			&\leq&O(1) \left(\sum_{i = 1 + (t-1)m}^{\min\{tm,T\}}\|\overline{U}_{i,m}-\overline{U}_{i,m-1}\|_4^2 \right)^{1/2} = O(1)\theta_{m,4}^{U}.
		\end{eqnarray*}
		and by $(\sum_{i=1}^m|a_i|)^{4}\leq m^{4-1}\sum_{i=1}^m|a_i|^4$ and the conditions on $x$, we have $I_{m,1}=O(1) \frac{T}{x^4} (m)^{4/2-1}\theta_{m,4}^{U,4}$. A similar bound holds for $I_{m,2}$. 
		
		For $I_{m,3}$, by Burkholder's inequality and Doob inequality, we have
		
		\begin{eqnarray*}
			&&E\left(\max_{1 + (t-1)m\leq j\leq \min\{tm,T\}}|\sum_{i=1 + (t-1)m}^{j}(\overline{U}_{i,m}-\overline{U}_{i,m-1})|^4\right)\\
			&\leq&2^{4-1}E\left(|Y_{t,m}|^4 + \max_{1 + (t-1)m\leq j\leq \min\{tm,T\}}|\sum_{i=j}^{\min\{tm,T\}}(\overline{U}_{i,m} -\overline{U}_{i,m-1} )|^4 \right)\\
			&\leq&2^{4-1} E\left(Y_{t,m}^4 \right) + 2^{4-1}(\frac{4}{4-1})^4E\left(Y_{t,m}^4\right).
		\end{eqnarray*}
		Hence, we have $\text{Pr}\left(\overline{X}_{T,m}\geq 3x_j x \right) =(m)^{4/2-1}\theta_{m,4}^{U,4}O(\frac{T}{x^4})$.
		
		By the above derivations and the classical Nagaev inequality for independent random variables again, we have
		\begin{eqnarray*}
			&&\text{Pr}\left(\max_{1\leq t\leq T}|S_{t}| \geq 5x\right) =\text{Pr}\left(\max_{1\leq t\leq T}|S_{t} - S_{t,\gamma} + S_{t,\gamma}- S_{t,0}+ S_{t,0}| \geq 5x\right)\\
			&\leq&\text{Pr}\left(\max_{1\leq t\leq T}|S_{t} - S_{t,\gamma}| \geq x\right) + \sum_{m=1}^{\gamma}\text{Pr}\left(X_{T,m} \geq 3 x_m x\right)+\text{Pr}\left(\max_{1\leq t\leq T}|S_{t,0}| \geq x \right)\\
			&\leq& 0 + \sum_{m=1}^\gamma(m)^{4/2-1}\theta_{m,4}^{U,4}O\left(\frac{T}{x^4}\right) + O\left(\frac{T}{x^4}\right)=  O\left(\frac{T}{x^4}\right).
		\end{eqnarray*}
		
		The proof is now completed. \hspace*{\fill}{$\blacksquare$}

		\bigskip

		\noindent \textbf{Proof of Lemma \ref{L5}:}
		\medskip
		
		\noindent (1). By Lemma \ref{L1}.3, we have $\overline{\kappa}_j^{3}=O(1/\sqrt{\gamma})$. Let $\{A_{j,t}\}_{t=1}^{\gamma}$ be i.i.d. random variables with continuous distribution function $F_j(\cdot)$ such that\footnote{Note that the existence of $E|A_{j,t}^8|$ is only needed in the proof for the third result of this lemma.} $E(A_{j,t}) = 0$, $E(A_{j,t}^2) = \overline{\sigma}_j^{2}$, $E(A_{j,t}^3) = \sqrt{\gamma} \overline{\kappa}_j^{3}$ and $E|A_{j,t}^8|=O(1)$. It follows that the first three moments of $\gamma^{-1/2}\sum_{t=1}^{\gamma}A_{j,t}$ coincide with $0$, $ \overline{\sigma}_j^{2}$ and $\overline{\kappa}_j^{3}$. In addition, by Lemma \ref{L1}.4 and Lemma \ref{L3}.6, we have 
		
		\begin{eqnarray*}
			&&E\left(|A_{j,t}|^4\right) - \overline{\tau}_j^{4} = O(1/\gamma),\nonumber \\
			&&E\left(|A_{j,t}|^4\right) - E(S_{j|\mathbb{N}}^{4}) = O(1/\gamma).
		\end{eqnarray*}
		Let $\overline{Z}_j =_D \gamma^{-1/2}\sum_{t=1}^{\gamma}A_{j,t}$, we can conclude that 
		
		\begin{eqnarray*}
			&&E(\overline{Z}_j) = 0,\quad E(\overline{Z}_j^2) - \overline{\sigma}_{j}^{2} = O(\gamma^{-1}),\quad E(\overline{Z}_j^3) - \overline{\kappa}_{j}^{3} = O(\gamma^{-1}),\quad E(\overline{Z}_j^4) - \overline{\tau}_{j}^{4} = O(\gamma^{-1}).
		\end{eqnarray*}
		
		Since the first four moments of $\overline{Z}_j$ (only three are necessary here) match those of $S_{j|\mathbb{N}}$ up to an error term of order $O(\gamma^{-1})$ by Lemmas \ref{L3}.2, \ref{L3}.4 and \ref{L3}.6, then the result follows directly from the classic Edgeworth expansion and Lemma \ref{L0}.
		
		\medskip
		
		\noindent (2). By Lemma \ref{L3}, the first four moments of $\overline{Z}+\widetilde{Z}$ match those of $S_{\mathbb{N}}$ up to an error of order $O_P(\gamma^{-1})$, so the second result can be justified in the same fashion as for the first result of this lemma.
		
		\medskip
		
		\noindent (3). By the construction of $\overline{Z}_j$ in the proof of part (1), and using the classical Fuk-Nagaev inequality (which is a simple version of Lemma \ref{L4}), the proof is the same as in the proof of Lemma \ref{L6}, so we omit it here for the time being.  \hspace*{\fill}{$\blacksquare$}

		\bigskip

		\noindent \textbf{Proof of Lemma \ref{L6}:}
		\medskip

		Before proving this lemma, we show that
		
		\begin{eqnarray}\label{eq.b.10}
			&&\|E_{\mathscr{F}_\gamma}(f(x_n\overline{V}_j)) - E(f(x_n\overline{Z}_j))\|_1\nonumber \\
			&=&(|x_n|^2 + \tau_T^5|x_n|^5)O(\gamma^{-1})+\sup_{x \in \mathbb{R}}|\Delta_{j,\gamma}(x)|O(\tau_T^5|x_n|^5+\tau_T^6|x_n|^6).
		\end{eqnarray}
		
		Then the results of this lemma then follow either immediately or in the same manner.
		
		\medskip
		
		By the following Taylor expansion:
		
		\begin{eqnarray*}
			f(x+h)-f(x) = \sum_{j=1}^{s+1}\frac{f^{(j)}(x)h^j}{j!} + \frac{h^{s+1}}{s!}\int_{0}^{1}(1-t)^s(f^{(s+1)}(th+x)-f^{(s+1)}(x))\mathrm{d}t
		\end{eqnarray*}
		and $E_{\mathscr{F}_\gamma}(\overline{V}_j) = E(\overline{V}_j)=0$, we have
		\begin{eqnarray*}
			&&E_{\mathscr{F}_\gamma}(f(x_n\overline{V}_j)) - E(f(x_n\overline{Z}_j))\\
			&=&\frac{x_n^2}{2}(\overline{\sigma}_{j|\gamma}^{2} - \overline{\sigma}_{j}^{2})f^{(2)}(0)+\frac{x_n^3}{6}(\overline{\kappa}_{j|\gamma}^{3} - \overline{\kappa}_{j}^{3})f^{(3)}(0)\\
			&&+\frac{1}{2}\int_{0}^{1}(1-t)^2E_{\mathscr{F}_\gamma}\left((x_n\overline{V}_j)^3(f^{(3)}(tx_n\overline{V}_j) - f^{(3)}(0))  \right)\mathrm{d}t\\
			&&-\frac{1}{2}\int_{0}^{1}(1-t)^2E\left((x_n\overline{Z}_j)^3(f^{(3)}(tx_n\overline{Z}_j) - f^{(3)}(0))  \right)\mathrm{d}t\\
			&:=&I_1 + I_2 + I_3,
		\end{eqnarray*}
		where the definitions of $I_1$ to $I_3$ are obvious. By Lemma \ref{L3}, we have $E|I_1|=O(x_n^2\gamma^{-1}+|x_n|^3\gamma^{-1})$. Thus, we just need to focus on $I_2$ and $I_3$ below.
		
		Let $h_T(x)$ be a three times continuously differentiable function such that $h_T(x)=1$ if $|x|\leq \tau_t/2$, $h_T(x)=0$ if $|x|\geq \tau_t$, and $h_T^{(s)}(x)\leq M$ for $s\in\{0,1,2,3\}$. Note that for a random variable $X$ and $q\geq1$, we have
		\begin{eqnarray}\label{eq.b.11}
			E\left[|X|^q I(|X|\geq \tau_T)\right] \leq q \tau_T^q \Pr(|X|\geq\tau_T) + q\int_{\tau_T}^{\infty}x^{q-1}\Pr(|X|\geq x)\mathrm{d}x.
		\end{eqnarray}
		
		Hence, by Lemma \ref{L4},
		\begin{eqnarray*}
			&& E[|\overline{V}_j|^3(1-h_T(\overline{V}_j))] \leq E[|\overline{V}_j|^3I(|\overline{V}_j|\geq \tau_T/2)] = O( \tau_T^{3-4} \gamma^{1-4/2} ) + O(\gamma^{1-4/2})\int_{\tau_T/2}^{\infty}x^{2-4}\mathrm{d}x\\
			&&=O( \tau_T^{-1} \gamma^{-1} ).
		\end{eqnarray*}
		
		Note that by $|f^{(3)}(x)|\leq 1$ and Jensen's inequality, we have
		
		\begin{eqnarray*}
			&&\|E_{\mathscr{F}_\gamma}\left[(x_n\overline{V}_j)^3(f^{(3)}(tx_n\overline{V}_j) - f^{(3)}(0)) (1-h_T(\overline{V}_j)) \right]\|_1\\
			&\leq&O(1)|x_n^3|\cdot \||\overline{V}_j|^3(1-h_T(\overline{V}_j))\|_1=O(|x_n^3|\tau_T^{-1} \gamma^{-1})=o(|x_n^3|\gamma^{-1}).
		\end{eqnarray*}
		
		Using Taylor expansion again, we have
		
		\begin{eqnarray*}
			&&E_{\mathscr{F}_\gamma}\left[(x_n\overline{V}_j)^3(f^{(3)}(tx_n\overline{V}_j) - f^{(3)}(0)) h_T(\overline{V}_j) \right]\\
			&=& tx_n^4 \int_{0}^{1}(1-s)E_{\mathscr{F}_\gamma}\left[\overline{V}_j^4f^{(4)}(stx_n\overline{V}_j) h_T(\overline{V}_j) \right]\mathrm{d}s.
		\end{eqnarray*}
		
		Let $g(x) = x^4 f^{(4)}(stx_nx) h_T(x)$. By Taylor expansion, we have
		
		\begin{eqnarray*}
			g(x) = h_T(x)\left(x^4 f^{(4)}(0) + stx_nx^5 f^{(5)}(0) + stx_nx^5\int_{0}^{1}(1-u)[f^{(5)}(stux_nx) - f^{(5)}(0)]\mathrm{d}u\right).
		\end{eqnarray*}
		Let $g_2(x) = h_T(x) x^5\int_{0}^{1}(1-u)[f^{(5)}(stux_nx) - f^{(5)}(0)]\mathrm{d}u$. As the derivatives of $f(\cdot)$ are uniformly bounded and by the definition of $h_T(x)$, we have $|g_2^{(s)}(x)|\leq O(1) \tau_T^5(1 + |x_n^3|)$ for $s\in \{0,1,2,3\}$. Then by Lemma \ref{L2}.1, we have
		
		\begin{eqnarray*}
			\|E_{\mathscr{F}_{\gamma}}(stx_n^5 g_2(\overline{V}_j) - stx_n^5g_2(V_{j,\mathrm{odd}}^*))\|_1 = O(\gamma^{-1} \tau_T^5(|x_n^5| + |x_n^8|) ).
		\end{eqnarray*}
		As $g_2(0) = 0$ and for any random variable $Y$ and differentiable function $f$
		
		\begin{eqnarray*}
			E[f(Y) - f(0)] = \int_{0}^{\infty}f^{(1)}(y) \Pr(Y\geq y)\mathrm{d}y - \int_{-\infty}^{0}f^{(1)}(y) \Pr(Y \leq y)\mathrm{d}y,
		\end{eqnarray*}
		and $h_T(x)$, $h_T^{(1)}(x)$ and $g_2^{(1)}(x)$ equal to zero for $|x|>\tau_T$, we obtain that
		
		\begin{eqnarray*}
			E_{\mathscr{F}_{\gamma}}(g_2(V_{j,\mathrm{odd}}^*)) &=& \int_{0}^{\tau_T}g_2^{(1)}(x)\mathrm{Pr}_{\mathscr{F}_{\gamma}}(V_{j,\mathrm{odd}}^*\geq x)\mathrm{d}x - \int_{-\tau_T}^{0}g_2^{(1)}(x)\mathrm{Pr}_{\mathscr{F}_{\gamma}}(V_{j,\mathrm{odd}}^* \leq x)\mathrm{d}x \\
			&=& \int_{0}^{\tau_T}g_2^{(1)}(x)\mathrm{Pr}(V_{j,\mathrm{odd}}^*\geq x)\mathrm{d}x - \int_{-\tau_T}^{0}g_2^{(1)}(x)\mathrm{Pr}(V_{j,\mathrm{odd}}^* \leq x)\mathrm{d}x,
		\end{eqnarray*}
		where we use the fact that $V_{j,\mathrm{odd}}^*$ is independent of $\mathscr{F}_{\gamma}$. In addition, we have
		
		\begin{eqnarray*}
			\left\|\int_{0}^{\tau_T}g_2^{(1)}(x)\left[\mathrm{Pr}(V_{j,\mathrm{odd}}^*\geq x) - \mathrm{Pr}(\overline{Z}_j\geq x)\right]\mathrm{d}x\right\|_1 \leq O(1) \sup_{ x\in \mathbb{R}}|\Delta_{j,\gamma}(x)|\left(\tau_T^5 + \tau_T^6|x_n| \right).
		\end{eqnarray*}
		
		Similarly, we have
		
		\begin{eqnarray*}
			\left\|\int_{-\tau_T}^{0}g_2^{(1)}(x)\left[\mathrm{Pr}(V_{j,\mathrm{odd}}^*\leq x) - \mathrm{Pr}(\overline{Z}_j\leq x)\right]\mathrm{d}x\right\|_1 \leq O(1) \sup_{ x\in \mathbb{R}}|\Delta_{j,\gamma}(x)|\left(\tau_T^5 + \tau_T^6|x_n| \right).
		\end{eqnarray*}
		
		Combing the above derivations, we obtain that 
		
		\begin{eqnarray*}
			&&\|E_{\mathscr{F}_{\gamma}}(stx_n^5g_2(\overline{V}_j) - stx_n^5g_2(\overline{Z}_{j}))\|_1 \nonumber \\
			&=& O(\gamma^{-1} \tau_T^5(|x_n^5| + |x_n^8|)) + O(1)\sup_{ x\in \mathbb{R}}|\Delta_{j,\gamma}(x)|\left(\tau_T^5|x_n|^5 + \tau_T^6|x_n|^6 \right).
		\end{eqnarray*}
		
		Using \eqref{eq.b.11} and similar arguments to the above, we obtain
		
		\begin{eqnarray*}
			\|E_{\mathscr{F}_\gamma}\left[(x_n\overline{V}_j)^4(1-h_T(\overline{V}_j))\right]\|_1=o(|x_n^4|\gamma^{-1}).
		\end{eqnarray*}
		
		By Lemma \ref{L3}.6, we have $\|E_{\mathscr{F}_\gamma}(\overline{V}_j^4 - \overline{Z}_j^4) \|_1=O(\gamma^{-1})$. Hence, putting every piece together, we have
		\begin{eqnarray*}
			&&\left\|E_{\mathscr{F}_\gamma}\left[(x_n\overline{V}_j)^3(f^{(3)}(tx_n\overline{V}_j) - f^{(3)}(0)) \right] - t\int_{0}^{1}(1-s)E_{\mathscr{F}_\gamma}\left[x_n^4\overline{Z}_j^4f^{(4)}(stx_n\overline{Z}_j) h_T(\overline{Z}_j) \right]\mathrm{d}s \right\|_1\\
			&=&O(|x_n^3|\tau_T^{-1} \gamma^{-1}) + o(|x_n^4|\gamma^{-1})+ O(\gamma^{-1} \tau_T^5(|x_n^5| + |x_n^8|)) + O(1)\sup_{ x\in \mathbb{R}}|\Delta_{j,\gamma}(x)|\left(\tau_T^5|x_n|^5 + \tau_T^6|x_n|^6 \right).
		\end{eqnarray*}
		
		Using the same arguments as the above and in the proof of Lemma \ref{L5}.3, we have 
		\begin{eqnarray*}
			&&\left\|E_{\mathscr{F}_\gamma}\left[(x_n\overline{Z}_j)^3(f^{(3)}(tx_n\overline{Z}_j) - f^{(3)}(0)) \right] - t\int_{0}^{1}(1-s)E_{\mathscr{F}_\gamma}\left[x_n^4\overline{Z}_j^4f^{(4)}(stx_n\overline{Z}_j) h_T(\overline{Z}_j) \right]\mathrm{d}s \right\|_1\\
			&=&O(|x_n^3|\gamma^{-2}).
		\end{eqnarray*}
		
		Finally, we obtain
		
		\begin{eqnarray*}
			&&\left\|E_{\mathscr{F}_\gamma}(f(x_n\overline{V}_j)) - E(f(x_n\overline{Z}_j))\right\|_1 \nonumber \\
			&=&(|x_n|^2 + \tau_T^5|x_n|^5)O(\gamma^{-1})+\sup_{x \in \mathbb{R}}|\Delta_{j,\gamma}(x)|O(\tau_T^5|x_n|^5+\tau_T^6|x_n|^6),
		\end{eqnarray*}
		which completes the proof of \eqref{eq.b.10}.
		
		\medskip
		
		\noindent (1). The proof is identical to the development of \eqref{eq.b.10}.
		
		\medskip
		
		\noindent (2). By Taylor expansion, we have
		\begin{eqnarray*}
			f(x_n V_{j,\mathrm{odd}}^{\diamond}) &=& f(x_n V_{j,\mathrm{odd}}^*) + \frac{x_n}{\sqrt{2\gamma}} f^{(1)}(x_n V_{j,\mathrm{odd}}^*)(H_{(2j-1)\gamma+1} - H_{(2j-2)\gamma+1}) \\
			&& + \frac{x_n^2}{4\gamma} f^{(2)}(x_n V_{j,\mathrm{odd}}^*)(H_{(2j-1)\gamma+1} - H_{(2j-2)\gamma+1})^2+o\left(\frac{x_n^2}{4\gamma} (H_{(2j-1)\gamma+1} - H_{(2j-2)\gamma+1})^2\right)
		\end{eqnarray*}
		and thus by using $E( H_t ) = 0$ and $f^{(s)}$ is bounded,
		
		\begin{eqnarray*}
			\|E(f(x_n V_{j,\mathrm{odd}}^{\diamond})) - E(f(x_n V_{j,\mathrm{odd}}^*))\|_1 \leq O(1)\frac{x_n^2}{\gamma}.
		\end{eqnarray*}
		Then, by triangle inequality and Lemma \ref{L2}.1, we have
		\begin{eqnarray*}
			&&\|E_{\mathscr{F}_\gamma}(f(x_n \overline{V}_{j})) - E(f(x_n V_{j,\mathrm{odd}}^{\diamond}))\|_1\\
			&\leq& \|E_{\mathscr{F}_\gamma}(f(x_n \overline{V}_{j})) - E(f(x_n V_{j,\mathrm{odd}}^*))\|_1 + \|E(f(x_n V_{j,\mathrm{odd}}^{\diamond})) - E(f(x_n V_{j,\mathrm{odd}}^*))\|_1\\
			& =& O(x_n^2/\gamma).
		\end{eqnarray*}
		
		The above bound also applies to $\overline{Z}_j$. Hence, the second result follows directly from the first result.
		
		\medskip
		
		(3)-(4). The proofs of these two results are almost identical to those for the first two results of this lemma. \hspace*{\fill}{$\blacksquare$}

		\bigskip

		\noindent \textbf{Proof of Lemma \ref{L7}:}
		\medskip
		
		Before proving the results of this lemma, we show that
		
		\begin{eqnarray}\label{eq.b.12}
			\sup_{w\in \mathbb{R}}|\Delta_T^{\diamond}(w)| = O(n/T) + (\log T)^{5/2} n^{-3/2}  \sup_{w\in\mathbb{R}}|\Delta_{j,\gamma}^{\diamond}(w)|.
		\end{eqnarray}
		
		Then the two results of this lemma follow immediately.
		
		\medskip
		
		By Berry's smoothing inequality (Lemma 2, XVI.3 in \citealp{feller2008introduction}), for $T^{+} \geq c\sqrt{T}$, we obtain
		
		\begin{eqnarray*}
			\sup_{w\in\mathbb{R}}|\Delta_T^\diamond(w)|\leq O(1)\left(\mathscr{U}_T^\diamond + \sup_{w\in\mathbb{R}}(\mathscr{T}_{c\sqrt{T}}^{T^+})^\diamond(w) + (T^+)^{-1}\right).
		\end{eqnarray*}
		Selecting $a>0$ of the density function of $H_t$  such that $c  > ab$ and using $E(e^{ix S_{\mathbb{N}}^{\diamond}}) = 0$ for $|x| >\sqrt{T} |ab|$, we have $\sup_{w\in\mathbb{R}}(\mathscr{T}_{c\sqrt{T}}^{T^+})^\diamond(w)=0$. In addition, by setting $T^{+} = +\infty$, we only need to focus on $\mathscr{U}_T^\diamond$.
		
		By $|e^{ix}| = 1$, $\overline{S}_{j|\mathbb{N}}^\diamond =_D V_{j,\mathrm{odd}}^{\diamond}$, $\widetilde{S}_{j|\mathbb{N}}^\diamond =_D V_{j,\mathrm{even}}^{\diamond}$ and the properties of conditional expectation, we have
		\begin{eqnarray*}
			&&\left|E(e^{ixS_{\mathbb{N}}^{\diamond}}) - E(e^{ix(\overline{Z}^{\diamond} + \widetilde{Z}^{\diamond})})\right|\\
			&=& \left|E\left[E_{\mathscr{F}_\gamma}\left[e^{ix\sum_{j=1}^n\overline{S}_{j|\mathbb{N}}^\diamond/\sqrt{n}}\right]e^{ix\sum_{j=1}^n\widetilde{S}_{j|\mathbb{N}}^\diamond/\sqrt{n}}\right] - E(e^{ix\overline{Z}^{\diamond}})E(e^{ix\widetilde{Z}^{\diamond}})\right|\\
			&\leq& \|E_{\mathscr{F}_\gamma}\left[e^{ix\sum_{j=1}^n\overline{S}_{j|\mathbb{N}}^\diamond/\sqrt{n}}\right] - E(e^{ix\overline{Z}^{\diamond}})\|_1 + |E\left[e^{ix\sum_{j=1}^n\widetilde{S}_{j|\mathbb{N}}^\diamond/\sqrt{n}}\right]-E(e^{ix\widetilde{Z}^{\diamond}})| \\
			& = &|E\left[e^{ix\sum_{j=1}^nV_{j,\mathrm{odd}}^\diamond/\sqrt{n}}\right] - E(e^{ix\overline{Z}^{\diamond}})| + |E\left[e^{ix\sum_{j=1}^nV_{j,\mathrm{even}}^\diamond/\sqrt{n}}\right]-E(e^{ix\widetilde{Z}^{\diamond}})| \\
			&:=&A_\gamma(x) + B_{\gamma}(x).
		\end{eqnarray*}
		
		Consider $A_\gamma(x)$. Define $\phi_j(x) = E(e^{ixV_{j,\mathrm{odd}}^\diamond})$, $\psi_j(x)=E(e^{ix\overline{Z}_{j}^\diamond})$ and $x_n=x/\sqrt{n}$. Since $\{V_{j,\mathrm{odd}}^\diamond\}$ is a sequence of random variables under $\Pr$, $|\phi_j(x)|\leq 1$, $|\psi_j(x)|\leq 1$, and by using the equality 
		
		\begin{eqnarray*}
			\prod_{i=1}^{n}a_j - \prod_{i=1}^{n}b_j = \sum_{i=1}^{n}\left( \prod_{j=1}^{i-1}b_j\right)(a_i-b_i)\left(\prod_{j=i+1}^{n}a_j\right),
		\end{eqnarray*}
		we have
		
		\begin{eqnarray*}
			A_\gamma(x) &=& \left|\prod_{i=1}^{n}\phi_j(x_n) - \prod_{i=1}^{n}\psi_j(x_n)\right|\\
			&\leq&\sum_{i=1}^{n}\left| \prod_{j=1}^{i-1}\psi_j(x_n)\right|\cdot|\phi_i(x_n)-\psi_i(x_n)|\cdot\left|\prod_{j=i+1}^{n}\phi_j(x_n)\right|.
		\end{eqnarray*}
		
		By using Eqn (5.8) and (5.9) in \cite{feller2008introduction}, XVI.5, we have $\phi_j(x_n)=O(e^{-x^2/n})$, $\psi_j(x_n)=O(e^{-x^2/n})$ and $|e^{-x^2/n}|^{n-1}\leq O(1)e^{-x^2}$. Hence, by Lemma \ref{L6}.1 and the condition on $n$, we have
		
		\begin{eqnarray*}
			\int_{-c\sqrt{T}}^{c\sqrt{T}}\frac{|A_{\gamma}(x)|}{|x|}\mathrm{d}x \leq O(\gamma^{-1}) + (\log T)^{5/2} n^{-3/2}  \sup_{w\in\mathbb{R}}|\Delta_{j,\gamma}^{\diamond}(w)|.
		\end{eqnarray*}
		
		Similarly, using Lemma \ref{L6}.3
		
		\begin{eqnarray*}
			\int_{-c\sqrt{T}}^{c\sqrt{T}}\frac{|B_{\gamma}(x)|}{|x|}\mathrm{d}x \leq O(\gamma^{-1}) + (\log T)^{5/2} n^{-3/2} \sup_{w\in\mathbb{R}}|\Delta_{j,\gamma}^{\diamond}(w)|.
		\end{eqnarray*}
		
		Hence, the proof of \eqref{eq.b.12} is completed.
		
		\medskip
		
		\noindent (1). Recall that we have let $T =2n\gamma$. Let further that $n = C(\log T)^5$ and $C > 0$ large enough. By using \eqref{eq.b.12} recursively, we have
		
		\begin{eqnarray*}
			\sup_{w\in \mathbb{R}}|\Delta_T^{\diamond}(w)| \leq O(1) T^{-1}n \sum_{k=0}^{\gamma+1}\left[(\log T)^5n^{-3/2} \right]^k(2n)^k + O(1)((\log T)^5 n^{-3/2})^{\gamma}=O(n/T).
		\end{eqnarray*}
		The proof of the first result is now completed.
		
		\medskip
		
		\noindent (2). The proof is similar to that of \eqref{eq.b.12}. The difference is that we use Lemmas \ref{L6}.2 and \ref{L6}.4. It yields that
		
		\begin{eqnarray*}
			\mathscr{U}_{T} \leq O(n/T) + (\log T)^{5/2} n^{-3/2}  \sup_{w\in\mathbb{R}}|\Delta_{j,\gamma}^\diamond(w)|.
		\end{eqnarray*}
		The result follows directly from the first result of this lemma. \hspace*{\fill}{$\blacksquare$}

\newpage


{
\renewcommand{\arraystretch}{0.9}
\setlength{\tabcolsep}{5pt}
\begin{table}[H]
\scriptsize
\caption{Results of Case 1 for Theorem \ref{Theo2.2} ($\rho_u=0.25$ and $\delta_\epsilon =0.25$)}\label{Table1}
\hspace*{0.5cm}
\begin{tabular}{llrrrrrrrrrrrr}
\hline\hline
 &  & \multicolumn{6}{c}{DWB} & \multicolumn{3}{c}{\multirow{2}{*}{MBB}} & \multicolumn{3}{c}{\multirow{2}{*}{Traditional}} \\
 &  & \multicolumn{3}{c}{Bartlett} & \multicolumn{3}{c}{Trapezoidal} & \multicolumn{3}{c}{} & \multicolumn{3}{c}{} \\
$N$ & $T$ & $0.8\ell_B$ & $\ell_B$ & $1.2\ell_B$ & $0.8\ell_T$ & $\ell_T$ & $1.2\ell_T$  &$0.8\ell_M$ & $\ell_M$ & $1.2\ell_M$ & $s_1^2$ & $s_2^2$ & $s_3^2$ \\
 50 & 50 & 0.021 & 0.014 & \multicolumn{1}{r|}{0.005} & 0.031 & 0.022 & \multicolumn{1}{r|}{0.013} & 0.161 & 0.158 & \multicolumn{1}{r|}{0.152} & 0.269 & 0.138 & 0.130 \\
 & 100 & 0.043 & 0.031 & \multicolumn{1}{r|}{0.024} & 0.043 & 0.036 & \multicolumn{1}{r|}{0.034} & 0.130 & 0.120 & \multicolumn{1}{r|}{0.115} & 0.252 & 0.125 & 0.125 \\
 & 200 & 0.049 & 0.039 & \multicolumn{1}{r|}{0.042} & 0.046 & 0.045 & \multicolumn{1}{r|}{0.045} & 0.106 & 0.099 & \multicolumn{1}{r|}{0.099} & 0.249 & 0.122 & 0.121 \\
 & 400 & 0.022 & 0.025 & \multicolumn{1}{r|}{0.024} & 0.029 & 0.026 & \multicolumn{1}{r|}{0.026} & 0.061 & 0.045 & \multicolumn{1}{r|}{0.041} & 0.256 & 0.112 & 0.120 \\
100 & 50 & 0.018 & 0.008 & \multicolumn{1}{r|}{0.005} & 0.024 & 0.016 & \multicolumn{1}{r|}{0.011} & 0.160 & 0.125 & \multicolumn{1}{r|}{0.126} & 0.234 & 0.111 & 0.103 \\
\hline \hline 
\end{tabular}
\end{table}
}

{
\renewcommand{\arraystretch}{0.9}
\setlength{\tabcolsep}{5pt}
\begin{table}[H]
\scriptsize
\caption{Results of Case 1 for Theorem \ref{Theo2.2} ($\rho_u=0.5$ and $\delta_\epsilon =0.25$)}\label{Table2}
\hspace*{0.5cm}\begin{tabular}{llrrrrrrrrrrrr}
\hline\hline
 &  & \multicolumn{6}{c}{DWB} & \multicolumn{3}{c}{\multirow{2}{*}{MBB}} & \multicolumn{3}{c}{\multirow{2}{*}{Traditional}} \\
 &  & \multicolumn{3}{c}{Bartlett} & \multicolumn{3}{c}{Trapezoidal} & \multicolumn{3}{c}{} & \multicolumn{3}{c}{} \\
$N$ & $T$ & $0.8\ell_B$ & $\ell_B$ & $1.2\ell_B$ & $0.8\ell_T$ & $\ell_T$ & $1.2\ell_T$  &$0.8\ell_M$ & $\ell_M$ & $1.2\ell_M$ & $s_1^2$ & $s_2^2$ & $s_3^2$ \\
 50 & 50 & 0.037 & 0.022 & \multicolumn{1}{r|}{0.010} & 0.045 & 0.040 & \multicolumn{1}{r|}{0.024} & 0.262 & 0.198 & \multicolumn{1}{r|}{0.200} & 0.388 & 0.138 & 0.274 \\
 & 100 & 0.043 & 0.038 & \multicolumn{1}{r|}{0.035} & 0.052 & 0.045 & \multicolumn{1}{r|}{0.039} & 0.174 & 0.127 & \multicolumn{1}{r|}{0.124} & 0.388 & 0.110 & 0.247 \\
 & 200 & 0.070 & 0.060 & \multicolumn{1}{r|}{0.055} & 0.071 & 0.071 & \multicolumn{1}{r|}{0.055} & 0.151 & 0.112 & \multicolumn{1}{r|}{0.125} & 0.382 & 0.126 & 0.274 \\
 & 400 & 0.077 & 0.066 & \multicolumn{1}{r|}{0.064} & 0.088 & 0.078 & \multicolumn{1}{r|}{0.070} & 0.129 & 0.095 & \multicolumn{1}{r|}{0.099} & 0.379 & 0.128 & 0.261 \\
100 & 50 & 0.033 & 0.021 & \multicolumn{1}{r|}{0.010} & 0.046 & 0.031 & \multicolumn{1}{r|}{0.024} & 0.237 & 0.174 & \multicolumn{1}{r|}{0.187} & 0.373 & 0.111 & 0.248 \\
\hline \hline 
\end{tabular}
\end{table}}

{
\renewcommand{\arraystretch}{0.9}
\setlength{\tabcolsep}{5pt}
\begin{table}[H]
\scriptsize 
\caption{Results of Case 1 for Theorem \ref{Theo2.2} ($\rho_u=0.5$ and $\delta_\epsilon =0.5$)}\label{Table3}
\hspace*{0.5cm}\begin{tabular}{llrrrrrrrrrrrr}
\hline\hline
 &  & \multicolumn{6}{c}{DWB} & \multicolumn{3}{c}{\multirow{2}{*}{MBB}} & \multicolumn{3}{c}{\multirow{2}{*}{Traditional}} \\
 &  & \multicolumn{3}{c}{Bartlett} & \multicolumn{3}{c}{Trapezoidal} & \multicolumn{3}{c}{} & \multicolumn{3}{c}{} \\
$N$ & $T$ & $0.8\ell_B$ & $\ell_B$ & $1.2\ell_B$ & $0.8\ell_T$ & $\ell_T$ & $1.2\ell_T$  &$0.8\ell_M$ & $\ell_M$ & $1.2\ell_M$ & $s_1^2$ & $s_2^2$ & $s_3^2$ \\
 50 & 50 & 0.035 & 0.024 & \multicolumn{1}{r|}{0.013} & 0.047 & 0.038 & \multicolumn{1}{r|}{0.024} & 0.263 & 0.190 & \multicolumn{1}{r|}{0.201} & 0.529 & 0.269 & 0.278 \\
 & 100 & 0.043 & 0.041 & \multicolumn{1}{r|}{0.038} & 0.053 & 0.043 & \multicolumn{1}{r|}{0.040} & 0.176 & 0.128 & \multicolumn{1}{r|}{0.128} & 0.525 & 0.252 & 0.253 \\
 & 200 & 0.067 & 0.059 & \multicolumn{1}{r|}{0.053} & 0.075 & 0.065 & \multicolumn{1}{r|}{0.054} & 0.150 & 0.119 & \multicolumn{1}{r|}{0.119} & 0.530 & 0.278 & 0.278 \\
 & 400 & 0.079 & 0.068 & \multicolumn{1}{r|}{0.064} & 0.088 & 0.075 & \multicolumn{1}{r|}{0.066} & 0.124 & 0.092 & \multicolumn{1}{r|}{0.097} & 0.502 & 0.267 & 0.271 \\
100 & 50 & 0.031 & 0.020 & \multicolumn{1}{r|}{0.011} & 0.045 & 0.034 & \multicolumn{1}{r|}{0.024} & 0.231 & 0.174 & \multicolumn{1}{r|}{0.190} & 0.500 & 0.242 & 0.244 \\
\hline \hline 
\end{tabular}
\end{table}}

 
 {
\renewcommand{\arraystretch}{0.9}
\setlength{\tabcolsep}{5pt}
\begin{table}[H]
\scriptsize
\caption{Results of Case 2 for Theorem \ref{Theo2.2} ($\rho_u=0.25$ and $\delta_\epsilon =0.25$)}\label{Table4}
\hspace*{0.5cm}\begin{tabular}{llrrrrrrrrrrrr}
\hline\hline
 &  & \multicolumn{6}{c}{DWB} & \multicolumn{3}{c}{\multirow{2}{*}{MBB}} & \multicolumn{3}{c}{\multirow{2}{*}{Traditional}} \\
 &  & \multicolumn{3}{c}{Bartlett} & \multicolumn{3}{c}{Trapezoidal} & \multicolumn{3}{c}{} & \multicolumn{3}{c}{} \\
$N$ & $T$ & $0.8\ell_B$ & $\ell_B$ & $1.2\ell_B$ & $0.8\ell_T$ & $\ell_T$ & $1.2\ell_T$  &$0.8\ell_M$ & $\ell_M$ & $1.2\ell_M$ & $s_1^2$ & $s_2^2$ & $s_3^2$ \\
50 & 50 & 0.028 & 0.024 & \multicolumn{1}{r|}{0.012} & 0.045 & 0.029 & \multicolumn{1}{r|}{0.019} & 0.191 & 0.177 & \multicolumn{1}{r|}{0.172} & 0.233 & 0.116 & 0.136 \\
 & 100 & 0.048 & 0.046 & \multicolumn{1}{r|}{0.032} & 0.047 & 0.046 & \multicolumn{1}{r|}{0.041} & 0.147 & 0.125 & \multicolumn{1}{r|}{0.117} & 0.241 & 0.126 & 0.129 \\
 & 200 & 0.059 & 0.050 & \multicolumn{1}{r|}{0.047} & 0.058 & 0.053 & \multicolumn{1}{r|}{0.052} & 0.110 & 0.103 & \multicolumn{1}{r|}{0.110} & 0.250 & 0.127 & 0.134 \\
 & 400 & 0.058 & 0.059 & \multicolumn{1}{r|}{0.057} & 0.060 & 0.058 & \multicolumn{1}{r|}{0.054} & 0.101 & 0.085 & \multicolumn{1}{r|}{0.093} & 0.231 & 0.132 & 0.136 \\
100 & 50 & 0.027 & 0.019 & \multicolumn{1}{r|}{0.010} & 0.034 & 0.029 & \multicolumn{1}{r|}{0.023} & 0.180 & 0.172 & \multicolumn{1}{r|}{0.159} & 0.245 & 0.137 & 0.137 \\
\hline \hline 
\end{tabular}
\end{table}
}

{
\renewcommand{\arraystretch}{0.9}
\setlength{\tabcolsep}{5pt}
\begin{table}[H]
\scriptsize
\caption{Results of Case 2 for Theorem \ref{Theo2.2} ($\rho_u=0.5$ and $\delta_\epsilon =0.25$)}\label{Table5}
\hspace*{0.5cm}\begin{tabular}{llrrrrrrrrrrrr}
\hline\hline
 &  & \multicolumn{6}{c}{DWB} & \multicolumn{3}{c}{\multirow{2}{*}{MBB}} & \multicolumn{3}{c}{\multirow{2}{*}{Traditional}} \\
 &  & \multicolumn{3}{c}{Bartlett} & \multicolumn{3}{c}{Trapezoidal} & \multicolumn{3}{c}{} & \multicolumn{3}{c}{} \\
$N$ & $T$ & $0.8\ell_B$ & $\ell_B$ & $1.2\ell_B$ & $0.8\ell_T$ & $\ell_T$ & $1.2\ell_T$  &$0.8\ell_M$ & $\ell_M$ & $1.2\ell_M$ & $s_1^2$ & $s_2^2$ & $s_3^2$ \\
50 & 50 & 0.029 & 0.011 & \multicolumn{1}{r|}{0.003} & 0.036 & 0.028 & \multicolumn{1}{r|}{0.019} & 0.195 & 0.180 & \multicolumn{1}{r|}{0.187} & 0.226 & 0.125 & 0.125 \\
 & 100 & 0.040 & 0.031 & \multicolumn{1}{r|}{0.030} & 0.046 & 0.039 & \multicolumn{1}{r|}{0.033} & 0.146 & 0.119 & \multicolumn{1}{r|}{0.119} & 0.242 & 0.124 & 0.131 \\
 & 200 & 0.056 & 0.058 & \multicolumn{1}{r|}{0.054} & 0.060 & 0.054 & \multicolumn{1}{r|}{0.053} & 0.116 & 0.098 & \multicolumn{1}{r|}{0.099} & 0.232 & 0.129 & 0.134 \\
 & 400 & 0.055 & 0.055 & \multicolumn{1}{r|}{0.054} & 0.061 & 0.051 & \multicolumn{1}{r|}{0.051} & 0.097 & 0.075 & \multicolumn{1}{r|}{0.087} & 0.239 & 0.116 & 0.131 \\
100 & 50 & 0.021 & 0.011 & \multicolumn{1}{r|}{0.012} & 0.032 & 0.023 & \multicolumn{1}{r|}{0.017} & 0.166 & 0.152 & \multicolumn{1}{r|}{0.148} & 0.229 & 0.121 & 0.114 \\
\hline \hline 
\end{tabular}
\end{table}}

{
\renewcommand{\arraystretch}{0.9}
\setlength{\tabcolsep}{5pt}
\begin{table}[H]
\scriptsize 
\caption{Results of Case 2 for Theorem \ref{Theo2.2} ($\rho_u=0.5$ and $\delta_\epsilon =0.5$)}\label{Table6}
\hspace*{0.5cm}\begin{tabular}{llrrrrrrrrrrrr}
\hline\hline
 &  & \multicolumn{6}{c}{DWB} & \multicolumn{3}{c}{\multirow{2}{*}{MBB}} & \multicolumn{3}{c}{\multirow{2}{*}{Traditional}} \\
 &  & \multicolumn{3}{c}{Bartlett} & \multicolumn{3}{c}{Trapezoidal} & \multicolumn{3}{c}{} & \multicolumn{3}{c}{} \\
$N$ & $T$ & $0.8\ell_B$ & $\ell_B$ & $1.2\ell_B$ & $0.8\ell_T$ & $\ell_T$ & $1.2\ell_T$  &$0.8\ell_M$ & $\ell_M$ & $1.2\ell_M$ & $s_1^2$ & $s_2^2$ & $s_3^2$ \\
50 & 50 & 0.020 & 0.011 & \multicolumn{1}{l|}{0.007} & 0.024 & 0.022 & \multicolumn{1}{l|}{0.016} & 0.180 & 0.174 & \multicolumn{1}{l|}{0.171} & 0.234 & 0.119 & 0.117 \\
 & 100 & 0.036 & 0.032 & \multicolumn{1}{l|}{0.028} & 0.041 & 0.037 & \multicolumn{1}{l|}{0.034} & 0.114 & 0.101 & \multicolumn{1}{l|}{0.099} & 0.223 & 0.122 & 0.108 \\
 & 200 & 0.055 & 0.047 & \multicolumn{1}{l|}{0.045} & 0.051 & 0.049 & \multicolumn{1}{l|}{0.046} & 0.104 & 0.096 & \multicolumn{1}{l|}{0.095} & 0.224 & 0.112 & 0.120 \\
 & 400 & 0.050 & 0.044 & \multicolumn{1}{l|}{0.043} & 0.046 & 0.043 & \multicolumn{1}{l|}{0.042} & 0.090 & 0.072 & \multicolumn{1}{l|}{0.075} & 0.252 & 0.132 & 0.137 \\
100 & 50 & 0.033 & 0.021 & \multicolumn{1}{l|}{0.007} & 0.045 & 0.036 & \multicolumn{1}{l|}{0.026} & 0.204 & 0.179 & \multicolumn{1}{l|}{0.179} & 0.252 & 0.142 & 0.141 \\
\hline \hline 
\end{tabular}
\end{table}}

 
 {
\renewcommand{\arraystretch}{0.9}
\setlength{\tabcolsep}{5pt}
\begin{table}[H]
\scriptsize 
\caption{Results for Corollary \ref{Coro1} ($\rho_u=0.25$ and $\delta_\epsilon =0.25$)}\label{Table7}
\hspace*{0.5cm}\begin{tabular}{llrrrrrrrrrrrr}
\hline\hline
 &  & \multicolumn{6}{c}{DWB} & \multicolumn{3}{c}{\multirow{2}{*}{MBB}} & \multicolumn{3}{c}{\multirow{2}{*}{Traditional}} \\
 &  & \multicolumn{3}{c}{Bartlett} & \multicolumn{3}{c}{Trapezoidal} & \multicolumn{3}{c}{} & \multicolumn{3}{c}{} \\
$N$ & $T$ & $0.8\ell_B$ & $\ell_B$ & $1.2\ell_B$ & $0.8\ell_T$ & $\ell_T$ & $1.2\ell_T$  &$0.8\ell_M$ & $\ell_M$ & $1.2\ell_M$ & $s_1^2$ & $s_2^2$ & $s_3^2$ \\
50 & 50 & 0.111 & 0.120 & \multicolumn{1}{r|}{0.138} & 0.099 & 0.109 & \multicolumn{1}{r|}{0.123} & 0.166 & 0.180 & \multicolumn{1}{r|}{0.184} & 0.080 & 0.078 & 0.070 \\
 & 100 & 0.079 & 0.086 & \multicolumn{1}{r|}{0.091} & 0.074 & 0.083 & \multicolumn{1}{r|}{0.081} & 0.126 & 0.127 & \multicolumn{1}{r|}{0.113} & 0.081 & 0.085 & 0.066 \\
 & 200 & 0.049 & 0.056 & \multicolumn{1}{r|}{0.057} & 0.048 & 0.053 & \multicolumn{1}{r|}{0.055} & 0.080 & 0.083 & \multicolumn{1}{r|}{0.084} & 0.065 & 0.063 & 0.041 \\
 & 400 & 0.053 & 0.050 & \multicolumn{1}{r|}{0.054} & 0.049 & 0.049 & \multicolumn{1}{r|}{0.051} & 0.078 & 0.072 & \multicolumn{1}{r|}{0.079} & 0.069 & 0.065 & 0.052 \\
100 & 50 & 0.103 & 0.114 & \multicolumn{1}{r|}{0.136} & 0.093 & 0.104 & \multicolumn{1}{r|}{0.124} & 0.159 & 0.173 & \multicolumn{1}{r|}{0.162} & 0.088 & 0.076 & 0.073 \\
\hline \hline 
\end{tabular}
\end{table}}

 {
\renewcommand{\arraystretch}{0.9}
\setlength{\tabcolsep}{5pt}
\begin{table}[H]
\scriptsize 
\caption{Results for Corollary \ref{Coro1} ($\rho_u=0.25$ and $\delta_\epsilon =0.5$)}\label{Table8}
\hspace*{0.5cm}\begin{tabular}{llrrrrrrrrrrrr}
\hline\hline
 &  & \multicolumn{6}{c}{DWB} & \multicolumn{3}{c}{\multirow{2}{*}{MBB}} & \multicolumn{3}{c}{\multirow{2}{*}{Traditional}} \\
 &  & \multicolumn{3}{c}{Bartlett} & \multicolumn{3}{c}{Trapezoidal} & \multicolumn{3}{c}{} & \multicolumn{3}{c}{} \\
$N$ & $T$ & $0.8\ell_B$ & $\ell_B$ & $1.2\ell_B$ & $0.8\ell_T$ & $\ell_T$ & $1.2\ell_T$  &$0.8\ell_M$ & $\ell_M$ & $1.2\ell_M$ & $s_1^2$ & $s_2^2$ & $s_3^2$ \\
50 & 50 & 0.116 & 0.127 & \multicolumn{1}{r|}{0.138} & 0.099 & 0.112 & \multicolumn{1}{r|}{0.126} & 0.181 & 0.203 & \multicolumn{1}{r|}{0.198} & 0.115 & 0.118 & 0.069 \\
 & 100 & 0.077 & 0.082 & \multicolumn{1}{r|}{0.093} & 0.068 & 0.078 & \multicolumn{1}{r|}{0.082} & 0.127 & 0.139 & \multicolumn{1}{r|}{0.119} & 0.101 & 0.112 & 0.070 \\
 & 200 & 0.048 & 0.054 & \multicolumn{1}{r|}{0.058} & 0.048 & 0.053 & \multicolumn{1}{r|}{0.058} & 0.087 & 0.093 & \multicolumn{1}{r|}{0.093} & 0.089 & 0.097 & 0.049 \\
 & 400 & 0.050 & 0.051 & \multicolumn{1}{r|}{0.055} & 0.044 & 0.043 & \multicolumn{1}{r|}{0.050} & 0.079 & 0.084 & \multicolumn{1}{r|}{0.076} & 0.097 & 0.104 & 0.060 \\
100 & 50 & 0.116 & 0.129 & \multicolumn{1}{r|}{0.141} & 0.105 & 0.119 & \multicolumn{1}{r|}{0.128} & 0.170 & 0.187 & \multicolumn{1}{r|}{0.176} & 0.117 & 0.101 & 0.076 \\
\hline \hline 
\end{tabular}
\end{table}}

{
\renewcommand{\arraystretch}{0.9}
\setlength{\tabcolsep}{5pt}
\begin{table}[H]
\scriptsize 
\caption{Results for Corollary \ref{Coro1} ($\rho_u=0.5$ and $\delta_\epsilon =0.25$)}\label{Table9}
\hspace*{0.5cm}\begin{tabular}{llrrrrrrrrrrrr}
\hline\hline
 &  & \multicolumn{6}{c}{DWB} & \multicolumn{3}{c}{\multirow{2}{*}{MBB}} & \multicolumn{3}{c}{\multirow{2}{*}{Traditional}} \\
 &  & \multicolumn{3}{c}{Bartlett} & \multicolumn{3}{c}{Trapezoidal} & \multicolumn{3}{c}{} & \multicolumn{3}{c}{} \\
$N$ & $T$ & $0.8\ell_B$ & $\ell_B$ & $1.2\ell_B$ & $0.8\ell_T$ & $\ell_T$ & $1.2\ell_T$  &$0.8\ell_M$ & $\ell_M$ & $1.2\ell_M$ & $s_1^2$ & $s_2^2$ & $s_3^2$ \\
50 & 50 & 0.119 & 0.134 & \multicolumn{1}{r|}{0.152} & 0.103 & 0.114 & \multicolumn{1}{r|}{0.131} & 0.178 & 0.207 & \multicolumn{1}{r|}{0.194} & 0.097 & 0.086 & 0.086 \\
 & 100 & 0.080 & 0.087 & \multicolumn{1}{r|}{0.088} & 0.082 & 0.084 & \multicolumn{1}{r|}{0.087} & 0.145 & 0.144 & \multicolumn{1}{r|}{0.127} & 0.098 & 0.085 & 0.077 \\
 & 200 & 0.058 & 0.059 & \multicolumn{1}{r|}{0.064} & 0.055 & 0.058 & \multicolumn{1}{r|}{0.059} & 0.100 & 0.097 & \multicolumn{1}{r|}{0.095} & 0.092 & 0.075 & 0.067 \\
 & 400 & 0.053 & 0.055 & \multicolumn{1}{r|}{0.056} & 0.055 & 0.052 & \multicolumn{1}{r|}{0.054} & 0.085 & 0.067 & \multicolumn{1}{r|}{0.079} & 0.092 & 0.069 & 0.073 \\
100 & 50 & 0.124 & 0.143 & \multicolumn{1}{r|}{0.157} & 0.119 & 0.128 & \multicolumn{1}{r|}{0.143} & 0.190 & 0.201 & \multicolumn{1}{r|}{0.194} & 0.106 & 0.091 & 0.093 \\
\hline \hline 
\end{tabular}
\end{table}}

{
\renewcommand{\arraystretch}{0.9}
\setlength{\tabcolsep}{5pt}
\begin{table}[H]
\scriptsize 
\caption{Results for Corollary \ref{Coro1} ($\rho_u=0.5$ and $\delta_\epsilon =0.5$)}\label{Table10}
\hspace*{0.5cm}\begin{tabular}{llrrrrrrrrrrrr}
\hline\hline
 &  & \multicolumn{6}{c}{DWB} & \multicolumn{3}{c}{\multirow{2}{*}{MBB}} & \multicolumn{3}{c}{\multirow{2}{*}{Traditional}} \\
 &  & \multicolumn{3}{c}{Bartlett} & \multicolumn{3}{c}{Trapezoidal} & \multicolumn{3}{c}{} & \multicolumn{3}{c}{} \\
$N$ & $T$ & $0.8\ell_B$ & $\ell_B$ & $1.2\ell_B$ & $0.8\ell_T$ & $\ell_T$ & $1.2\ell_T$  &$0.8\ell_M$ & $\ell_M$ & $1.2\ell_M$ & $s_1^2$ & $s_2^2$ & $s_3^2$ \\
50 & 50 & 0.126 & 0.143 & \multicolumn{1}{r|}{0.155} & 0.119 & 0.129 & \multicolumn{1}{r|}{0.139} & 0.211 & 0.225 & \multicolumn{1}{r|}{0.214} & 0.150 & 0.134 & 0.104 \\
 & 100 & 0.084 & 0.091 & \multicolumn{1}{r|}{0.096} & 0.079 & 0.087 & \multicolumn{1}{r|}{0.085} & 0.159 & 0.153 & \multicolumn{1}{r|}{0.139} & 0.152 & 0.129 & 0.099 \\
 & 200 & 0.067 & 0.065 & \multicolumn{1}{r|}{0.074} & 0.063 & 0.063 & \multicolumn{1}{r|}{0.068} & 0.111 & 0.112 & \multicolumn{1}{r|}{0.105} & 0.136 & 0.121 & 0.079 \\
 & 400 & 0.053 & 0.054 & \multicolumn{1}{r|}{0.056} & 0.055 & 0.054 & \multicolumn{1}{r|}{0.050} & 0.092 & 0.082 & \multicolumn{1}{r|}{0.085} & 0.135 & 0.113 & 0.076 \\
100 & 50 & 0.134 & 0.150 & \multicolumn{1}{r|}{0.165} & 0.125 & 0.136 & \multicolumn{1}{r|}{0.148} & 0.189 & 0.226 & \multicolumn{1}{r|}{0.205} & 0.150 & 0.127 & 0.107 \\
\hline \hline 
\end{tabular}
\end{table}}

}
\end{document}